\documentclass[aps,prd,preprintnumbers,nofootinbib,12pt]{revtex4-1}
\usepackage{stackrel}
\usepackage{soul}
\usepackage{bm}
\usepackage{latexsym}
\usepackage{dcolumn}
\usepackage{amsmath,amsfonts,amssymb}
\usepackage{graphicx,epsfig}
\usepackage{xcolor}
\usepackage{psfrag}
\usepackage{dblfloatfix}    
\usepackage{amsthm}
\usepackage[utf8x]{inputenc}  
\usepackage[normalem]{ulem}
\usepackage{tikz}
\usepackage{oplotsymbl}
\usepackage{cancel}
\usepackage{setspace}
\usepackage{graphicx,color}
\usepackage{dcolumn}
\usepackage{bm}
\usepackage[T1]{fontenc}
\definecolor{ForestGreen}{RGB}{50,240,50}
{}{}{}
\def\be{\begin{equation}}
\def\ee{\end{equation}}
\def\bea{\begin{eqnarray}}
\def\eea{\end{eqnarray}}

\def\gby{g_{_{BY}}}
\def\gyb{g_{_{YB}}}
\def\gbl{g_{_{BL}}}

\def\vevs{{\it vev's}}
\def\hsm{h_{\mathrm{SM}}}
\def\mzprime{M_{Z'}}
\def\ntrli{\chi_i^0}

\def\ntrlone{\chi_1^0}

\def\gammazprime{\Gamma_{Z'}}
\def\gammazprimell{\Gamma_{Z'}^{(\ell^+ \ell^-)}}
\def\gammazprimeff{\Gamma_{Z'}^{(f \bar{f})}}
\def\gammazprimeww{\Gamma_{\scriptscriptstyle{Z'}}^{\scriptscriptstyle{(WW)}}}

\def\thetaprime{\theta'}
\def\sthetaprime{s_{\theta'}}
\def\cthetaprime{c_{\theta'}}
\def\thetaprimemax{\theta'_{\mathrm{max}}}
\def\lsim{\raise0.3ex\hbox{$\;<$\kern-0.75em\raise-1.1ex\hbox{$\sim\;$}}}
\def\gsim{\raise0.3ex\hbox{$\;>$\kern-0.75em\raise-1.1ex\hbox{$\sim\;$}}}

\usepackage{pstricks}
\begin{document}
\setlength{\abovedisplayskip}{7pt}
\setlength{\belowdisplayskip}{7pt}
%
%
\title{The $Z'$-boson of the $B$--$L$ Supersymmetric Standard Model \\
and its Large Hadron Collider Searches}
\author{Waleed Abdallah}
\email{awaleed@sci.cu.edu.eg}
\affiliation{Department of Mathematics, Faculty of Science, Cairo University, Giza 12613, Egypt}
\author{AseshKrishna Datta}
\email{asesh@hri.res.in}
\affiliation{Harish-Chandra Research Institute, A CI of Homi Bhabha National Institute (HBNI), Chhatnag Road, Jhunsi, Prayagraj (Allahabad), UP, 211019, India}
\author{Stefano Moretti}
\email{s.moretti@soton.ac.uk}
\affiliation{Department of Physics and Astronomy, Uppsala University, Box 516, SE-751 20 Uppsala, Sweden}
\affiliation{School of Physics and Astronomy, University of Southampton, Highfield, Southampton SO17 1BJ, UK}
\date{\today}
%
\begin{abstract}
We discuss how the $Z'$-boson of the $B$--$L$ Supersymmetric (SUSY) Standard 
Model (BLSSM) could evade the current lower bound of around 5~TeV on the mass 
of such a resonance (of sequential nature) from the Large Hadron Collider (LHC) by a significant 
margin. This happens when the experimental sensitivities are critically 
impaired as the $Z'$-boson becomes `fat' or develops some leptophobia or 
possesses an optimally large decay Branching Ratio (BR) to BLSSM-specific states 
(including the SUSY ones) or when some or all of these are at play simultaneously. 
We describe how such a $Z'$-boson could acquire there features while still 
respecting the non-negotiable precision constraints from the LEP and the 
SLC experiments running at the $Z$-pole as well as those from the multi-purpose  
experiments at the LHC that search for such a resonance. We explore the 
interplay of the aforementioned phenomena and identify the regions of the BLSSM 
parameter space that give rise to the described situation by carrying out a thorough 
scan of it. We find that $\mzprime$ masses as low as 2.24~TeV may still be 
allowed in the BLSSM under favorable circumstances.
\end{abstract}
\vspace*{-1cm}
\maketitle
%
\section{Introduction}
The search for new resonances has been a primary physics programme at 
 colliders for a long time. The Large Hadron Collider (LHC) is no exception. Of such possible new resonances, the hunt for a neutral, heavy vector boson, a 
so-called $Z'$-boson, a cousin to the Standard Model (SM) $Z$-boson in the Electro-Weak (EW) sector, merits a 
particular mention. For, such an excitation has a generic presence in scenarios 
Beyond the Standard Model (BSM) of particle physics that are augmented with an extra $U(1)$ 
symmetry. A widely adopted prototype benchmark scenario is known as the 
Sequential SM (SSM) in which the $Z'$-boson has the same couplings to the SM 
fermions as the SM $Z$-boson~\cite{Altarelli:1989ff}.

A comprehensive list of scenarios in which such (or similar) resonances are 
present includes the ones motivated by extended gauge models like the
$E_6$-based ones~\cite{London:1986dk, Hewett:1988xc, King:2020ldn}, Composite 
Higgs Models (CHMs)~\cite{DeCurtis:2011yx,Barducci:2012kk} and those with Heavy Vector Triplets 
(HVT)~\cite{delAguila:2010mx,deBlas:2012qp,Pappadopulo:2014qza}, the minimal 
$Z'$ models~\cite{Salvioni:2009mt}, the Generalized Sequential Model 
(GSM)~\cite{Accomando:2010fz}, frameworks with extra spatial dimensions like 
the Randall-Sundrum 
(RS) scenario~\cite{Kelley:2010ap,Randall:1999ee,Randall:1999vf} and the non-Minimal 
Universal Extra-Dimensions 
(nMUED) framework~\cite{Flacke:2008ne,sniyogi-thesis,Datta:2015zra}. The list further
includes various types of Grand Unified Theories (GUTs), Supersymmetric (SUSY) or not, viz., scenarios with a generalized Left-Right ($LR$) 
symmetry or with a $B$--$L$ symmetry~\cite{Leike:1998wr,Langacker:2008yv}.   

It is important to note here that both phenomenological and experimental 
analyses (until very recently) of such a massive $Z'$ resonance have been 
predominantly carried out by assuming it to be a narrow one, i.e., one which  
(total decay) width-to-mass ratio 
($\gammazprime \over \mzprime$), a conventional measure of broadness of a 
resonance, is within a few percent ($\lesssim 5\%$). This is true for quite a 
few of the aforementioned scenarios (e.g., the SSM, $E_6$, HVT and GUT-inspired $LR$-
symmetry based ones, etc.). Hence the well-known Narrow Width Approximation 
(NWA), which simplifies the analysis and interpretation of the results, has 
been extensively adopted not only in phenomenological studies  but also in 
 experimental searches at the LHC (for such a resonance in these scenarios), 
in the Drell-Yan (DY) dilepton ($e^+e^-$ and $\mu^+\mu^-$) and diboson ($W^+W^-$) final 
states (see, for 
example,
Refs.~\cite{ATLAS:2019erb, ATLAS:2019nat, ATLAS:2019fgd, ATLAS:2020fry, CMS:2018ipm, CMS:2018mgb, CMS:2019gwf, CMS:2021ctt, CMS:2021klu}%
\footnote{Notable exceptions are a few recent LHC analyses, in
Refs.~\cite{ATLAS:2019erb, ATLAS:2019fgd, CMS:2018ipm, CMS:2018mgb, CMS:2019gwf}, 
in which resonances with ${\gammazprime \over \mzprime} \geq 10\%$, and up to 
55\%~\cite{CMS:2019gwf}, have been considered.}).
In the SSM, these have put a (model-dependent) lower bound of 
$\mzprime \gtrsim 5$~TeV, by analyzing the dilepton and $W^+W^-$ final 
states, while the bound obtained by studying the background-prone dijet final 
state is naturally much weaker.

However, it has been shown in the 
literature~\cite{Berdine:2007uv,Kauer:2007zc,Uhlemann:2008pm,Accomando:2010fz} 
that even when the width of a resonance is in the NWA regime, the Finite Width 
(FW) effects should be taken into account to draw correct conclusions, in 
particular, when these could potentially broaden the relevant (viz., dilepton, dijet, etc.) invariant mass distributions (about the resonant pole) 
to the extent that those fall well beyond experimental resolutions. Further, it was 
realized that the final states (viz., dilepton, dijet, etc.), in which a BSM 
resonance is predominantly searched for, might receive contributions from SM 
processes (with $\gamma, Z, W^\pm$ mediation) through interference, and hence, 
generally speaking, the new physics contribution is not independent of the 
known (read, SM) physics~\cite{Accomando:2013sfa,Accomando:2015cfa}.
In particular, the CMS collaboration~\cite{CMS:2018ipm, CMS:2021ctt}, by 
adopting specific prescriptions advocated in
Refs.~\cite{Accomando:2010fz, Accomando:2013sfa} to avoid a model-dependent 
interference effect between the $Z'$ mediated process and the SM background 
channels mediated by the photon and the $Z$-boson, has reached a set of 
conclusions (including a lower bound on $\mzprime$) that are somewhat
model-independent. A model-independent approach to constrain the $Z'$-sector 
has also been advocated in Ref.~\cite{Bandyopadhyay:2018cwu} using unitarity 
and direct searches of the $Z'$-boson at the LHC.

Interestingly enough, $\gammazprime \over \mzprime$ could turn large in various 
new physics scenarios. This happens when the $Z'$-boson either possesses 
stronger couplings to states it decays to, or has a larger set of states to 
decay to, or both. It is well-known that for such `fat' resonances, the NWA 
ceases to hold, and experimental strategies that target narrow resonances 
generically lose sensitivity when deployed in search for such broader ones. 
Phenomenological studies of broad resonances have been taken up in well-known 
theoretical schemes, like, for example,  Technicolor 
scenarios~\cite{Belyaev:2008yj}, 
frameworks with colored resonances like the axigluon and coloron 
states~\cite{Choudhury:2011cg},  CHMs~\cite{Barducci:2012kk}, the RS
scenario~\cite{Kelley:2010ap} or the nMUED one~\cite{sniyogi-thesis,Datta:2015zra} as well as the generic frameworks where 
the $Z'$-boson couples to the third generation fermions in a way different from 
the other fermions and are not constrained as strongly as the latter ones 
(and, hence, can be large)~\cite{Hayreter:2019dzc}.

In the present work, we discuss some phenomenological aspects of the $Z'$-boson 
of the well-known $B$--$L$ SUSY SM (BLSSM)
scenario~\cite{Khalil:2007dr, Kikuchi:2008xu,FileviezPerez:2010ek,Fonseca:2011vn},
which is the minimal $B$--$L$ extension of the once very popular Minimal 
SUSY SM (MSSM), with the SM gauge group being augmented by a 
$U(1)_{B-L}$ symmetry. The scenario finds its major motivations in its ability 
to explain the non-vanishing neutrino 
masses~\cite{Khalil:2006yi,Emam:2007dy,Abbas:2007ag,Huitu:2008gf} 
along with a host of other phenomenologically interesting implications. 
In addition, the scenario exhibits better compatibility with the available 
experimental data compared to the MSSM.

We note that the combined decay width of the $Z'$-boson of the BLSSM scenario, 
when its major (canonical) decay modes, viz., dijet,  
dilepton (including the neutrinos) and  $W^+W^-$, are included, generally never becomes sufficiently large to make the 
$Z'$-boson broad enough (i.e., ${\gammazprime \over \mzprime} \gtrsim 5\%$), which would
undermine the applicability of the NWA. However, given that the BLSSM scenario 
features a mixing between the SM $Z$-boson and the $Z'$-boson and that the 
partial width of the $Z'$-boson decaying to $W^+W^-$ ($\gammazprimeww$) is 
directly proportional to the mixing angle ($\thetaprime$), the same can be 
enhanced in certain regions of the BLSSM parameter space, where the mixing 
$\thetaprime$ is large, but still respects the inescapable upper bound it 
draws from various analyses of the precision data from the LEP and SLC
experiments~\cite{ALEPH:2005ab,Erler:2009jh,ALEPH:2013dgf,ParticleDataGroup:2024cfk,ALEPH:2006bhb} at the $Z$-pole. Such an enhanced $\gammazprimeww$ 
could eventually turn the $Z'$-boson resonance significantly broad. This would not 
only help evade the reference lower bounds on $\mzprime$ (in the SSM) from the 
LHC experiments that assume a narrow $Z'$-boson resonance, but would also relax the 
upper bounds on $\thetaprime$ that are extracted from those analyses by various 
phenomenological studies~\cite{Pankov:2019yzr, Osland:2020onj, Osland:2022ryb}.

Further, we discuss that the combined (partial) width of the $Z'$-boson in all 
other modes that include the SM and BLSSM Higgs bosons and neutrinos as well as  
various BLSSM-specific SUSY excitations could never become large enough to 
render the $Z'$-boson broad enough. However, in certain regions of the BLSSM 
parameter space, the $Z'$-boson could still evade the current LHC bounds by 
exploiting other possibilities that the scenario offers, in particular, when 
the $Z'$-boson can still be considered a narrow resonance. Of these, the 
presence of a certain degree of possible leptophobia in the $Z'$-boson 
immediately weakens the yield in the dilepton final state at the LHC and 
thus would relax the current lower bound (the strongest one, in the SSM) on 
$\mzprime$ derived by studying the same. Also, a sizable intrinsic (partial) 
width of the $Z'$-boson to the BLSSM-specific states (in particular,
EWinos) could sufficiently erode the rates in both the  dilepton and 
the $W^+W^-$ final states, thus relaxing the lower(upper) bounds on
$\mzprime (\thetaprime)$ obtained in these final states.

It should also be noted that the already known FW interference 
effects~\cite{Berdine:2007uv,Kauer:2007zc,Uhlemann:2008pm,Accomando:2010fz}, 
which could be destructive as well, might now turn somewhat more drastic in 
altering the crucial kinematic distributions, like the invariant mass of the 
particles to which the putative $Z'$ resonance decays. In addition, there may 
be other dynamical effects and their possible interplays with the kinematics, 
which should be taken into account. These include the presence of other 
possible decays of the resonance to different BSM states that a given scenario 
(here, the BLSSM) offers, various new couplings that, independently or in 
conjunction, might enhance the resonant decay width in the first place, 
hence, they could critically alter the production rate in a given final state (say, 
dilepton or diboson), a secondary (but somewhat compensating) effect on the 
said production rate coming from an altered~\cite{Datta:2015zra} enhanced 
propagator drawing on a larger total width, etc. 

Operationally, our goal is then to look for an even larger $\gammazprime$ in a 
legitimate manner for a given $\mzprime$ and to study the combined 
implications of the effects described above. In short, on the one hand, 
broadening of a resonant peak would erode the efficiency of the specifically 
tailored invariant mass cuts (loss of acceptance)%
\footnote{A related issue is that the signal from a broad resonance would 
affect the background estimation using the sliding-window fit that is adopted 
in searches for a narrow resonance~\cite{ATLAS:2019fgd}.}
used in searches for a narrow resonance in the conventional final states (viz., 
 dilepton and dijet ones). Also, such a broadening would result in decreased 
cross sections for the said final states due to an enhanced propagator effect. 
These effects, when working in tandem, might significantly weaken  the current 
lower bound on the mass of such a resonance ($\mzprime \gtrsim 5$~TeV) or, for 
that matter, the overall reach in $\mzprime$ in the respective final states. On 
the other hand, the rate in the $W^+W^-$ final state, which is directly related 
to the dominant contribution to an enhanced $\gammazprime$, i.e., from 
$\Gamma_{Z' \to W^+W^-}$ ($\gammazprimeww$), would be reinforced. As 
$\gammazprime$ increases, this could, under favorable circumstances, 
overcompensate for the loss effects mentioned earlier, and could take the lead 
role in putting the most stringent (even though, relaxed) lower bound on 
$\mzprime$ via the $W^+W^-$ final state.

The paper is organized as follows. In Sec.~\ref{Sec:2}, we discuss the 
theoretical setup of the BLSSM scenario with an emphasis on its $Z'$ sector.
We follow this up with a discussion on the existing stringent bounds on the 
$Z'$ sector derived from various experimental results. In Sec.~\ref{Sec:3}, we 
present in detail the dependencies of various key observables on the input 
parameters of the scenario, followed by studies of how the cross sections of 
various final states involving a resonant $Z'$-boson vary with $\mzprime$ as 
$\gammazprime \over \mzprime$ varies. These are then used to estimate how the 
current experimental bounds on $\mzprime$ might get relaxed, in particular, 
when the $Z'$-boson possesses a modest decay BR to SUSY states. In Sec.~\ref{Sec:4}, we conclude. 
%
\section{The theoretical setup}
\label{Sec:2}
In this section, we first briefly present the general setup of the BLSSM 
scenario based on the gauge group
$SU(3)_c \times SU(2)_L \times U(1)_Y \times U(1)_{B-L}$. This is followed by 
discussions on its $Z'$ sector, the tiny but all-important (kinetic) mixing 
of the SM $U(1)_Y$ and  $U(1)_{B-L}$ gauge bosons and its dependence on various inputs
parameters. Thus, the strength of various interaction vertices of the
$Z'$-boson that are crucial for the present study.
%
\subsection{The general setup of the BLSSM scenario}
The BLSSM scenario under consideration contains the following fields in 
addition to the MSSM ones: three chiral right-handed superfields ($\hat{N}_i$), 
`$i$' being the fermion generation index, a vector superfield, $\hat{Z}'$, 
associated with gauge group $U(1)_{B-L}$, plus  two chiral SM-singlet Higgs 
superfields, $\hat{\eta}_1$, $\hat{\eta}_2$. The superpotential of this model 
is given by
\bea {W_{\mathrm{BLSSM}}} &=&
W_{\rm MSSM} + (Y_{\nu})_{ij}\hat{L}_i \hat{H}_2 \hat{N}^c_j
+ (Y_N)_{ij}\hat{N}^c_i\hat{\eta}_1\hat{N}^c_j + \mu'\hat{\eta}_1\hat{\eta}_2 \; ,
\label{super-potential-b-l}
\eea
where the second and third terms on the right-hand side are the BLSSM-specific 
trilinear terms involving the chiral superfields
$\hat N_i$, with $i,j \in \{1,2,3\}$ while  the last addend is the
$\mu^\prime$-term (analogous to the well-known $\mu$-term contained in 
$W_\mathrm{MSSM}$) involving $\hat{\eta}_1$ and $\hat{\eta}_2$. The
$SU(2)_L \times U(1)_Y$ and $U(1)_{B-L}$ charges of various superfields 
appearing in $W_{\mathrm{BLSSM}}$ are shown in Table~\ref{ub-l-charge} with the 
$SU(2)$ doublet chiral quark superfields $\hat{Q}_i$ and the $SU(2)$ singlet ones, 
$\hat{U}_i$ and $\hat{D}_i$ (already appearing in $W_\mathrm{MSSM})$, along with 
the $SU(2)$ doublet chiral lepton superfield, $\hat{L}_i$, and the Higgs 
superfields $\hat{H}_1$ and $\hat{H}_2$.
\vskip 5pt
\begin{table}[h]
\begin{center}
\bgroup
\def\arraystretch{1.5}
\begin{tabular}{|c|c|c|c|c|c|c|c|c|c|c|} \hline
 & $\hat{L}_i$ & $\hat{N}^c_i$ & $\hat{E}^c_i$ & $\hat{Q}_i$ & $\hat{U}^c_i$ & $\hat{D}^c_i$ & $\hat{H}_1$ & $\hat{H}_2$ & $\hat{\eta}_1$  & $\hat{\eta}_2$
  \\ \hline
{ $SU(2)_L\times U(1)_Y$}
 & $ ({\bf 2}, -\frac{1}{2})$  &  $({\bf 1}, 0)$  &  $({\bf 1}, 1)$
 & $({\bf 2}, \frac{1}{6})$&  $({\bf 1},-\frac{2}{3})$
 &  $({\bf 1}, \frac{1}{3})$& $({\bf 2}, -\frac{1}{2})$& $({\bf 2}, \frac{1}{2})$ &
$({\bf 1}, 0)$ &  $({\bf 1}, 0)$
  \\ \hline
 $U(1)_{B-L}$ & $-\frac{1}{2}$  &  $\frac{1}{2}$  &  $\frac{1}{2}$
 & $\frac{1}{6}$&  $-\frac{1}{6}$&  $-\frac{1}{6}$& 0& 0 & $-1$ & 1
   \\ \hline
\end{tabular}
\egroup
\caption{
\label{ub-l-charge}
The $SU(2)_L \times U(1)_Y$ and $U(1)_{B-L}$ charges of the chiral superfields 
of the BLSSM.
}
\end{center}
\end{table}
\noindent
The corresponding soft SUSY-breaking Lagrangian density can be written as the 
sum of the MSSM component, $-{\cal L}^{\rm (soft)}_\mathrm{MSSM}$, and the 
BLSSM-specific terms in the following way:
\begin{eqnarray}
- {\cal L}^{\rm (soft)}_\mathrm{BLSSM} &=&
- {\cal L}^{\rm (soft)}_\mathrm{MSSM} + {{m}}_{\tilde{N}ij}^{2}{\tilde{N}}_{i}^{c*}{\tilde{N}}_{j}^{c} +  m^2_{\eta_1} \vert{\eta_1}\vert^2 +
 m^2_{\eta_2}\vert{\eta_2}\vert^2 \\ \nonumber 
 &+& \left[ Y_{\nu ij}^{A}{\tilde{L}}_{i}
{\tilde{N}^c}_{j} H_2 + Y_{N ij}^{A}{\tilde{N}}_i^{c}
{\tilde{N}}_j^{c}\eta_{1}
+ B \mu^\prime \eta_1 \eta_2 \,
+ \frac{1}{2} M_{B'} \lambda_{\tilde{B}'} \lambda_{\tilde{B}'}
+ M_{BB'} \lambda_{\tilde{B}} \lambda_{\tilde{B}'}
+ \mathrm{h.c.}
\right],%
\label{Lsoft}%
\end{eqnarray}%
where an overhead `\textasciitilde' for any given state denotes the scalar 
component of the corresponding chiral matter superfield. The scalar components
of the Higgs superfields $\hat{H}_{1,2}$ and $\hat{\eta}_{1,2}$ are denoted as 
$H_{1,2}$ and $\eta_{1,2}$ (the so-called `bileptons'), respectively. The field
$\lambda_{\tilde{B}'}$ is the fermionic component of the vector superfield 
which is $\tilde{B}^\prime$ ($B'$ino~\cite{Abdallah:2017gde} or 
BLino~\cite{OLeary:2011vlq}), in addition to the bino and wino fields of the 
MSSM. Also, $Y_{f ij}^{A}\equiv(Y_f A_f)_{ij}$, where $f=\nu,{\cal N}$ (i.e., SM and heavy neutrinos).

The gauge group $SU(3)_c \times SU(2)_L \times U(1)_Y \times U(1)_{B-L}$ breaks 
down to the coveted group $SU(3)_c \times U(1)_{\rm em}$ as the real components 
of the electrically neutral Higgs and bilepton fields acquire vacuum expectation values \big(\vevs,
$\{v_1,v_2 \}$ and $\{ v'_1, v'_2\}$\big), as given below, thus breaking both 
EW and $B$--$L$ symmetries, respectively:
\begin{subequations}
\begin{align}
H_1 &= {1 \over \sqrt{2}} \big(H_{1R} + v_1 + i H_{1I} \big)
\hskip 5pt & H_2 &= {1 \over \sqrt{2}} \big(H_{2R} + v_2 + i H_{2I} \big) \\
\eta_1 &= {1 \over \sqrt{2}} \big(\eta_{1R} + v'_1 + i \eta_{1I} \big)
\hskip 5pt & \eta_2 &= {1 \over \sqrt{2}} \big(\eta_{2R} + v'_2 + i \eta_{2I} \big) \;,
\end{align}
\end{subequations}
where the suffix $R(I)$ represents the real(imaginary) components of the 
respective fields.
%
\vspace{-0.3cm}
\subsection{Gauge kinetic mixing and the $Z'$ sector}
\label{subsec:zprime-sector}
The gauge kinetic mixing present between the two Abelian gauge groups, $U(1)_Y$ 
and $U(1)_{B-L}$, could have non-trivial consequences for the phenomenology of 
the $Z'$-boson of the BLSSM. To study these, it is convenient to work with 
modified covariant derivatives which absorb such mixing effects. This approach
is equivalent to~\cite{Fonseca:2011vn} explicitly dealing with the field 
strength tensors appearing in the gauge kinetic mixing terms
$-\chi_{ab} \hat{F}^{a,\mu\nu} \hat{F}_{b,\mu\nu}$ with $a \neq b$. The 
relevant covariant derivative would be of the form
$D_\mu =\partial_\mu - i Q^T_\phi G B$, with $Q_\phi$ being an array comprising 
the charges of the field `$\phi$' under the two Abelian gauge groups whereas  `$G$' 
is the coupling matrix given by
\be
G= 
\begin{pmatrix}
g_{_{YY}} \quad \gyb \\
\gby \quad g_{_{BB}}
\end{pmatrix} \, ,
\label{eqn:g-couplings}
\ee
with the array `$B$' containing the corresponding gauge fields
$B \equiv (B_\mu^{(Y)}$ and $B_\mu^{'(B-L)})^T$. As observed in 
Ref.~\cite{Fonseca:2011vn}, as long as these Abelian gauge groups are not 
broken, one has the freedom to rotate the gauge fields to any convenient basis 
without affecting their charge assignments while absorbing the rotation in a 
redefinition of the gauge couplings of Eq.~(\ref{eqn:g-couplings}). Such a 
convenient basis is the one where the $\{2,1 \}$ element of the transformed 
matrix `$\widetilde{G}$' vanishes, such that only the doublet Higgs bosons 
contribute to the $SU(2)_L \otimes U(1)_Y$ block of the $3 \times 3$ gauge 
boson mass matrix and the influence of the singlet Higgs bosons of the BLSSM, 
$\eta_1$ and $\eta_2$, on this sector arises only through the off-diagonal 
terms of said matrix. Operationally, such a basis proves to be convenient 
in addressing EW precision data. The relevant transformation 
takes the form
\be
G=
\begin{pmatrix}
g_{_{YY}} \quad \gyb \\
\gby \quad g_{_{BB}}
\end{pmatrix}
~~ \Longrightarrow ~~ 
\widetilde{G} =
\begin{pmatrix}
  g'_{_{YY}} & g'_{_{YB}} \\
  g'_{_{BY}} & g'_{_{BB}} \, \\
\end{pmatrix}
=
\begin{pmatrix}
  g_1& \tilde{g} \\
  0 & \gbl  \\
\end{pmatrix} \, ,
\ee
where $g_1$ is identified as the usual hypercharge coupling of the SM, which is 
now given by 
\be 
g_1 = \frac{g_{_{YY}} g_{_{BB}} - \gyb \gby}{\sqrt{g_{_{BB}}^2 + \gby^2}} \, ,
\label{eqn:g1}
\ee
while
\be
\gbl =\sqrt{g_{_{BB}}^2 + \gby^2}\, , \qquad ~~\tilde{g} = \frac{\gyb g_{_{BB}} + \gby g_{_{YY}}}{\sqrt{g_{_{BB}}^2 + \gby^2}} \, .
\label{eqn:gbl-gtilde}
\ee
Note that, in Eq.~(\ref{eqn:g1}), the numerator on the right-hand side is 
always positive since the product $\gyb \gby$ is a very small quantity, with 
$g_{_{YY}}$ and $g_{_{BB}}$ both being positive, by definition. Thus, $g_1$ is 
also a positive quantity given that the denominator, which is identified as 
$\gbl$ in Eq.~(\ref{eqn:gbl-gtilde}), is, by construction, a positive definite 
quantity.
Note that, for $\gby \to 0$, one finds 
$g_1 \to g_{_{YY}}$, \, $\gbl \to g_{_{BB}}$ \, and \, $\tilde{g}=\gyb$, a 
limit that has been frequently adopted in the phenomenological analyses of the 
BLSSM scenario (see, for example
Ref.~\cite{OLeary:2011vlq,Krauss:2012ku,Abdallah:2014fra}).
In this context, it is important to note that in the present work we deviate 
from this scheme and explore the role of a finite $\gby$ in triggering a large 
decay width for the $Z'$-boson, which would have important phenomenological 
implications for its search at the LHC. Towards this, guided by the relations 
presented in Eqs.~(\ref{eqn:g1}) and~(\ref{eqn:gbl-gtilde}), we fix our choice 
of the fundamental gauge couplings from the BLSSM sector as $\gbl$, $\gby$ and 
$\gyb$, which we will vary to explore the scenario.%
\footnote{Nonetheless, given the small values (${\cal O}(10^{-3})$) to which 
$\gby$ is restricted on phenomenological grounds (which can still significantly 
modify the decay width of the $Z'$-boson), identifying $\gbl$ with $g_{_{BB}}$ 
and $\gyb$ with $\tilde{g}$ would be acceptable.}  

As for the mass-squared matrix for the neutral gauge bosons, it is given, in
the interaction basis $\{B^\mu, W_3^\mu\,, B'^\mu \}$, by
\bea {\cal M}^2_0&=&\left(
  \begin{array}{ccc}
    \frac{1}{4}g_1^2 v^2 + g^2_{_{BY}} v'^2 ~&~ -\frac{1}{4}g_1 g_2 v^2 ~&~ \frac{1}{4} g_1 g_{_{YB}} v^2+\gbl \gby v'^2 \\
    -\frac{1}{4}g_1 g_2 v^2 ~&~ \frac{1}{4}g_2^2 v^2 ~&~ -\frac{1}{4} g_2 \gyb v^2 \\
    \frac{1}{4} g_1 \gyb v^2+\gbl \gby v'^2 ~&~ -\frac{1}{4} g_2 \gyb v^2 ~&~ \frac{1}{4}\gyb^2 v^2+\gbl^2 v'^2\\
  \end{array}
\right) \, .
\eea
The rotation matrix connecting the interaction and the physical
(\{$A^\mu$, $Z^\mu$, $Z'^\mu$ \}) bases can be written in terms of two mixing 
angles, $\theta_W$ (the Weinberg angle) and $\thetaprime$ (the $Z$--$Z'$ mixing angle), as
 \be \left(
      \begin{array}{c}
        A^\mu \\
        Z^\mu \\
        Z'^\mu \\
      \end{array}
    \right)
=\left(
  \begin{array}{ccc}
    c_{\theta_W} & s_{\theta_W} & 0 \\
    -s_{\theta_W} c_{\thetaprime} & c_{\theta_W} c_{\thetaprime} & s_{\thetaprime} \\
    s_{\theta_W} s_{\thetaprime} & -c_{\theta_W} s_{\thetaprime} & c_{\thetaprime} \\
  \end{array}
\right)\left(
      \begin{array}{c}
        B^\mu \\
        W_3^\mu \\
        B'^\mu \\
      \end{array}
    \right),
    \label{matrix}
\ee
\vskip 5pt
\noindent
where $s_{\theta_W} (\sthetaprime) \equiv \sin\theta_W (\sin\thetaprime)$ and
$c_{\theta_W} (\cthetaprime) \equiv \cos\theta_W (\cos\thetaprime)$ with \\
\be \tan{2\thetaprime}\simeq\frac{2 \gyb\sqrt{g_1^2+g_2^2}+16 \gby \gbl (\frac{v'}{v})^2 s_{\theta_W}        }{\gyb^2+4(\frac{v'}{v})^2 \gbl^2 -g_1^2-g_2^2} \, .
\label{eqn:thetaprime1}
\ee
\vskip 5pt
\noindent
The angle $\thetaprime$ refers to the mentioned `mass mixing' between the 
states $Z$ and $Z'$, which is to be contrasted with the gauge kinetic mixing 
discussed earlier. However, note that the effects of the latter kind of mixing 
sneak into $\thetaprime$ through the quantities $\gyb$ and $\gby$, as is seen 
in Eq.~(\ref{eqn:thetaprime1}).
In the physical basis, one finds the masses of the $Z$- and $Z'$-bosons, $M_Z$ 
and $\mzprime$, respectively, to be as follows:
\be
M_Z^2 = \frac{1}{4} \, (g_1^2 +g_2^2) \, v^2 \quad \mathrm{and} \quad M_{Z'}^2 = \gbl^2 v'^2 + \frac{1}{4} \,\gyb^2 v^2,
\label{eqn:mzmzprime}
\ee
with
\be
v=\sqrt{v_1^2+v_2^2} \; (\simeq 246 ~ \text{GeV}) \quad \mathrm{and} \quad v'=\sqrt{v'^2_1+v'^2_2}
\label{eqn:vvprime}
\ee
being the \vevs~of the neutral Higgs fields specific to the 
MSSM and the BLSSM sectors, respectively, where
$\langle H_{1,2}\rangle=\frac{v_{1,2}}{\sqrt2}$ and 
$\langle\eta_{1,2}\rangle=\frac{v'_{1,2}}{\sqrt2}$.
Note that, from Eq.~(\ref{eqn:mzmzprime}), $\mzprime$ goes essentially as
$\gbl v'$, given that the quantity $\gyb v$ is much smaller. Requiring
$\mzprime \sim {\cal O}(\mathrm{TeV})$ and that
$v' \sim {\cal O}(\mathrm{TeV})$ as well~\cite{Khalil:2007dr}, $\gbl$ should be 
of ${\cal O}(0.1)$.

Further, using the relations in Eq.~(\ref{eqn:mzmzprime}), the expression for
$\tan 2\thetaprime$ in Eq.~(\ref{eqn:thetaprime1}) can be rewritten in terms of $M_Z$ and $\mzprime$ as
\be
\tan{2\thetaprime}\simeq\frac{\gyb M_Z v+4g_{_{BY}} \gbl^{-1}s_{\theta_W} M_{Z'}^2 }{M_{Z'}^2-M_Z^2} \, .
\label{eqn:thetaprime2}
\ee
The pure BLSSM limit (i.e., vanishing gauge kinetic mixing) is trivially 
realized for $\gby$=$\gyb$=0 as $\thetaprime$ (i.e., the $Z$--$Z'$ mixing) 
vanishes. The converse is generally not true. For, a cancellation between the 
terms in Eq.~(\ref{eqn:thetaprime2}) for finite $\gyb$ and $\gby$ cannot be 
ruled out, since, in general, both $\gyb$ and $\gby$ could take any sign. 
However, to comply with the generic upper bound on $\thetaprime$ ($<$ a few 
$10^{-3}$~radian), as derived from the precision studies on the properties of 
the SM $Z$-boson by the LEP and SLC 
experiments~\cite{ALEPH:2005ab, ALEPH:2006bhb, Erler:2009jh, ALEPH:2013dgf, ParticleDataGroup:2024cfk}, one requires, in Eq.~(\ref{eqn:thetaprime2}),
$|\gby \gbl^{-1}|<{\cal O} (10^{-4})$.%
\footnote{For $\mzprime \sim {\cal O}(100 \, \text{GeV})$, a cancellation 
between the terms in the numerator of Eq.~(\ref{eqn:thetaprime2}) is required 
to obtain a sufficiently small $\thetaprime$. Note that, $\gbl$ being positive, 
this would then require a relative sign between $\gyb$ and $\gby$, while both 
of them could now take suitably (and simultaneously) large values. We exploit 
this fact subsequently in the present work.}
In Sec.~\ref{subsec:bounds}, we take a critical look into these bounds, as they
are central to the present work. Further, note that the parameter $\gby$ 
actively controls the magnitude of $\thetaprime$, given as 
$\gbl \sim {\cal O}(0.1)$.
%
\subsection{Other sectors of the BLSSM}
\label{subsec:other-sectors}
Given that the non-SM decays of a resonant $Z'$-boson to various states of the 
BLSSM scenario are central to the present work, we outline them in this 
subsection.
\subsubsection{The Higgs sector}
\label{subsubsec:higgs-sector}
It has been shown in Refs.~\cite{Khalil:2007dr,Burell:2011wh} that both EW and 
$B$--$L$ symmetries can be broken radiatively in SUSY theories and that can 
happen at the TeV scale.

The $CP$-odd scalar sector of the BLSSM scenario under consideration has no
mixing between the $SU(2)$ doublets and the bileptons at tree level. Thus, once 
the EW and BLSSM symmetries break, the mass-squared matrix for this sector, in 
the basis
$\big\{ H_{1I}, H_{2I}, \eta_{1I}, \eta_{2I} \big\}$, takes a block-diagonal 
form and is given by
{\small
\begin{equation}
    {\cal M}^2_{\mathrm{odd}}=\left(
  \begin{array}{cccc}
    B\mu\tan\beta & B\mu & 0 & 0 \\
    B\mu & B\mu\cot\beta & 0 & 0 \\
    0 & 0 & B\mu'\tan\beta' & B\mu' \\
    0 & 0 & B\mu' & B\mu'\cot\beta'
  \end{array}
\right)
\quad \mathrm{with} \;\, \beta (\beta') = \tan^{-1} \frac{v_2}{v_1} (\tan^{-1} \frac{v'_2}{v'_1}) \, .
\label{eqn:cpodd-matrix}
\end{equation}
}%
The masses of the resulting two physical $CP$-odd Higgs states of the MSSM and 
BLSSM varieties, $A$ and $A'$, respectively, are given by
\bea
m_A^2 =\frac{2B\mu}{\sin 2\beta} \; , \quad m_{A'}^2=\frac{2B\mu'}{\sin 2\beta'} \; ,
\eea
while the two neutral Nambu-Goldstone bosons end up providing masses to the $Z$ 
and $Z'$-bosons.

Furthermore, in the $CP$-even sector, the $SU(2)$ doublet (i.e., MSSM) Higgs 
fields mix with the BLSSM Higgs (i.e., bileptons) fields thanks to the gauge kinetic terms. The mass-squared matrix for the $CP$-even scalars, in the basis
$\big\{ H_{1R}, H_{2R}, \eta_{1R}, \eta_{2R} \big\}$, is given by
{\footnotesize
\begin{align}
 M_{\mathrm{even}}^2 = \left( \begin{array}{cc} 
 			     M^2_{hH} ~ &~  M^2_{hh'} \\
                              M^{2^{{T}}}_{hh'} ~ &~  M^2_{h'{H}'} \end{array} \right)
 & \;\, \mathrm{with} \; \left\{
    {\tiny
    \begin{array}{l}
        M^2_{hH}=  
                  \left( 
              \begin{array}{cc}
               m^2_{A} s^2_{\beta} + \frac{1}{4}(g_1^2+g_2^2+\gyb^2) v^2_1 &-\frac{1}{2} m^2_{A} s_{2\beta} - \frac{1}{4}(g_1^2+g_2^2+\gyb^2) v_1 v_2 \\
                   \\
              -\frac{1}{2} m^2_{A} s_{2\beta} - \frac{1}{4}(g_1^2+g_2^2+\gyb^2) v_1 v_2 ~~&~~ m^2_{A} c^2_{\beta} + \frac{1}{4}(g_1^2+g_2^2+\gyb^2) v^2_2
             \end{array} 
                \right) \nonumber  \\ \\
        M^2_{hh'}=  \frac{1}{2} \big(\gyb \gbl+\gby g_1 \big) \left(\begin{array}{cc}
              v_1 v'_1 & -  v_1 v'_2\\
                    \\
               - v_2 v'_1 &  ~  v_2 v'_2 
              \end{array}\right) \\ \\
        M^2_{h'{H}'}=\left( \begin{array}{cc}
               m^2_{A'} c^2_{\beta'} + (\gbl^2+\gby^2) v'^2_1 &-\frac{1}{2} m^2_{A'} s_{2\beta'} - (\gbl^2+\gby^2) v'_1 v'_2 \\
                    \\
              -\frac{1}{2} m^2_{A'}s_{2\beta'} - (\gbl^2+\gby^2) v'_1 v'_2 ~~&~~ m^2_{A'} s^2_{\beta'} + (\gbl^2+\gby^2) v'^2_2 
              \end{array}\right) \; ,
    \end{array} }
   \right . \\
   \label{eqn:cpeven-matrix}
\end{align}
}%
where $M^2_{hH}$, $M^2_{h'H'}$ and $M^2_{hh'}$ are identified as the 
corresponding one in the MSSM sector, the one specific to the BLSSM scenario 
and the one that induces mixing among the MSSM and BLSSM Higgs fields, 
respectively. Thus, there are now four electrically neutral $CP$-even Higgs 
bosons of which two each are of the MSSM and BLSSM types, modulo the mixing.
%
\subsubsection{The EWino sector}
\label{subsubsec:ewino-sector}
The EWino sector of the BLSSM comprises the neutralino and chargino sectors 
of which the former is an augmented version of its MSSM counterpart in the 
presence of a new $U(1)_{B-L}$ gaugino ($\tilde{B}'$, or the BLino) and two new 
higgsino-like states ($\tilde{\eta}_1$ and $\tilde{\eta}_2$, or the bileptinos) 
in the BLSSM scenario. The $7\times 7$ neutralino mass matrix, in the basis
$\Big\{ \{\tilde{B},\tilde{W}^0, \tilde{H}_1, \tilde{H}_2\}, \{\tilde{B}', \tilde{\eta}_1, \tilde{\eta}_2 \} \Big\}$,
is then given in compact form by 
{\footnotesize
\begin{align}\label{neutralino-mass-matrix}
M_N=
\left(
\begin{array}{cccc|ccc}
 & & & & & & \\
 & \mathbf{X^{(4 \times 4)}_\mathbf{MSSM}} & & & & \mathbf{X_{Mix}^{(4\times 3)}} ~&~  \\
 & & & & & & \\
\hline
 & & & & M_{B'} & -\gbl v'_1 & \gbl v'_2 \\
 & \mathbf{{\Big( \mathbf{X_{Mix}^{(4\times 3)}} \Big)^\mathbf{T}}} & & & 
 -\gbl v'_1 & 0 & -\mu' \\
 & & & & \gbl v'_2 & -\mu' & 0
\end{array}
\right)
& \quad \mathrm{with} \quad
\mathbf{X_{Mix}^{(4\times 3)}} =
\left(
\begin{array}{ccc}
M_{BB'} & -\gby v'_1 & \gby v'_2 \\
 0 & 0 & 0 \\
-\frac{1}{2} \gyb v_1 & 0 & 0 \\
 \frac{1}{2} \gyb v_2 & 0 & 0
\end{array}
\right) \, ,
\end{align}
}%
where $M_{B'}$ and $M_{BB'}$ are the soft SUSY-breaking mass parameter for the 
$U(1)_{B-L}$ gaugino and the off-diagonal gaugino mass term appearing in the 
presence of gauge kinetic mixing between the gauge groups $U(1)_Y$ and
$U(1)_{B-L}$. On diagonalization of $M_N$ using a $7 \times 7$ unitary (mixing) 
matrix $N$, i.e., $N^* M_N N^{-1}=M_{\mathrm{diag}}$, one obtains the physical 
neutralinos ($\ntrli, \, i=1,2, \ldots,7$) and their mass-eigenvalues 
($m_{\ntrli}$, with $m_{\ntrlone}$ being the mass of the lightest neutralino, 
which is also the Lightest SUSY Particle (LSP) in our scenario. The elements 
$N_{ij}$, with $i,j \in \{1,2, \ldots , 7 \}$, dictate the admixture of
`$j$'th gaugino/higgsino in `$i$'th neutralino.

For the present work, it is important to take note of the  fact that the 
($4\times 4$) MSSM block and the $(3 \times 3)$ BLSSM block in $M_N$ couple 
only for a non-vanishing gauge kinetic mixing. The null entries in the
$(4 \times 3)$ off-diagonal block ensure that there are no direct couplings 
among the $SU(2)$ gaugino and higgsinos of the MSSM type with the bileptinos 
of the BLSSM scenarios. For a large enough $M_{1,2}$ and
$M_{B'}$ ($\mu, \mu' \ll M_{1,2}, M_{B'}$), $M_{1,2}$ being the $U(1)_Y$
and $SU(2)$ gaugino masses appearing in the ($4 \times 4$) MSSM block, the two 
lightest neutralino states, i.e., the LSP and next-to-LSP (NLSP),  turn out to 
be a pure bileptino and two higgsino states and the heavier ones are mixed 
states of the BLino and the other bileptino, one being BLino-dominated, having 
an intermediate mass, while the bileptino-dominated one has a significant
mass-splitting (unlike in the MSSM) from its lighter pure cousin 
state~\cite{Abdallah:2017gde}.

For $M_{1,2} \ll \mu, \mu', M_{B'}$, the LSP and the NLSP neutralinos are the 
MSSM-like gauginos. The mixing among the bileptinos and the BLino results in 
three heavier physical states: a light bileptino state with mass $\sim \mu'$ 
followed by two mixed states, one dominated by the other bileptino and the 
heaviest one dominated by the BLino. Note that, for such a hierarchy of EWino 
parameters, the neutralino mass matrix ensures that a large, negative $M_{B'}$ 
($\sim -5$~TeV) drives the mixed bileptinos-BLino state much lighter. Also, to 
keep our scenarios cleaner for our purpose, we choose to work with a small 
enough off-diagonal soft parameter, $M_{BB'}$ (${\cal O}(M_{BB'}) \ll v'$) 
which strongly restrains any mixing between the bino and the BLino.

Furthermore, since the BLSSM does not offer any new $SU(2)$ gauge boson or any 
$SU(2)$ doublet Higgs state, the chargino sector of the BLSSM remains identical 
to the MSSM and is given by
\begin{equation}\label{chargino-mass-matrix}
 M_C=
\left(
\begin{array}{ccc}
M_2 & {1 \over \sqrt{2}} g_2 v_2 \\
{1 \over \sqrt{2}} g_2 v_1   & \mu
\end{array}
\right)\,,
\end{equation}
which, being an asymmetric $2 \times 2$ matrix, can be diagonalized by two
$2 \times 2$ unitary matrices, $U$ and $V$, such that $U^* M_C V^\dagger$ is 
diagonal, with mass eigenvalues for the two chargino eigenstates being 
$m_{\tilde{\chi}^\pm_1} < m_{\tilde{\chi}^\pm_2}$.

For the present study, we take all the sfermions (squarks, sleptons, and 
sneutrinos) of the scenario to be decoupled. Hence, we do not discuss the 
sfermion sector explicitly. Nevertheless, for the sake of completeness, we list 
their couplings with the $Z'$-boson (which is done for the first ever time) in 
the Appendix. We refer the reader to Ref.~\cite{OLeary:2011vlq} for a 
description of the same (as well as all the sectors that we have discussed 
above).
%
\subsection{Couplings of the $Z'$-boson}
\label{subsec:couplings}
We now move on to discuss the tree-level couplings of the $Z'$-boson of the 
BLSSM with particles that are immediately relevant for its production 
and/or (primary) decays that constitute the final states in which a $Z'$-boson 
has been searched for at the LHC. These are the SM fermions (the quarks and the 
leptons) and gauge bosons (the $W^\pm$'s). Their couplings to the $Z'$-boson are necessarily different from the 
corresponding ones in the SSM~\cite{Leike:1998wr} and are given by:
\begin{subequations}
\label{eqn:zprime-couplings}
\begin{align}
Z'u_L\bar{u}_L &: \; -\frac{i}{6} \bigg[ \Big\{ \big(g_1+\gby \big) s_{\theta_W}-3 g_2 c_{\theta_W} \Big\} s_{\theta'}+\big(\gbl+\gyb \big) c_{\theta'} \bigg],
\label{eqn:zprime-ululbar} \\
Z'u_R\bar{u}_R &: \; -\frac{i}{6} \bigg[\big(4 g_1+\gby \big) s_{\theta_W} s_{\theta'}+\big(\gbl+4 \gyb \big) c_{\theta'}\bigg],
\label{eqn:zprime-ururbar} \\
Z'd_L\bar{d}_L &: \; -\frac{i}{6}\bigg[ \Big\{ \big(g_1+\gby \big) s_{\theta_W}+3 g_2 c_{\theta_W} \Big\} s_{\theta'}+\big(\gbl+\gyb \big) c_{\theta'}\bigg],
\label{eqn:zprime-dldlbar}\\
Z'd_R\bar{d}_R &: \; \frac{i}{6}\bigg[ \big(2 g_1-\gby \big) s_{\theta_W} s_{\theta'}-\big(\gbl-2 \gyb \big) c_{\theta'}\bigg],
\label{eqn:zprime-drdrbar} \\
Z'\ell_L^+ \ell_L^- &: \; \frac{i}{2} \bigg[ \Big\{ \big(g_1+\gby \big) s_{\theta_W}-g_2 c_{\theta_W} \Big\} s_{\theta'}+\big(\gbl+\gyb \big) c_{\theta'} \bigg],
\label{eqn:zprime-llllbar} \\
Z'\ell_R^+ \ell_R^- &: \; \frac{i}{2} \bigg[ \big(2g_1+\gby \big) s_{\theta_W} s_{\theta'}+\big(\gbl+2\gyb \big) c_{\theta'} \bigg],
\label{eqn:zprime-lrlrbar}\\
\hskip 45pt Z' W^+W^- &: \;  -i g_2 c_{\theta_W} \sthetaprime.
\label{eqn:zprime-ww}
\end{align}
\label{eqn:couplings}
\end{subequations}
\hspace*{-0.2cm}Furthermore, it is now possible that the $Z'$ decays to various Higgs states
and the lighter SUSY  states of the BLSSM scenario, in particular, the 
neutralinos and charginos, whenever those are kinematically accessible. Decays of the $Z'$-boson to Higgs boson(s) may involve a pair of Higgs bosons (i.e., $Z' \to A H$) or a single ($CP$-even) Higgs boson accompanied by a $Z$-boson (i.e., $Z' \to Z H$); the latter occurring only in the presence of gauge kinetic mixing. We 
present the interactions of these states with the $Z'$-boson in the Appendix. 
Note that, as we have already mentioned at the end of the last section, we 
consider the sfermions to be rather heavy and hence decoupled for all practical 
purposes. Still, for the sake of completeness, we present their interactions 
with the $Z'$-boson also in the Appendix.
It should be noted that, while the couplings of the $Z'$-boson with the quarks 
(see Eqs.~(\ref{eqn:zprime-ululbar})--(\ref{eqn:zprime-drdrbar})) show up at 
both production (at the LHC) and decay level, the other couplings in
Eq.~(\ref{eqn:couplings}) appear only through decay vertices. These couplings 
reduce to the corresponding ones for pure BLSSM (i.e., without a gauge 
kinetic mixing) for $\gby=\gyb=0$ when $\thetaprime$ (i.e., the $Z$--$Z'$ 
mixing) vanishes. However, note that a vanishing $\thetaprime$ by itself may 
not guarantee a pure BLSSM scenario since a cancellation between the terms 
proportional to $\gby$ and $\gyb$ in Eq.~(\ref{eqn:thetaprime2}) is a 
possibility for finite values of these two couplings. Furthermore, for small
$\gby$, $\sthetaprime (\approx \thetaprime)$ is small. Hence, the first terms 
in the expressions for all the fermionic couplings of the $Z'$-boson can be 
ignored. These couplings then effectively depend only on $\gbl$ and $\gyb$.

Furthermore, the coupling $Z'W^+W^-$ (see Eq.(\ref{eqn:zprime-ww})) goes as 
$\sthetaprime \approx \thetaprime$, thus indicating that this is solely induced 
by the SM $Z$-boson admixture in the $Z'$-boson. Even though this coupling 
$Z'W^+W^-$ is $\thetaprime$-suppressed, given its direct dependence on 
$\sthetaprime$, an appropriately small value of the same could still render 
the partial width $\gammazprimeww$, and hence the total decay width of the
$Z'$-boson, optimally large for our purposes. This can be understood by 
studying the expression for $\gammazprimeww$ given by~\cite{Altarelli:1989ff}
\begin{equation}
\gammazprimeww= \frac{\alpha}{48}\cot^2\theta_W\, s_{\theta'}^2\,
\mzprime\left(\frac{\mzprime}{M_W}\right)^4\left(1+\frac{20 M^2_W}{\mzprime^2}
 +\frac{12 M_W^4}{\mzprime^4}\right)
 \left(1-\frac{4M_W^2}{\mzprime^2}\right)^{3/2},
\label{eqn:gammaz}
\end{equation}%
where, $\alpha$ is the fine structure constant and $M_W$ is the mass of the SM
$W^\pm$-boson. For $\mzprime \gg M_{Z,W}$, when the product of the last two
terms in parentheses approaches unity,
$\gammazprimeww \propto \sthetaprime^2 \mzprime^5/M_Z^4$, which showcases its 
well-known quintic dependence on $\mzprime$. This could more than compensate 
for the $\sthetaprime$ suppression and make $\gammazprimeww$ pretty large and dominate $\gammazprime$, thereby turning the $Z'$-boson `fat'. A fat $Z'$-boson
(with ${\gammazprime \over \mzprime} \gtrsim 10\%$) would invariably make the
current experimental analyses grossly inefficient in probing its properties, 
as those analyses assume the $Z'$-boson to be a narrow resonance. This would thus 
lead to relaxed experimental bounds on the observables of the $Z'$ sector. A 
non-vanishing $\thetaprime$, even when it is small, therefore, plays a central role in the present study.
%
\subsection{Bounds on $\thetaprime$ and $\mzprime$}
\label{subsec:bounds}
The key observables in the $Z'$ sector are $\thetaprime$, $\mzprime$ and
$\gammazprime$. Bounds on $\thetaprime$ and $\mzprime$ have long been routinely 
derived in a model-dependent way from data obtained by various lepton 
and hadron colliders. As mentioned in Sec.~\ref{subsec:zprime-sector}, a 
stringent upper bound of $\thetaprime \lesssim 10^{-3}$~radian at 95\% 
CL was placed by the LEP and SLC
experiments~\cite{ALEPH:2005ab, ALEPH:2006bhb, Erler:2009jh, ALEPH:2013dgf, ParticleDataGroup:2024cfk}
via precision studies of BR[$Z \to \ell^+ \ell^-$] and the forward-backward 
asymmetries at the SM $Z$-pole as well as by studying the $Z$-boson line-shape, all of which 
constrain the virtual effects originating from the interference with a 
hypothetical $Z'$-boson and/or from a possible $Z$--$Z'$ mixing. These are now 
surpassed by the LHC experiments, which tend to place stronger bounds of 
$\thetaprime \lesssim {\cal O}(10^{-4})$~radian (again, at 95\% 
CL)~\cite{Pankov:2019yzr, Osland:2020onj, Osland:2022ryb}%
\footnote{Generally, from  experimental data, model-dependent bounds on the 
mixing parameter $\xi=\sthetaprime (\simeq \thetaprime)$, as functions of 
$\mzprime$, are derived.}.
However, it should be noted that, until very 
recently~\cite{CMS:2019gwf,CMS:2021ctt}%
\footnote{In fact, Ref.~\cite{CMS:2019gwf} studies the spin-1 resonance in the 
form of a $Z'$-boson with $\gammazprime \over \mzprime$ up to as large as 55\% 
and demonstrated how, with an increasing $\gammazprime \over \mzprime$, the 
upper limit on the quantity $\sigma \times {\rm BR}$ in the dijet final state 
gets weakened. However, note that more stringent bounds on such a resonant 
sector are generically obtained in the dilepton and $W^+W^-$ final states 
for which the maximum $\gammazprime \over \mzprime$ considered by the 
experimental searches have been mere 10\%~\cite{ATLAS:2019erb, CMS:2021ctt} and 
5\%, respectively.}, 
such LHC analyses in searches for a resonant $Z'$-boson have been exclusively 
carried out assuming the $Z'$-boson to be a narrow resonance, i.e., 
$\gammazprime \over \mzprime$ to be only of a few percent, such that the 
intrinsic width of the $Z'$-boson is much smaller than the experimental 
resolution, i.e., ${\gammazprime \over \mzprime} < \Delta \mzprime/\mzprime$ 
(customarily taken to be 5\%)\, when the NWA could be safely adopted. This is 
the case for the analyses undertaken by various experimental 
collaborations~\cite{ALEPH:2005ab, ALEPH:2006bhb, Erler:2009jh, ALEPH:2013dgf, ParticleDataGroup:2024cfk}
as well as the ones that are carried out outside the
same~\cite{Pankov:2019yzr, Osland:2020onj, Osland:2022ryb} to derive limits on 
$\mzprime$ and $\thetaprime$ from experimental results. Here, we argue that 
larger values of $\thetaprime$, which could turn the $Z'$ resonance broad, 
would thus tend to escape the current LHC bounds on its properties.

Such a weakening of the upper bound on $\thetaprime$ (as could be obtained from 
the current direct searches of a resonant
$Z'$-boson~\cite{Pankov:2019yzr, Osland:2020onj, Osland:2022ryb}) with the 
broadening of the $Z'$ resonance is primarily due to the fact that such 
searches quickly lose sensitivity as $\gammazprime\over \mzprime$ grows. This is due to (but 
may not be limited to), viz.,
\begin{itemize}
\item
loss of optimality of the invariant mass windows that are chosen for the
final-state particles which a $Z'$-boson, which has been assumed to be a narrow resonance, decays to, thus leading to a diminished signal-to-background ratio,
\item 
off-shell contributions to the relevant cross sections gaining importance 
with increasing                    $\gammazprime\over \mzprime$, thus modifying the effects of Parton Distribution Functions (PDFs) and higher-order QCD and/or EW effects, both potentially altering the 
normalization of the cross section  and/or the resonance shape,
\item 
the $Z$--$Z'$ interference effects having more enhanced model-dependencies 
as the $Z'$ resonance gets broader (e.g., as more decay channels contribute).
\end{itemize}

In contrast, the bound on $\thetaprime$ ($\lesssim 10^{-3}$~radian), as derived 
by the LEP and the SLC
experiments~\cite{ALEPH:2005ab, ALEPH:2006bhb, Erler:2009jh, ALEPH:2013dgf, ParticleDataGroup:2024cfk},
is pretty robust in nature thanks to the fact that a `fatter' $Z'$-boson could 
hardly affect the estimation of $\thetaprime_\mathrm{max}$ from the $Z$-pole 
observables which, for $\mzprime \gg M_Z$, are essentially devoid of those 
physics effects (for large $\gammazprime \over \mzprime$) that are pronounced in  the vicinity of the $Z'$ pole. Thus, we need to be particularly vigilant about 
its values that we use to find 
$\gammazprime$ (or, rather, $\gammazprime \over \mzprime$) large enough for our 
purposes. This compels us to critically assess the true nature of the said bound 
and hence its possible limitations.

In this context, it is important to also note that such a bound on $\thetaprime$ is 
known to get relaxed for a somewhat leptophobic
$Z'$-boson~\cite{Erler:2009jh, Babu:1997st, Umeda:1998nq}. This is apparent 
from the expressions for the leptonic couplings of the $Z'$-boson presented in 
Eqs.~(\ref{eqn:zprime-llllbar})~and~(\ref{eqn:zprime-lrlrbar}). Given that 
$\thetaprime$ would anyway be on the smaller side even when it evades the said 
bound, such couplings would, by large, be determined by terms proportional to 
$c_{\thetaprime}$. Further, note that the bounds from the experiments mentioned 
above at the $Z$-pole concern the $Z'$-boson admixture in the $Z$-boson. Thus, 
what matters is the effective leptonic coupling of the $Z'$-boson, which goes 
as $\sim(g_{Z' \ell^+\ell^-}) \times \thetaprime$. Strictly speaking, the
LEP and SLC data on various $Z$-pole observables require this quantity to be 
smaller than a few $10^{-4}$ to comply with the experimental bounds.\footnote{This translates to $\thetaprime \lesssim 10^{-3}$~radian for 
$g_{Z'\ell^+\ell^-}$ being of a sequential type, i.e., SM-like.}
Thus, allowing for a smaller $g_{Z'\ell^+\ell^-}$ (i.e., resorting to a 
somewhat `leptophobic' $Z'$-boson) could make $\thetaprime \gtrsim 10^{-3}$~radian viable.

To exploit this caveat systematically, one now requires an appropriate 
definition of the quantity $g_{Z'\ell^+\ell^-}$. However, this is not 
straightforward for the $Z'$-boson couples differently to the left and right 
chiral components of the leptons (as can be seen from
Eqs.~(\ref{eqn:zprime-llllbar})~and~(\ref{eqn:zprime-lrlrbar})). In addition, 
it matters what combinations of those specific coupling factors enter the 
expressions of various $Z$-pole observables like the resonant production cross 
section of the $Z$-boson ($\sigma_{e^+e^- \to Z}$), $\Gamma_Z$, the
forward-backward asymmetries of the final state leptons, etc. Towards this, two 
such different combinations could be identified:
\begin{equation}
g^\ell_{\mathrm{avg}} = \sqrt{g^2_{_{Z'L}} + g^2_{_{Z'R}}}
\qquad
\mathrm{and}
\quad
g^\ell_{_{V}} = \frac{1}{2} \Big( g_{_{Z'L}} + g_{_{Z'R}} \Big) \, ,
\end{equation}%
where $g_{_{Z'_L}} = (\gbl + \gyb)/2$ and $g_{_{Z'R}} = (\gbl + 2\gyb)/2$, 
$g_{_{Z'_{L,R}}}$ are the couplings of the $Z'$-boson to the left- and
right-handed leptons (electrons), as those appear in
Eqs.~(\ref{eqn:zprime-llllbar})~and~(\ref{eqn:zprime-lrlrbar}), respectively. 
While $g^\ell_\mathrm{avg}$ can be seen as an average coupling of the
$Z'$-boson to the SM leptons, $g^\ell_{_{V}}$ is the vector coupling of the 
$Z'$-boson to these leptons. In any case, these couplings enter the expressions 
for the $Z$-pole observables in a rather involved manner and,  ultimately, their 
precise values are extracted from the global fits only. For our purposes, a 
simpler consideration would suffice. Hence, we adopt $g^\ell_\mathrm{avg}$ as 
the reference leptonic coupling of the $Z'$-boson, although we keep track of 
$g_V^\ell$.
%

\begin{figure}[!t]
\begin{center}
 \includegraphics[width=7.5cm,height=6cm]{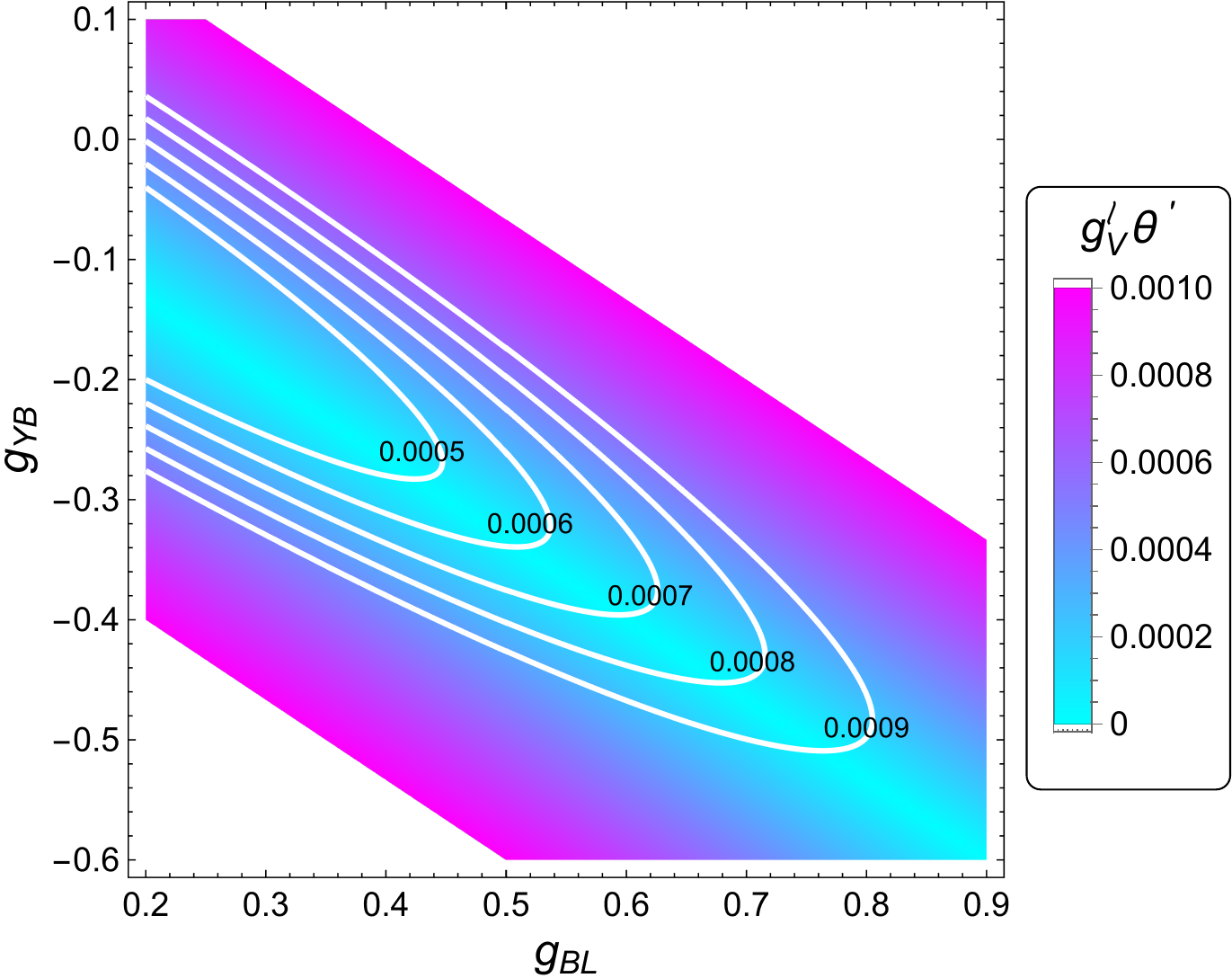}~~~~~~
\includegraphics[width=7.5cm,height=6cm]{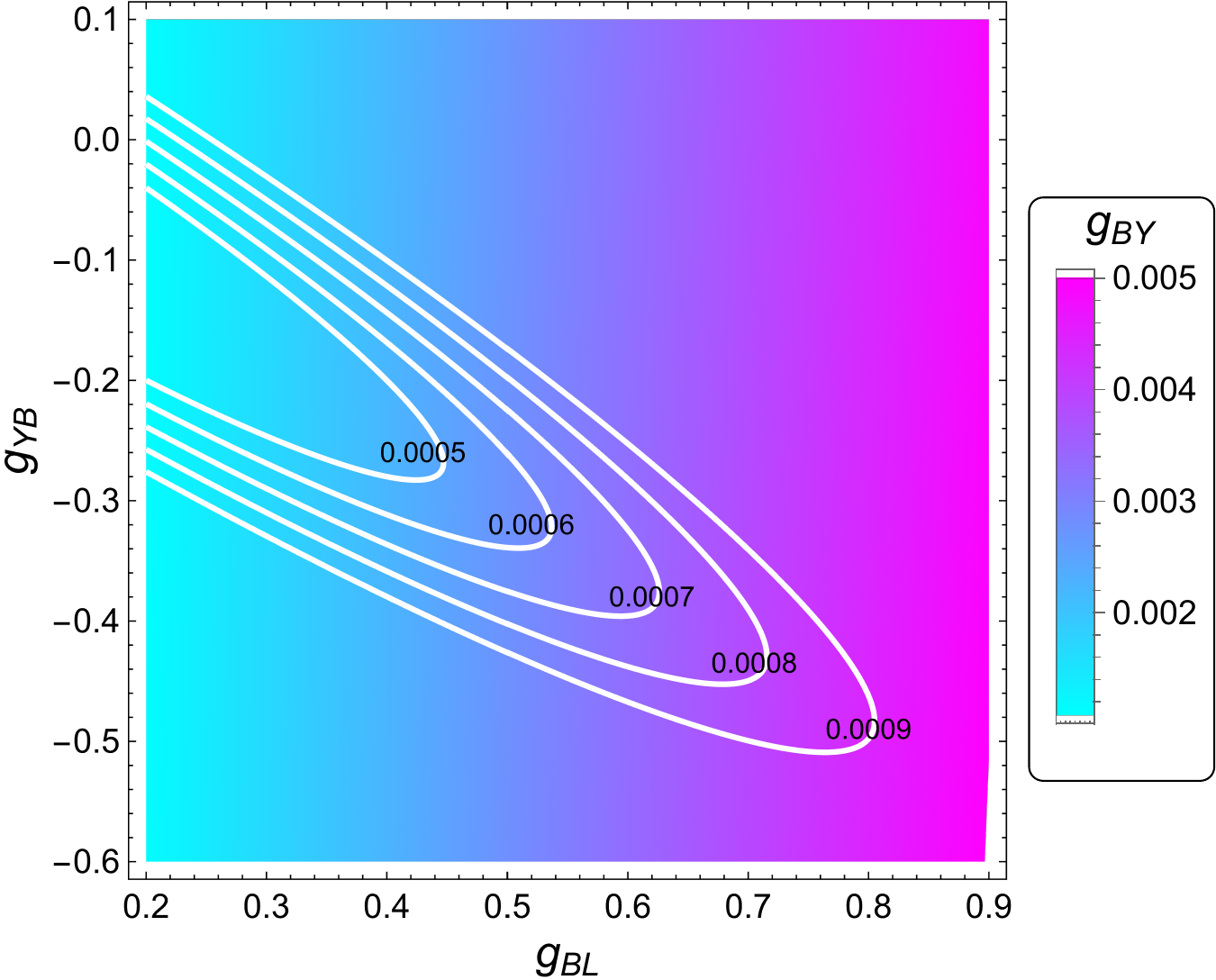}
\caption{
Contours of fixed values of $g_{\rm avg}^\ell . \thetaprime$ in the plane 
$\gbl$--$\gyb$ for $\thetaprime=0.005$~radian with colors (from the palette) in 
the background indicating the magnitude of $g^\ell_V . \thetaprime$ (left) and 
$\gby$ (right). See text for details.
}
\label{Fig:gBL-gYB-contours}
\end{center}
\end{figure}
In the left plot of Fig.~\ref{Fig:gBL-gYB-contours} we present contours of 
fixed values of $g^\ell_\mathrm{avg} . \thetaprime \sim {\cal O}(10^{-4})$ 
radian in the $\gbl$--$\gyb$ plane, for a modest (fixed) value of $\thetaprime$ 
(=0.005~radian)\, where the colors in the background indicate the size of 
$g^\ell_V$, as presented in the attached palette. This plot demonstrates how,
a priori, we are free to use relatively larger values of $\thetaprime$ than 
what is loosely adopted in the literature, and can still comply with the data 
pertaining to various $Z$-pole observables, thanks to the leptophobic couplings 
of the $Z'$-boson that we take advantage of.%
\footnote{Note that a certain degree of generic leptophobia in the $Z'$-boson 
would mean its eroded coupling to not only electrons but also muons. Together, 
these contribute to a possible weakening of the constraints on the $Z'$ sector 
as derived from its searches in the dilepton final state that involve both 
these leptons.}
It can also be gleaned from this plot that any given range of 
$g^\ell_\mathrm{avg}$ generically guarantees a range with even smaller values 
of $g^\ell_V$. Thus, our choice of $g^\ell_\mathrm{avg}$ as the reference 
leptonic coupling of the $Z'$-boson can be considered as one that allows 
exploration of larger ranges of $\gbl$ and $\gyb$ without getting into a conflict with the $Z$-pole data from the LEP and SLC experiments.
The right plot of Fig.~\ref{Fig:gBL-gYB-contours} is the same as on its left, 
but now reveals the range of $\gby$ involved, via the color palette attached to 
it. (Note that $\gby$ controls the magnitude of $\thetaprime$.) In this work, we 
choose to revisit the recent LHC bounds on the $Z'$ sector by (conservatively) 
confining  ourselves to a rather limited region of the BLSSM parameter space with
$g^\ell_\mathrm{avg} . \thetaprime \lesssim 5 \cdot 10^{-4}$~radian.

A more robust (and inevitable) upper bound on $\thetaprime$ ($\thetaprimemax$) 
would arise exclusively from the measured value of the oblique $T$-parameter~\cite{Peskin:1990zt,Degrassi:1993kn,Altarelli:1990zd,Grojean:2006nn} 
at the LEP and SLC experiments which limits the violation of the global 
custodial $SU(2)$ symmetry of the SM (through the allowed shift in the SM $Z$-
boson mass), as quantified from the allowed deviation of the $\rho$-parameter 
from its SM value of $\rho=1$. In the case of a leptophobic $Z'$-boson, 
$\thetaprimemax$ derived from the global fit to $Z$-pole precision data is 
driven by the measured value of the $T$-parameter and can be as high as 
$\thetaprimemax \sim 0.01$~radian for
$\mzprime \sim 1$~TeV~\cite{Erler:2009jh}, and smaller for larger $\mzprime$. 
However, since we deal with a somewhat heavier $Z'$-boson in this work, we take 
a conservative approach and stick to the choice of $\thetaprimemax=0.005$~radian in the present work.

Furthermore, we comply with the bounds obtained from the consideration of 
generalized oblique parameters, which translate into constraints on the scale of (heavy) new physics from the EW precision data measured below, at and above the $Z$-peak, and, in the context of the BLSSM scenario, is given by
\begin{equation}
  {\mzprime \over {Q^{BL}_e \left(\gbl+\gyb\right)}} >6.7  \; \mathrm{TeV}
\label{eqn:oblique}
\end{equation}
at 99\%  CL~\cite{Carena:2004xs, Cacciapaglia:2006pk}, where  $Q^{BL}_e$ is the 
charge of the involved leptons under $U(1)_{B-L}$.
It should also be noted that, for $\gyb <0$, the cancellation that takes place 
in the denominator of the above relation ($\gbl$ being taken to be positive) 
would make it easier to satisfy the indicated bound. At the same time, for 
small $\thetaprime$, the coupling constants appearing in the denominator (i.e., 
$\gbl$ and $\gyb$) govern the interaction strengths at the $Z'f\bar{f}$ 
vertices (see Eq.~(\ref{eqn:couplings})) where similar cancelations might 
occur. Hence, production rates for the dilepton pairs would be suppressed. 
This could result in a relaxation of the reported lower bound(s) on $\mzprime$ 
from the LHC to an extent that could survive its downgraded lower bound 
obtained earlier from the consideration of the oblique parameters above.

Note further that, given Eq.~(\ref{eqn:zprime-couplings}), the conditions to find a somewhat leptophobic $Z'$-boson, as mentioned above, may set in some 
hadrophobia in the $Z'$-boson as well, thus driving some of the $Z'q\bar{q}$
couplings smaller. This could affect the resonant production rate of the
$Z'$-boson at the LHC and, hence, could further relax the lower bound on 
$\mzprime$, in particular, in the dilepton final state.
%
\section{Results}
\label{Sec:3}
In this section, we discuss our results in detail by first looking into the
dependencies of some of the key observables like $\mzprime$, $\thetaprime$,
$\gammazprime \over \mzprime$, the relevant decay BRs of the
$Z'$-boson (except for the ones to various BLSSM-specific states), etc., on the 
fundamental input parameters of the scenario. We follow this up with the study 
of cross sections for the dilepton and $W^+W^-$ final states, mediated by a 
resonant $Z'$-boson, at the 13~TeV LHC, for some representative values of 
$\mzprime$ and $\gammazprime \over \mzprime$. For the purpose at hand, we do 
not yet impose the precision bound on $\thetaprime$, as discussed in 
Sec.~\ref{subsec:bounds}. Hence, the discussion offers an insight into  a 
broader range of $\gammazprime$ than what the precision data could actually 
allow. Finally, we bring into the picture the BLSSM-specific states, in 
particular, the lighter EWinos and the additional Higgs bosons, to 
consolidate the phenomenological possibilities involving the $Z'$-boson at the 
13~TeV LHC, by imposing a somewhat conservative upper bound on $\thetaprime$, 
as dictated by 
precision data. 

The study uses a suitable implementation of the BLSSM scenario in the 
package {\tt SARAH-v4.9.0}~\cite{Staub:2015kfa}. Further, we use the
{\tt SPheno}~\cite{Porod:2003um, Porod:2011nf} code that it generates the BLSSM particle spectra, the mixing matrices involved, the coupling 
strengths of the $Z'$-boson to various BLSSM excitations and its decay partial widths
and BRs to such states. Production cross sections for all 
relevant processes are computed at the Leading Order (LO)%
\footnote{For the dilepton final states, the $k$-factor is known to be 
consistent with unity~\cite{CMS:2018ipm} while for the $W^+W^-$ final state we 
restrict ourselves to the LO cross sections following 
Ref.~\cite{ATLAS:2020fry}.},
at the 13~TeV LHC, via {\tt MG5\_aMC@NLO-v2.4.3}~\cite{Alwall:2014hca}, with 
its default setting%
\footnote{We treat a broader $Z'$ resonance in the same way as a narrow one, 
following the usual experimental practice (see, for example, 
Ref.~\cite{CMS:2019gwf}). A better way to deal with broad resonances would be 
to consider an energy-dependent Breit-Wigner
propagator~\cite{sniyogi-thesis,Altarelli:1989hv,Sjostrand:2006za} for improved 
descriptions of resonance shapes.}
that employs the {\tt NNPDF2.3-LO} PDFs~\cite{Ball:2012cx}.
%
\subsection{The correlation among $\thetaprime$, $\mzprime$ and
$\gammazprime \over \mzprime$}
\label{subsec:mixing-mass-width}
The phenomenology of the resonantly produced $Z'$-boson of the BLSSM scenario introduced 
is primarily governed by three input coupling parameters, 
i.e., $\gbl$, $\gby$ and $\gyb$ (for a given $v'$), and broadly concerns the 
observables like $\mzprime$, $\thetaprime$ and $\gammazprime \over \mzprime$, 
which are functions of the parameters mentioned above. As can be seen from 
Eq.~(\ref{eqn:couplings}) and the Appendix, these coupling parameters appear 
explicitly at the $Z'f\bar{f}$, $Z'Z H_i$ and $Z'$-sparticle pair interaction 
vertices and hence would directly control the yield in the dilepton, $ZH_i$ 
and sparticle pair productions, respectively, at the LHC via a resonant
$Z'$-boson. However, for these interactions, $\thetaprime$, by its presence at 
the involved vertices (in the form of $\sthetaprime$ and $\cthetaprime$) 
moderates the contributions of the three coupling parameters to the overall 
interaction strengths.

It may be gleaned from Eq.~(\ref{eqn:couplings}) that, with  $\thetaprime$ being 
small, terms proportional to $c_{\thetaprime}$ would dominate and, hence, the 
coupling parameters $\gbl$ and $\gyb$ appearing in those terms would play the 
dominant roles in the dilepton processes. Conversely, production of a
$W^\pm$-pair via a resonant $Z'$-boson faces an a priori suppression from the 
$Z'W^+W^-$ vertex (see Eq.~(\ref{eqn:zprime-ww})) due to a small $\thetaprime$. 
As pointed out in the previous section, this could be more than compensated for 
as $\mzprime$ grows when, in spite of a suppressed coupling, $\gammazprimeww$, 
and hence BR$\left[ Z' \to W^+W^- \right]$, could grow rapidly with $\mzprime$.

As for the various production processes, their rates tend to suffer not only
due to the resonant $Z'$-boson becoming heavier, but also because it becomes 
fatter (broadening of the resonance) as its mass grows. However, for the 
$W^+W^-$ final state, such a suppression could be compensated for by an
increase in BR$\left[Z' \to W^+ W^- \right]$, as $\mzprime$ increases. Thus, 
the phenomenology of such a $Z'$-boson at the LHC has intricate dependencies on 
various basic input parameters and some derived observable quantities, as 
listed above.

On the operational side, we choose to vary both $\mzprime$ and $\gbl$ 
simultaneously since this would allow us to exploit both the kinematic and 
dynamical possibilities to their fullest extent in a direct manner. Given that 
$\mzprime \simeq \gbl v'$ (see Eq.~(\ref{eqn:mzmzprime})), this amounts to 
varying~$v'$. Furthermore, we find that $\gbl \sim {\cal O} (0.1)$ tends to 
keep the fermionic (quarks and leptons) couplings to the $Z'$-boson on the 
smaller side. These smaller couplings eventually help us to evade the 
experimental bounds on $\mzprime$ derived via the dilepton final state.
The value of $\gbl$ mentioned above falls in the correct range, as we are interested in $\mzprime, v' \sim {\cal O} (\mathrm{TeV})$.
%

\begin{figure}[!t]
\begin{center}
\includegraphics[width=5cm,height=4cm]{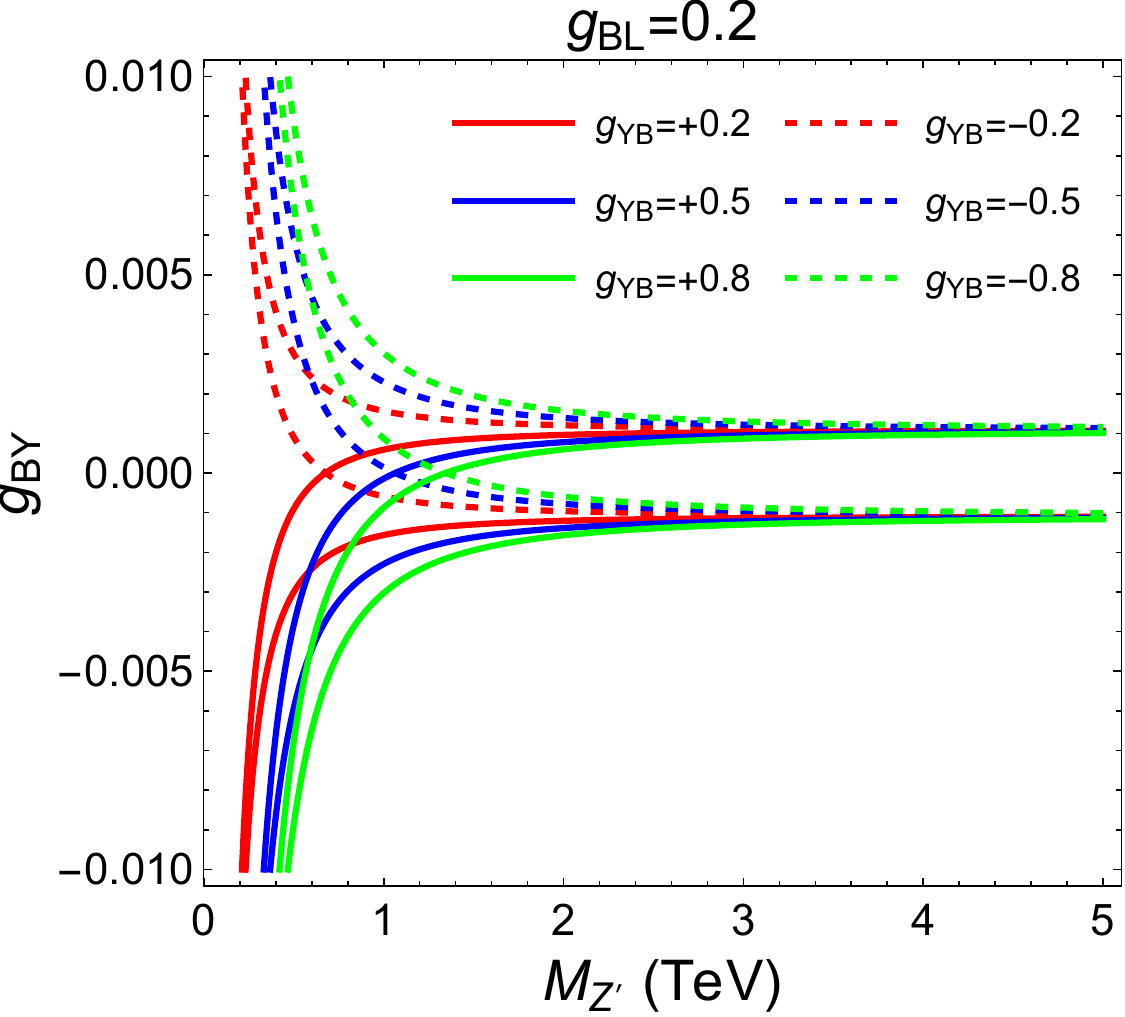}~~~~~
\includegraphics[width=5cm,height=4cm]{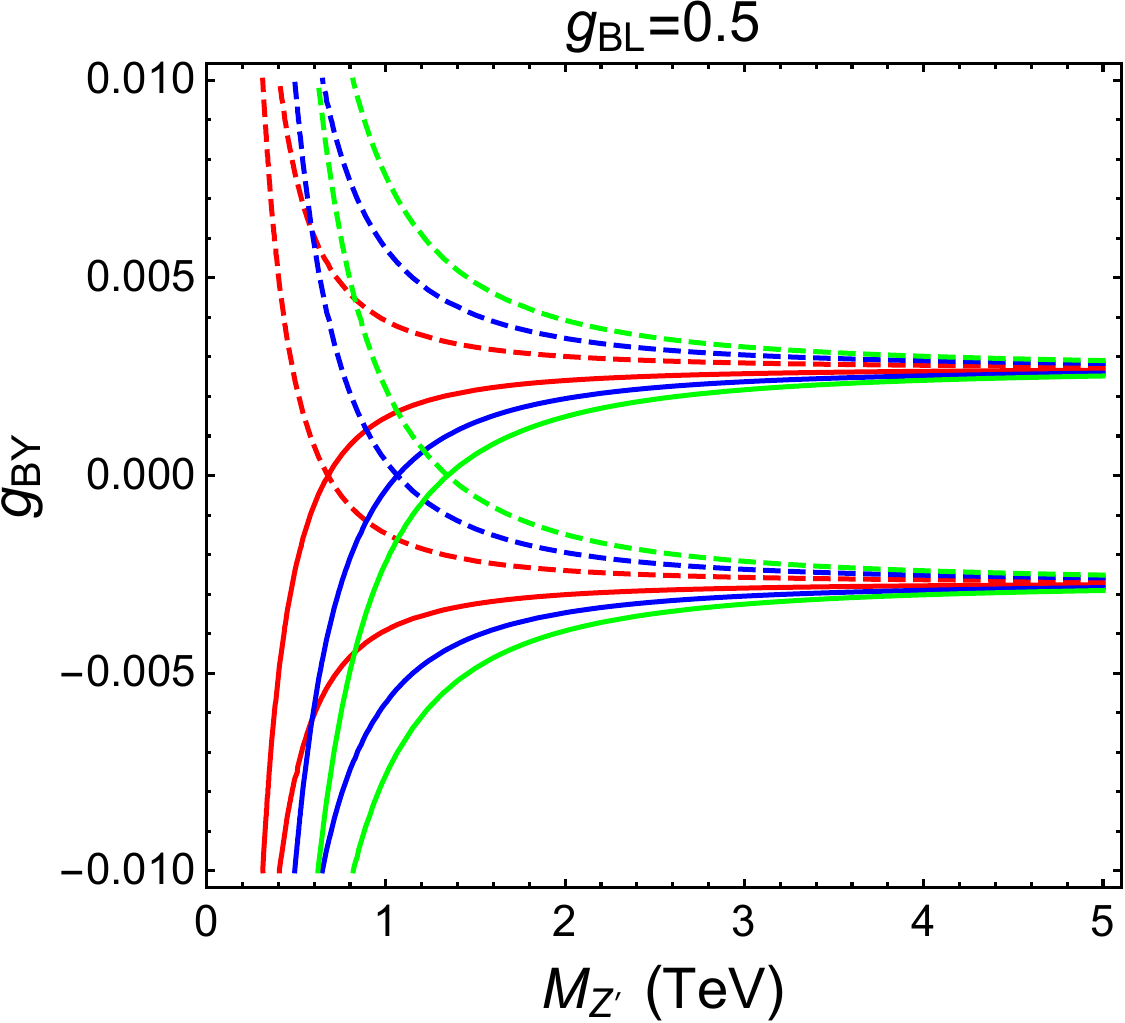}~~~~~
\includegraphics[width=5cm,height=4cm]{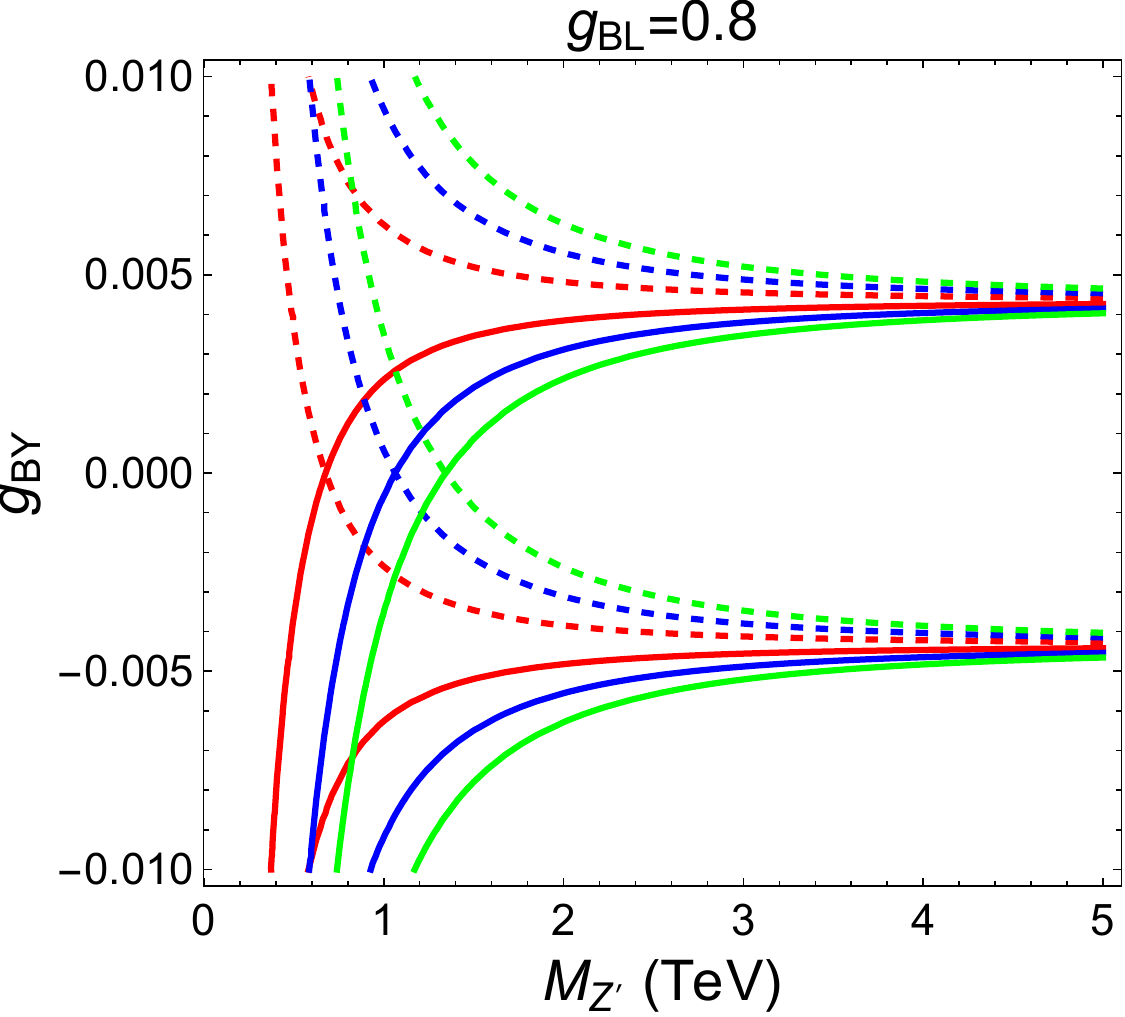}\\[0.5cm]
\caption{
Contours of $|\thetaprimemax|$ = 0.005~radian in the plane $\mzprime$--$\gby$ 
for various values and signs of $\gyb$ (the colored solid and dashed lines) and 
three different values of $\gbl$ (0.2, 0.5, and 0.8, from left to right). See 
text for details.
}
\label{Fig:MZp-thetap-analytical}
\end{center}
\end{figure}
Given the important role that $Z$--$Z'$ mixing is going to play in our present 
study, we first illustrate in Fig.~\ref{Fig:MZp-thetap-analytical} what the 
contours of $\thetaprimemax = 0.005$~radian look like in the $\mzprime$--$\gby$ 
plane for a set of representative values of $\gyb$ and $\gbl$. These are 
obtained using the expression for $\tan2\thetaprime$ in
Eq.~(\ref{eqn:thetaprime2}). The choice of variables, $\mzprime$ and $\gby$, 
that define the chosen plane is guided by how non-trivially $\thetaprime$ 
depends on them and the crucial roles these two variables play in the 
phenomenology we study in this work. Note that, for smaller values of 
$\mzprime$, when the first term on the right-hand side of
Eq.~(\ref{eqn:thetaprime2}) assumes substantial importance, $\gby$ could assume 
much larger values without being in conflict with the LEP and SLC  constraint on 
$\thetaprime$. Note that for each value of $\gyb$, two branches appear that 
respect $|\thetaprime|=0.005$~radian, the top (bottom) corresponding to 
positive (negative) $\thetaprime$. The area enclosed between such pairs of the 
same color and type (bold or dashed) indicates the ranges of $\mzprime$ and 
$\gby$ over which $|\thetaprime| \leq 0.005$~radian. However, note that these 
plots are for demonstration purposes only to probe the theoretical dependencies 
of various key observables on the BLSSM-specific input parameters (the gauge 
couplings $\gby, \gyb, \gbl$ and $\mzprime$) and, in some cases, they may not 
have mutually compatible values of these parameters that comply with the 
constraints from the oblique observables. In Sec.~\ref{subsec:susy}, we present the 
results of a full scan of these parameters taking into account the said constraints.
%
\begin{figure}[!t]
\begin{center}
\includegraphics[width=5cm,height=4cm]{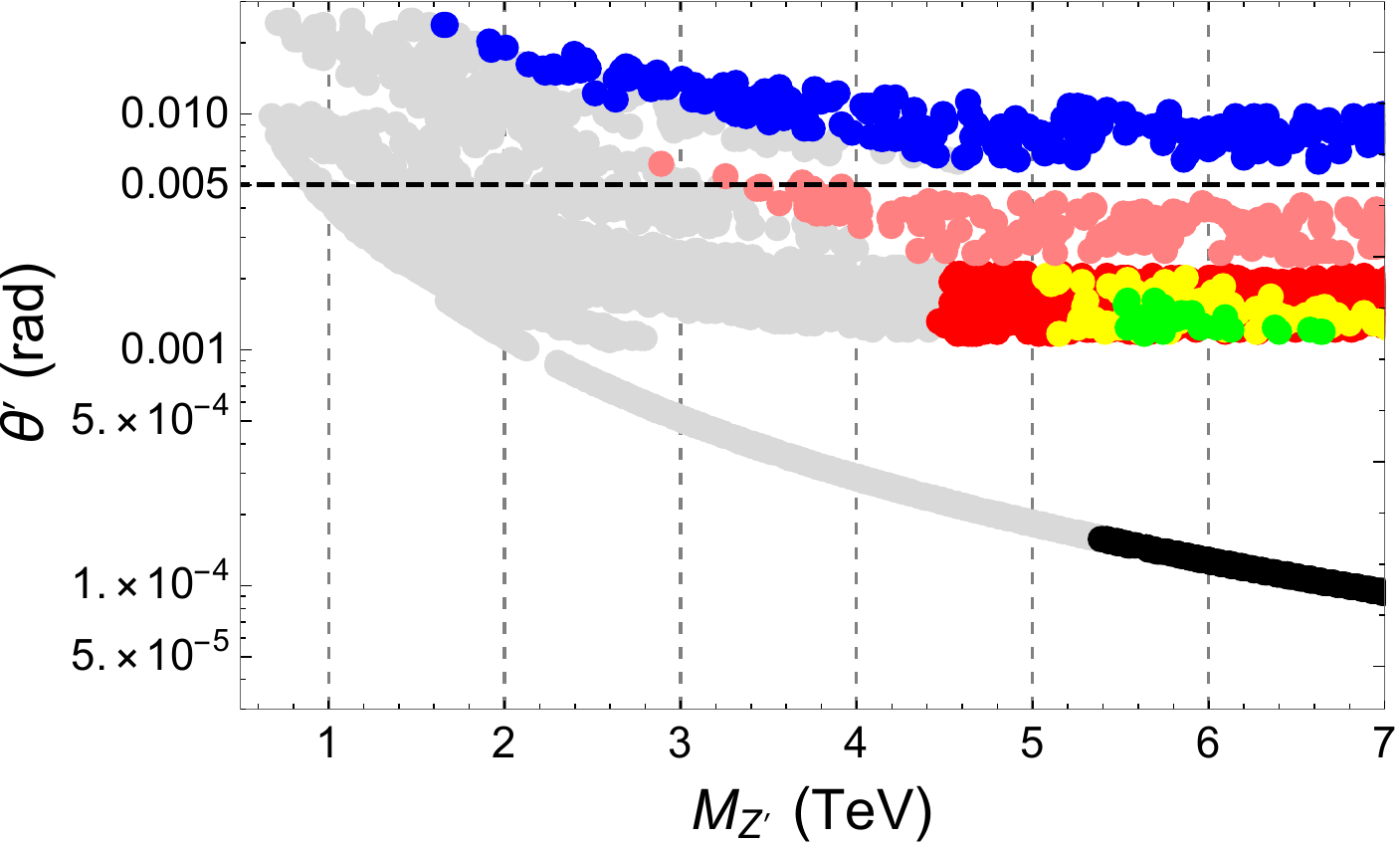}~~~~~
\includegraphics[width=5cm,height=4cm]{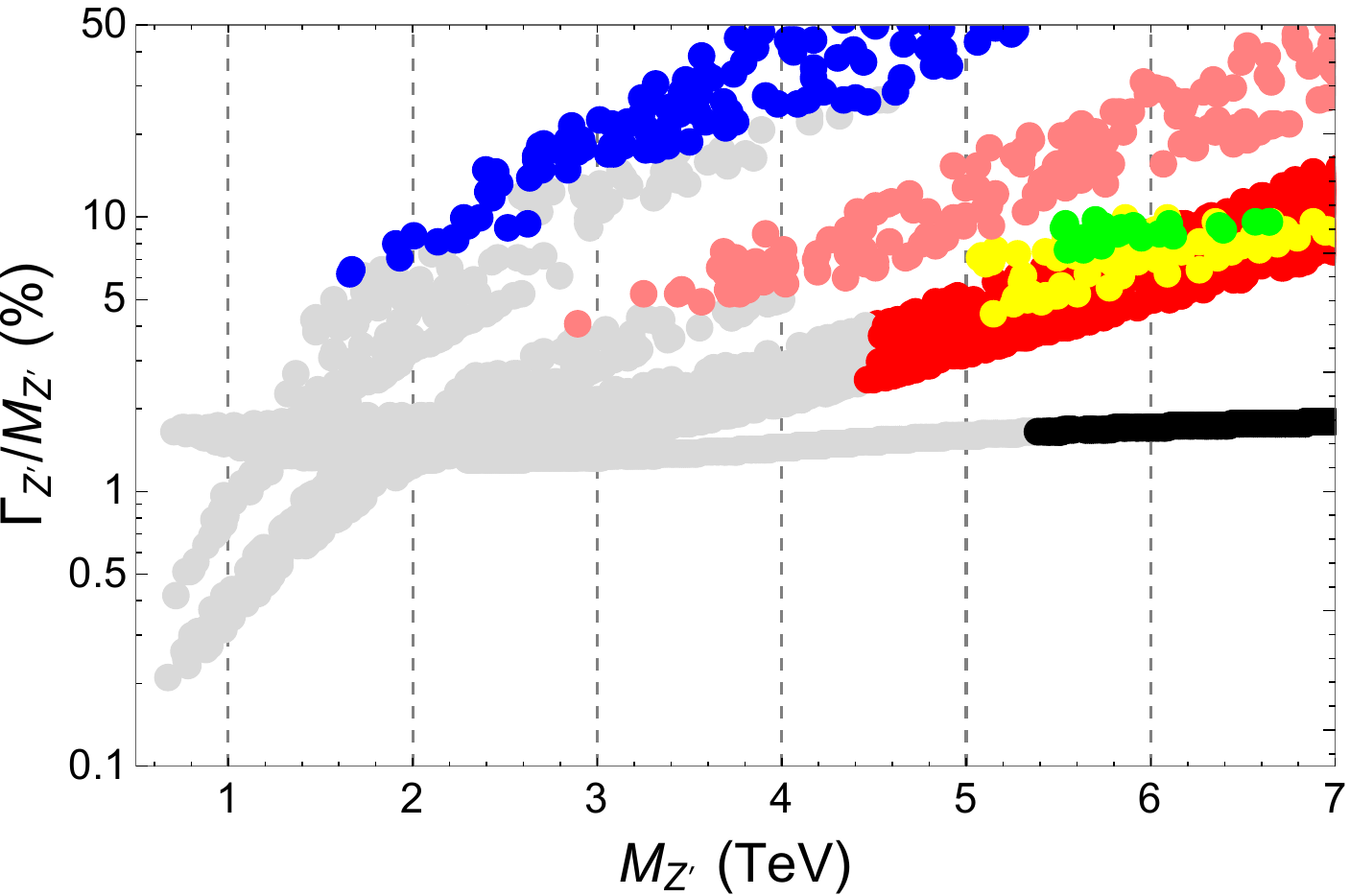}~~~~~
\includegraphics[width=5cm,height=4cm]{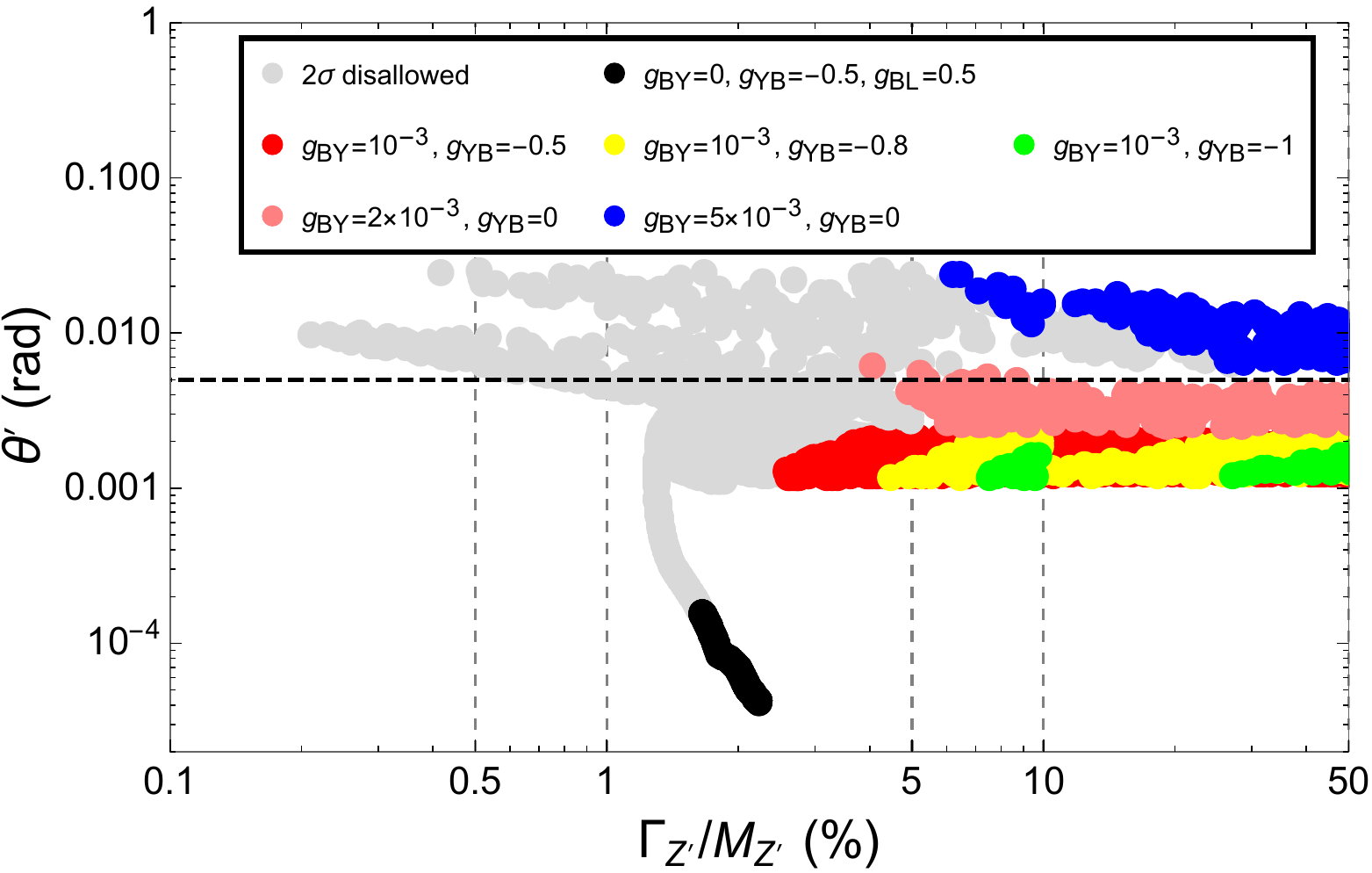}
\caption{Variations of $\thetaprime$ (left), $\gammazprime \over \mzprime$ 
(middle) as functions of $\mzprime$ and the projection of the simultaneous 
(parametric) variations of $\gammazprime \over \mzprime$ and $\thetaprime$ 
(right) for various representative combinations of the coupling parameters 
$\gby$ and $\gyb$. Here, $\gbl$ is varied over the range $\gbl \in [0.2,0.8]$ except 
for the case of the bottom-most curve, for which $\gbl = 0.5$ is chosen. Further, $v'$ 
is varied over the range $v'\in [2~\text{TeV}, 25~\text{TeV}]$. Grey regions 
are excluded at the $2\sigma$ level, which are obtained by simulating searches 
for the $Z'$-boson in the dilepton final state at the 13~TeV LHC, with an 
integrated luminosity of 150~fb$^{-1}$ (see text for details).
}
\label{Fig:MZp-thetap-GZpOMZp}
\end{center}
\end{figure}
%

The mutual relationships  among $\thetaprime$, $\mzprime$ and
$\gammazprime \over \mzprime$ are illustrated by the scatter plots in 
Fig.~\ref{Fig:MZp-thetap-GZpOMZp}, again for demonstration purposes. These 
are generated by a random scan of the BLSSM parameter space and simulating the 
LHC search of Ref.~\cite{ATLAS:2019erb} for the $Z'$-boson (with $\gammazprime$ as 
predicted by the BLSSM scenario) in the dilepton final state at the 13~TeV 
LHC and for $150$~fb$^{-1}$ of data. The grey regions are ruled out at 
$2\sigma$-level, where $\sigma=2(\sqrt{S+B}-\sqrt{B})$, following 
Poissonian statistics, which is appropriate for the case in hand, in which the 
involved numbers of signal ($S$) and background ($B$) events in each 
analysis bin of our simulation is small~\cite{Abdallah:2015uba}. The change of $\mzprime$ is achieved by varying both 
$v'\in [2~{\rm TeV},\, 25~{\rm TeV}]$ and $\gbl \in [0.2,0.8]$. The color 
convention reflecting the representative combinations of $\gby$ and $\gyb$ is 
given by the plot legends conveniently placed in the rightmost plot of this 
figure, but holds for the first two plots as well.

The plot on the left of Fig.~\ref{Fig:MZp-thetap-GZpOMZp} shows the variation of
$\thetaprime$ (in radians) with $\mzprime$ and closely follows from the relation
in Eq.~(\ref{eqn:thetaprime2}).  The plot clearly indicates that $\thetaprime$ 
is dominantly controlled by $\gby$, for three different values 
($(1, 2, 5) \times 10^{-3}$~radians) of which we find three distinct bands and 
that $\thetaprime$ increases with increasing $\gby$. The widths of the bands in 
all three cases are primarily dictated by the variation of $\gbl$ over the 
aforementioned range. Note that for $\gby \neq 0$ and, for large $\mzprime$, the 
second term in the numerator dominates, hence, 
$\tan 2\thetaprime \approx 2\thetaprime \sim 4 g_{_{BY}} g^{-1}_{_{BL}}s_{\theta_W}$
and is rendered independent of $\mzprime$, with $\thetaprime$ being constrained to 
be small by the LEP and SLC experiments ($|\thetaprimemax| \simeq 0.005$~radian, which 
we consider here). This explains the flattening of its variation for larger 
$\mzprime$ values.

The appearance of grey regions in these plots for smaller values of $\mzprime$ 
and larger $\theta^\prime$ is expected since such combinations of the two quantities  
lead to stronger deviations from the SM 
expectations and, hence, such regions are the first to get ruled out by an 
experiment. However, it is interesting to note that, for the red, pink and blue 
bands for which $\gby$ increases progressively ($\gby=(1,2,5)\times 10^{-3}$, 
respectively), even increasing values of $\thetaprime$, accompanied by further 
decreasing $\mzprime$, are getting allowed, thus intruding into what should 
already have been forbidden regions. This is because for a given $\mzprime$, 
$\thetaprime$ and, hence, $\gammazprime \over \mzprime$ increase with increasing 
$\gby$ rendering the $Z'$-boson `fatter', which results in a drop in 
sensitivity of the tailored invariant mass window chosen by the experiments for 
a given $\mzprime$, and, in addition, an enhanced propagator that slows down 
the rate at which the dilepton rate would have grown otherwise, this causing 
a further drop in the overall experimental sensitivity. The variations of 
$\gammazprime \over \mzprime$ with $\thetaprime (\gby)$, for a given 
$\mzprime$, are illustrated in the middle plot of
Fig.~\ref{Fig:MZp-thetap-GZpOMZp} and will be discussed below. 
%

Furthermore, the sensitivity of $\thetaprime$ to $\gyb$ appears to be 
comparatively smaller. This is demonstrated only in the multi-colored band 
(third from the top) carrying $\gby=0.001$. However, it should be noted that,  
with increasing $\gyb$, the allowed regions shrink. This is because, with 
increasing $\gyb$, the coupling $Z'\bar{f}f$ increases, leading to an enhanced 
dilepton cross section, thus reducing the allowed region. The thinner 
hyperbolic strip with a black leading portion that runs down to the bottom of 
the plot is for $\gby=0$. Along this strip, $\gammazprime$ remains small.
This is representative of a `minimally non-trivial' $Z'$ scenario that deviates 
from  $Z'_{\rm SSM}$ in its couplings to fermions
$\big(\sim \gbl+\gyb \neq g^{Z\bar{f}f}_{_\mathrm{SM}} \big)$ 
while the $Z'$-boson still remains a narrow resonance.

The variation $\gammazprime \over \mzprime$ (in \%) is illustrated in the 
middle plot of Fig.~\ref{Fig:MZp-thetap-GZpOMZp} using the same set of data as 
for the left plot. The plot shows a rapid growth in
$\gammazprime \over \mzprime$ with increasing $\mzprime$ which is driven by the 
partial width of the $Z'$-boson,
$\gammazprimeww \propto s_{\thetaprime}^2 \mzprime^5/M_Z^2$
in the presence of a non-vanishing $\thetaprime$. Another rapid growth in
$\gammazprime \over \mzprime$, mentioned in a previous paragraph, is seen when 
it jumps from the narrow width regime (with $\sim 3\%$) to a rather broad-resonance one of ($\gtrsim 50\%$) in sync with a growing $\thetaprime$ as 
$\gby$ varies (see Eq.~(\ref{eqn:thetaprime2})) between 0.001~radian
(the multi-color band) and 0.005~radian (the blue band), for a fixed value of 
$\mzprime=5$~TeV. Even for a moderate $\gby$ (=0.002) and $\mzprime=5$~TeV, one 
can find $\gammazprime \over \mzprime$ values as large as 20\%. These would 
already grossly invalidate the NWA and hence a bound on $\mzprime$ which 
assumes this. The rightmost plot is simply a parametric rendition of the same 
data in the plane ${\gammazprime \over \mzprime}$--$\thetaprime$, where the 
magnitude of $\mzprime$ is indicated by the palette. This summarizes the 
contents of the previous two plots and serves as a snapshot of the 
possibilities that exist.
%
%
\begin{figure}[t!]
\begin{center}
\hskip -0.20cm
\includegraphics[width=5.1cm,height=4.7cm]{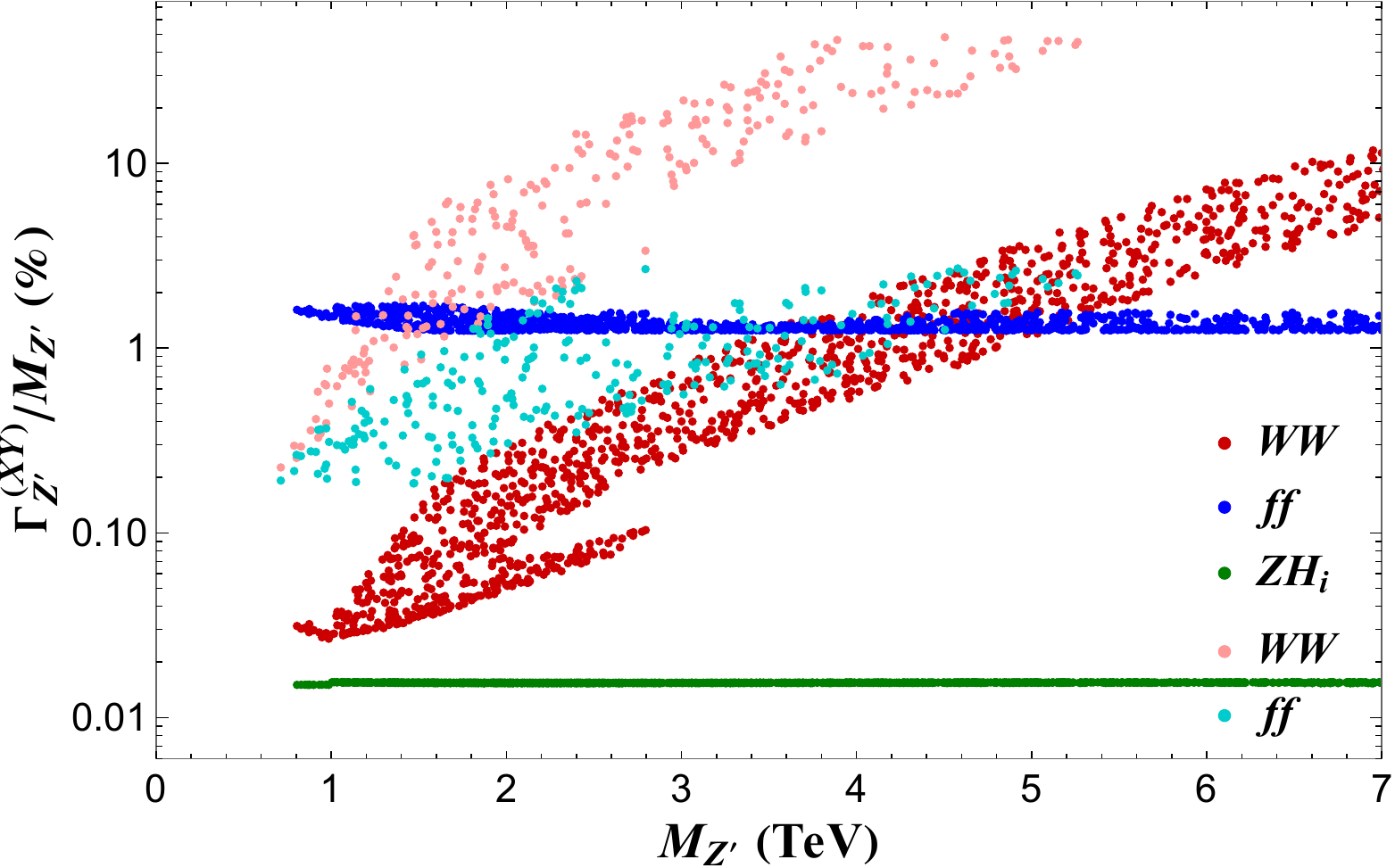}~~~~~~
\hskip -0.30cm
\includegraphics[width=5.1cm,height=4.7cm]{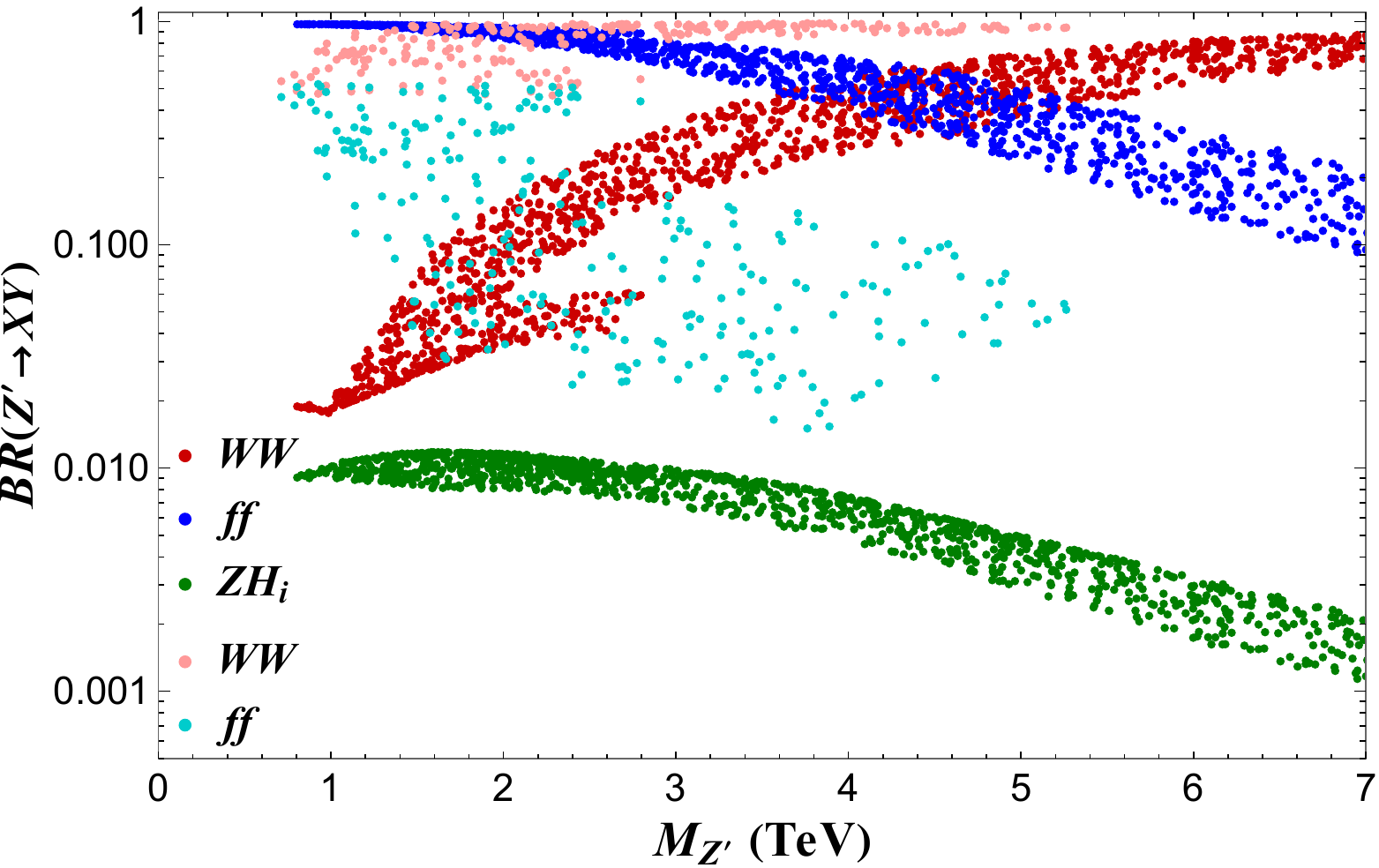}~~~~~~
\hskip -0.30cm
\includegraphics[width=5.1cm,height=4.7cm]{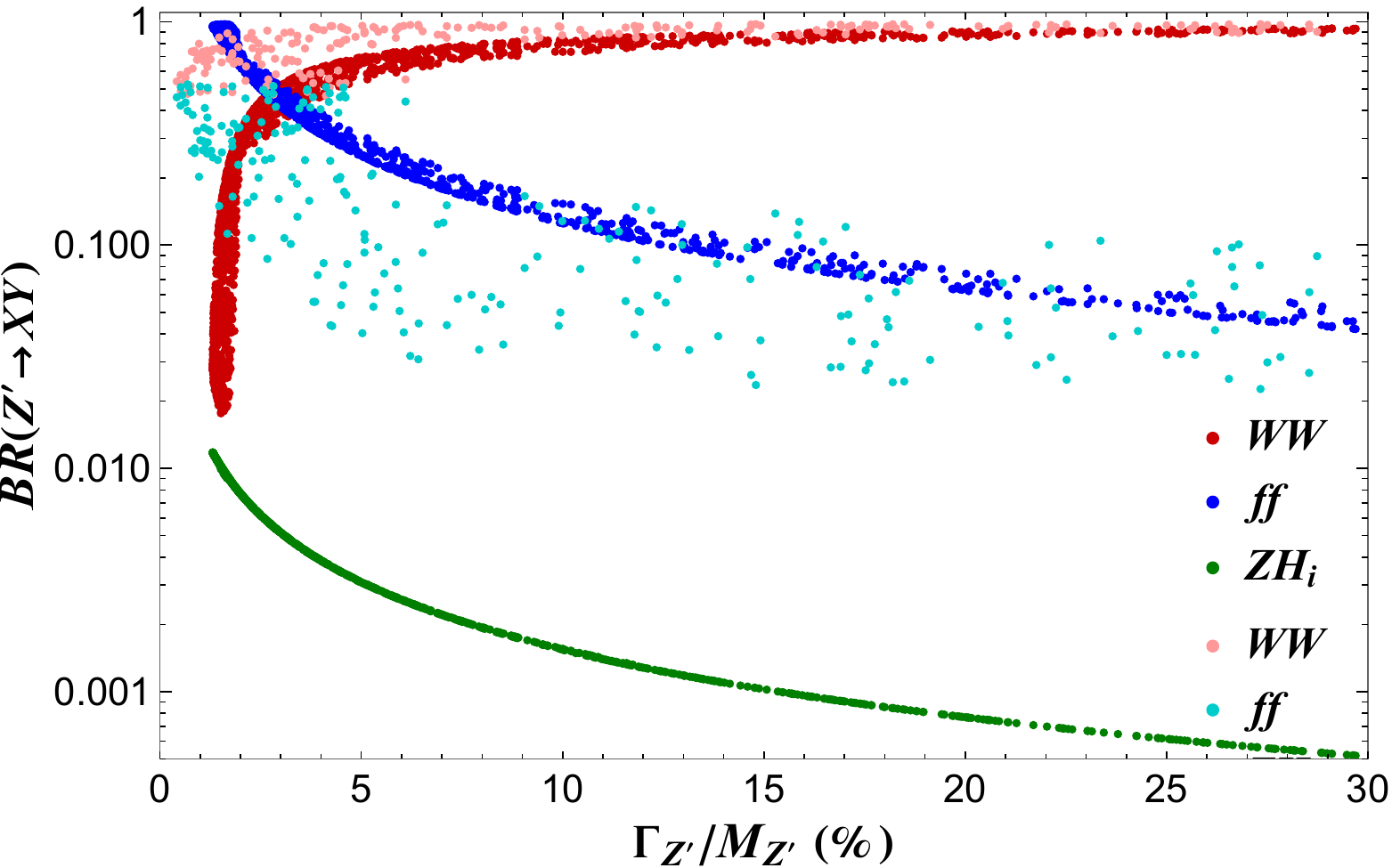}
\caption{
Variations of the ratios of various partial widths of the $Z'$-boson and its 
mass, $\Gamma_{Z'}^{(XY)} \over M_{Z'}$ (left), variations of corresponding 
BRs (middle) as functions of $\mzprime$, and parametric 
variations of these BRs and $\gammazprime \over \mzprime$ while 
$\mzprime$ varies over the range shown in the previous two plots (right). 
The following choices are made for the ranges or values of various coupling 
parameters: $\gbl\in [0.2,0.8]$, $\{\gby, \gyb \}= \{ 0.001, -0.5 \}$ (in 
darker shade) and $\{ 0.005, 0 \}$ (in lighter shade). Note that points in 
light green shade do not appear at all since the relevant observables have very 
small values.
}
\label{Fig:MZp-BR-GZpOMZp}
\end{center}
\end{figure}
%

Plots in Fig.~\ref{Fig:MZp-BR-GZpOMZp} illustrate how the contribution of 
$\gammazprimeww$ to $\gammazprime \over \mzprime$ compares with partial widths 
for the only other somewhat prominent decay modes of the $Z'$-boson, viz.,
$Z' \to f \bar{f}$ (given by $\gammazprimell$) and $Z' \to Z H_i$ (given by 
$\Gamma_{Z'}^{(ZH_i)}$, with $H_i$, $i=1,2,3,4$, denoting two $CP$-even
Higgs-like states from each of the MSSM and BLSSM sectors), for representative 
choices of $\gbl$ ($\in [0.2,0.8]$) and for two representative sets (in 
blue and red in Fig.~\ref{Fig:MZp-thetap-GZpOMZp}) of values for $\gby$ and 
$\gyb$, viz., $\{\gby, \gyb \}= \{ 0.001, -0.5 \}$ (in darker shade) and
$\{ 0.005, 0 \}$ (in lighter shade). Note that SUSY excitations could appear in 
the decay of the $Z'$-boson whenever these are allowed kinematically. However, 
such decays are not prominent here and, hence, we do not present these. We study 
their implications in detail in Sec.~\ref{subsec:susy}.

From the left plot of Fig.~\ref{Fig:MZp-BR-GZpOMZp}, it is seen that, as 
$\mzprime$ varies, $\gammazprimeff \over \mzprime$ stays at a relatively
insignificant level while $\gammazprimeww \over \mzprime$ grows from a similar 
minuscule level to 10\% and beyond for $\mzprime \geq 6 (3)$~TeV for
$\gby=0.001 \,(0.005)$. In particular, note that for $\gby=0.005$,
$\gammazprime \over \mzprime$ attains a value $\sim 100$\% already at 
$\mzprime\approx 6$~TeV; a situation where the dilepton $M_{\ell^+ \ell^-}$ 
distribution would appear as a continuum. In contrast,
$\Gamma_{Z'}^{(ZH_i)}\over \mzprime$ always remains inconsequent, and therefore 
we only plot its contribution when $\gby=0.001$ for which its magnitudes are 
greater than for $\gby=0.005$. The variations of the corresponding BRs are illustrated in the middle plot, which most importantly 
demonstrate how rapidly BR[$Z' \to W^+W^-$] takes off with increasing 
$\mzprime$ and $\gby$. When combined with the observation from the left plot, 
it can be gleaned from this one that by the time the ratio 
$\gammazprime \over \mzprime$ attains a value $\sim 5$\%, BR[$Z' \to W^+W^-$] 
has started to dominate over BR[$Z' \to f \bar{f}$] irrespectively of the 
magnitude of $\gby$. Furthermore, as $\gammazprime \over \mzprime$ reaches a 
level of 10\%, a complete dominance of BR[$Z' \to W^+W^-$] sets in. The plot on 
the right, which is a parametric projection of the first two, confirms this.
%
\subsection{Dilepton and $W^+W^-$ rates via a resonant $Z'$-boson}
\label{subsec:prod-xsec}
In this subsection, we study the production rates of the $Z'$-boson of the 
BLSSM in the dilepton ($e^+e^-$ and $\mu^+ \mu^-$) and the $W^+W^-$ final 
states from which the strongest bounds on the $Z'$ sector are found to arise. 
We do not consider the dijet final state, to which the LHC experiments 
are comparatively much less sensitive, more so when the resonance is 
broad~\cite{CMS:2019gwf}. However, we touch upon the $Z\hsm$ final state to 
check if it could have a substantial rate in the BLSSM scenario. In any case, 
this final state is found to be only weakly constrained at the 
LHC~\cite{ATLAS:2020qiz}.

Given that a broad $Z'$-boson is one of the focus topics of the present study, 
for which the NWA does not hold, we straightaway compute the cross sections for 
the processes $pp \to \ell^+ \ell^-$ with
$\ell \equiv e,\mu$ and $pp \to W^+W^-$. For the dilepton invariant mass ($M_{\ell^+\ell^-}$) 
distribution and for a broad $Z'$ resonance, in particular, when it is not too heavy, there may be a 
collective effect from the SM $\gamma + Z$-mediated process and the
$Z'$-mediated one that is most visible over the tail of the SM contribution at smaller values of $M_{\ell^+\ell^-}$, with which there is now a significant overlap of the shoulder of the
$Z'$-mediated distribution. Hence, throughout, we keep all three diagrams in our computations, while, for the dilepton final state, in addition, we also keep track of the exclusive contributions from the SM (photon- and $Z$-mediated diagrams) and  BLSSM ($Z'$-mediated diagram). We further track the overall BLSSM contribution as the sum of the contributions coming from the
$Z'$-mediated diagram and its interference with the two SM diagrams. This can simply be found by subtracting the SM ($\gamma+Z$) contribution from the total ($\gamma+Z+Z'$) contribution.

A few generic (model-independent) issues should be kept in mind while studying 
the above mentioned cross sections. Apart from on $\mzprime$, the total (uncut) 
cross sections would depend simultaneously on the strengths of the production 
and decay vertices of the $Z'$-boson. Also, an increasing $\gammazprime$ would 
suppress the said cross sections. Further, note that the strengths of the 
participating vertices would affect $\gammazprime$ itself. Thus, if a certain 
decay mode of the $Z'$-boson contributes dominantly to $\gammazprime$, the 
cross section in the corresponding final state tends to get reinforced, 
however, to be eventually pulled back (to a certain extent) by the increased
$\gammazprime$ appearing in the $Z'$ propagator. As we shall see, this tension 
can be of some phenomenological importance. In contrast, the shapes of the differential distributions of the cross section in the invariant mass of the final state particles (in particular, the dilepton final state) are broadly characterized by the quantity $\gammazprime \over \mzprime$.

Before we get into a detailed study of how $\gammazprime \over \mzprime$
systematically affects the reach in $\mzprime$ (or, for that matter, alters the 
lower bound on $\mzprime$) in the dilepton final state within  LHC 
experiments, it would be worthwhile to have some snapshots of their invariant 
mass distributions as compared to that for the corresponding SM background. These would provide insight on how and why the experimental sensitivities of these final states drop with increasing $\gammazprime \over \mzprime$.

\begin{figure}[t]
\begin{center}
\includegraphics[width=6cm,height=4cm,keepaspectratio]{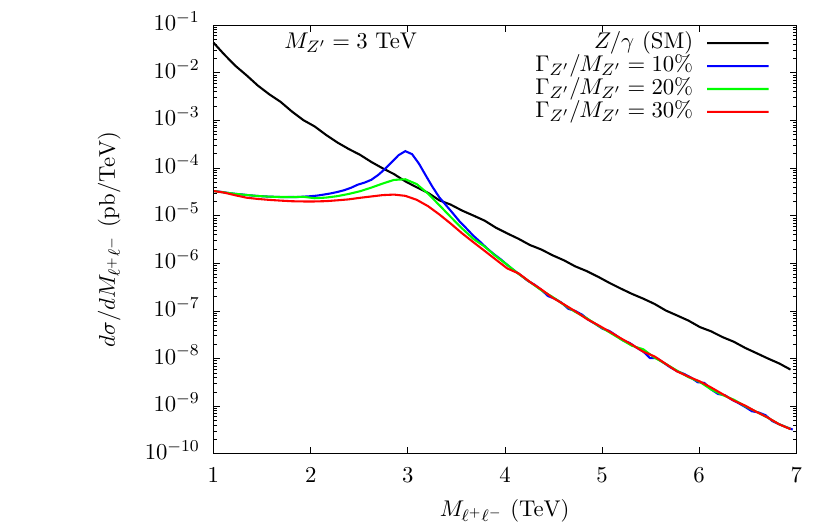}\!\!\!\!\!\!\!\!\!\includegraphics[width=6cm,height=4cm,keepaspectratio]{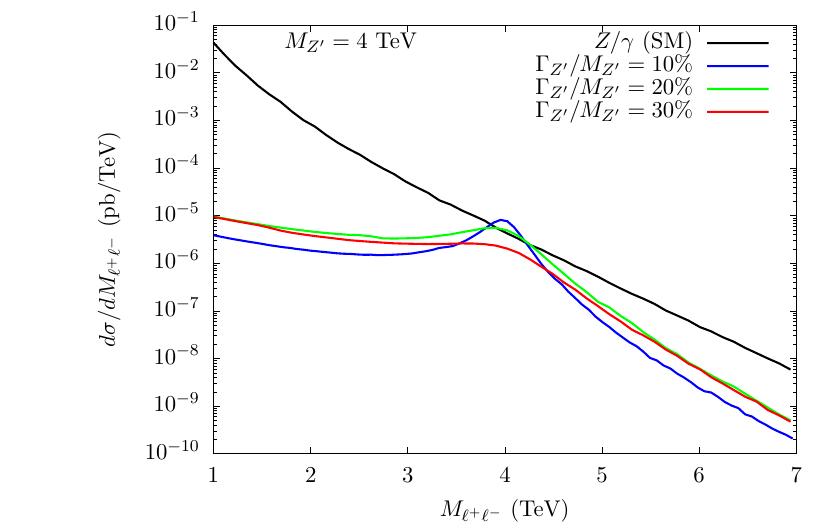}\!\!\!\!\!\!\!\!\!\includegraphics[width=6cm,height=4cm,keepaspectratio]{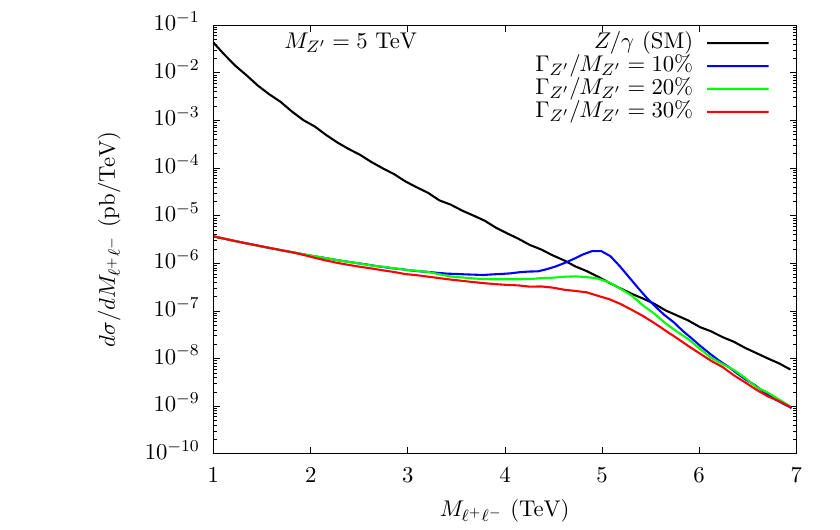} \\
\includegraphics[width=6cm,height=4cm,keepaspectratio]{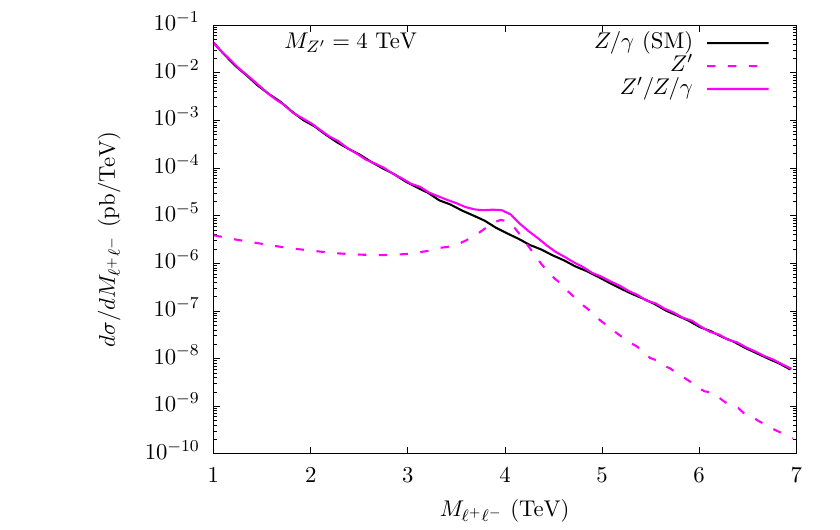}\!\!\!\!\!\!\!\!\!\includegraphics[width=6cm,height=4cm,keepaspectratio]{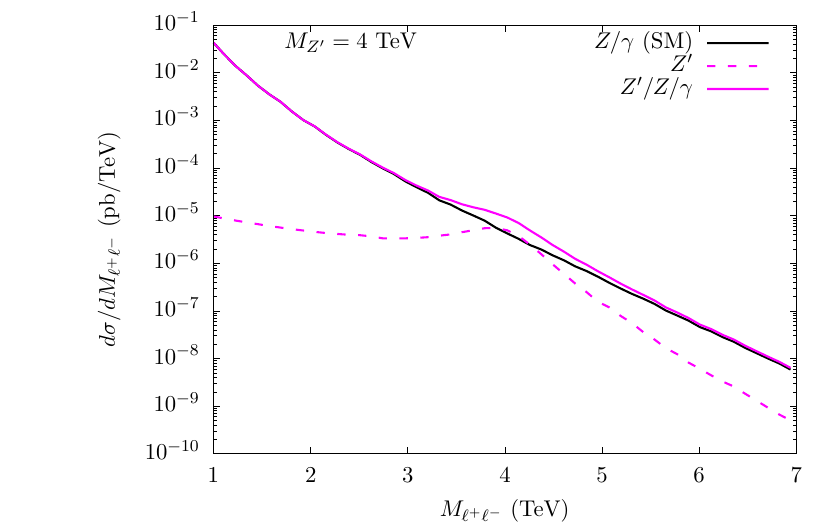}\!\!\!\!\!\!\!\!\!\includegraphics[width=6cm,height=4cm,keepaspectratio]{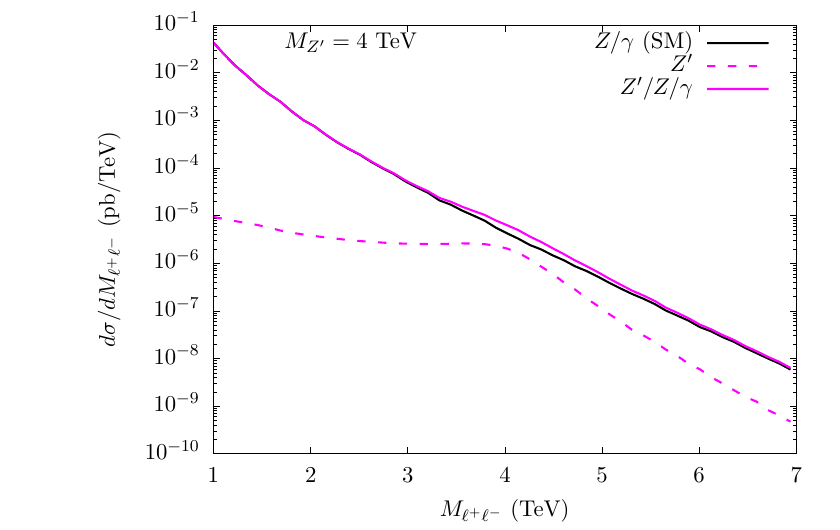}\\[0.5cm]
\caption{Differential (cross section) distributions in $M_{\ell^+\ell^-}$ for 
the process $pp\to \ell^+\ell^-$ via the $Z'$-boson as the only mediator, with 
$\mzprime$ set to $3$~TeV~(left), $4$~TeV~(middle) and $5$~TeV~(right) for 
${\gammazprime \over \mzprime}=10\%$ (in blue), 20\% (in green) and 30\% (in 
red) (top panel), and the same distributions for $\mzprime=4$~TeV with the
$Z'$-boson as the only mediator (dashed magenta) and with the complete set of 
diagrams mediated by photon, $Z$- and $Z'$-bosons (magenta), with
${\gammazprime \over \mzprime}=10\%$ (left), 20\% (middle) and 30\% (right) 
(bottom panel).
}
\label{fig:mee-dist}
\end{center}
\end{figure}

In the top panel of Fig.~\ref{fig:mee-dist}, we compare the theoretical
$M_{\ell^+ \ell^-}$ distributions for the SM (in black) with those deriving 
contributions exclusively from the $Z'$-mediated diagrams, and having three 
different values of $\gammazprime \over \mzprime$ (10\% (blue), 20\% (green) 
and 30\% (red), for $\mzprime=$ 3~TeV (left), 4~TeV (middle) and 5~TeV 
(right). As expected, with increasing $\mzprime$, the peaks of the 
distributions shift towards larger $\mzprime$, and in each plot (i.e., for a 
given $\mzprime$), the distributions flatten out as
$\gammazprime \over \mzprime$ increases.

The plots in the bottom panel of Fig.~\ref{fig:mee-dist} illustrate how the 
physical $M_{\ell^+ \ell^-}$ distributions (in magenta) drawing on the 
coherently combined contributions from diagrams mediated by the photon, $Z$- 
and $Z'$-bosons, turn out to be for a representative value of $\mzprime$ 
($=4$~TeV), and when ${\gammazprime \over \mzprime}=$ 10\% (left), 20\% (middle) 
and 30\%~(right). These are again overlaid onto the SM distributions shown in  
black.
Also, in each plot, the corresponding distributions for only the $Z'$-mediated 
process (in dashed magenta) are presented. Note that these are the very same 
curves shown already in the middle plot of the top panel (which corresponds to 
$\mzprime=4$~TeV). It is clearly seen that the physical $M_{\ell^+ \ell^-}$ 
distributions become increasingly indistinguishable from the SM continuum as 
$\gammazprime \over \mzprime$ increases. Note, further, that the new physics 
contribution is not independent of the SM one, as is pointed out in 
the Introduction.
%

\begin{figure}[t!]
\begin{center}
\hskip -20pt
\includegraphics[scale=0.60]{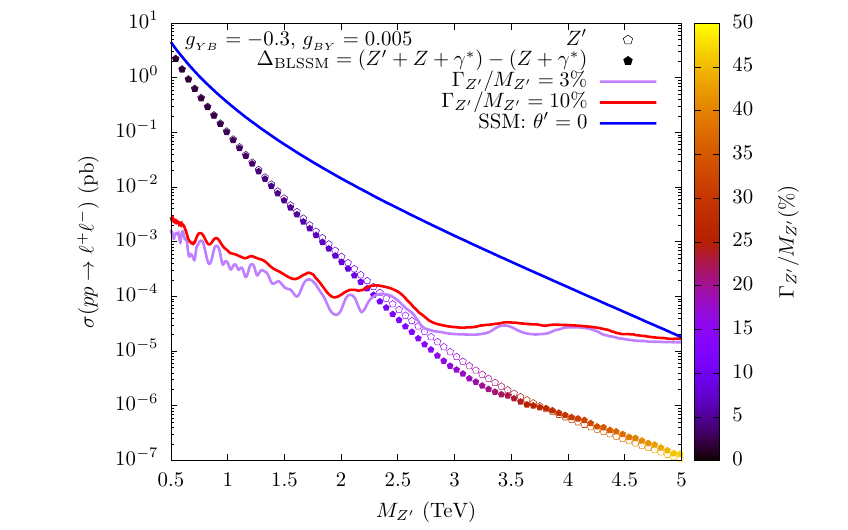}
\hskip -15pt
\includegraphics[scale=0.60]{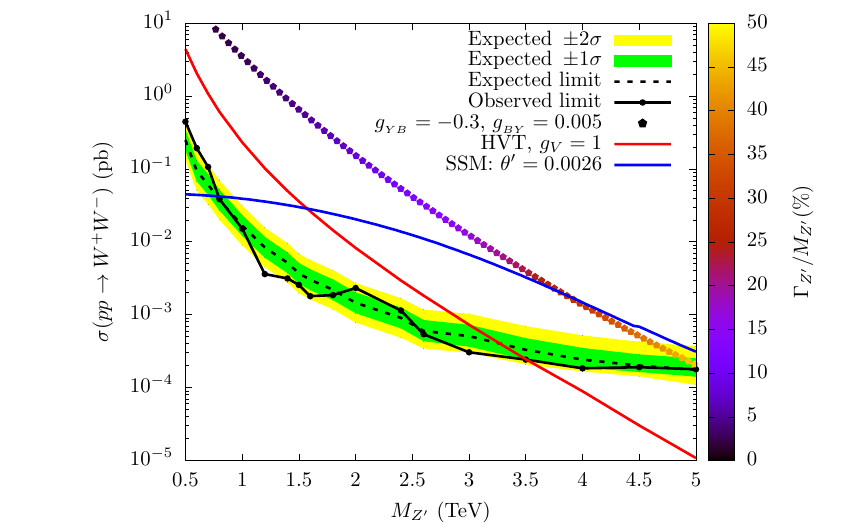}
\caption{Variations of the fiducial dilepton cross sections for
$pp \to \ell^+ \ell^-$, $\ell \equiv e, \mu$,
with $M_{\ell^+ \ell^-} > \mzprime - 2\gammazprime$ (left) and the total cross 
sections for $pp \to W^+W^-$  (right) as functions of $\mzprime$, for various 
different setups/situations, at the 13~TeV LHC and contrasted against the 
corresponding upper limits~\cite{ATLAS:2019erb, ATLAS:2020fry} on the cross 
sections in reference at 95\% CL ($\sim 2\sigma$). See text for details.
}
\label{fig:xsec-relative}
\end{center}
\end{figure}

In Fig.~\ref{fig:xsec-relative} we illustrate the variations of the fiducial 
dilepton cross sections for $pp \to \ell^+ \ell^-$ 
(left) and the total cross sections for $pp \to W^+W^-$ (right), for various 
different setups as functions of $\mzprime$ at the 13~TeV LHC and 
compare those against the respective upper limits 
on the cross sections in reference at 95\% CL ($\sim 2\sigma$), as obtained by 
the ATLAS collaboration~\cite{ATLAS:2019erb}. 
For demonstration purposes only, we use $\{\gyb,\gby\} = \{-0.3,0.005 \}$, for 
both plots. This choice is guided by the fact that a negative $\gyb$ would help 
realize some degree of leptophobia, while the value of $\gby$ would ensure a 
somewhat large $\thetaprime$, and hence a moderately large 
$\gammazprime \over \mzprime$. Although, together, these would still help 
achieve some degree of compliance with the precision data from the LEP and SLC 
experiments, we postpone the discussion about requiring a  stricter compliance to Sec.~\ref{subsec:susy}. 
Thus, the analysis that follows in the current section is essentially of a 
demonstrative nature as to what kind of drastic effect larger $\thetaprime$ 
values (corresponding to a relatively larger value of $\gby=0.005$), 
and hence a `fatter' $Z'$-boson, could inflict on its phenomenology, if they 
survive  said precision constraints. Thus, in this plot, the variation in 
$\gammazprime \over \mzprime$ arises from the variation of $\mzprime$ only.
The variation in $\mzprime$ is achieved by varying $\gbl$ for a fixed value of  
$v'=\sqrt{v_1'^2+v_2'^2}$, with
\{$v'_1, v_2'\} \equiv \{5~\mathrm{TeV},~5.5~ \mathrm{TeV} \}$. The color 
palettes indicate the magnitudes of $\gammazprime \over \mzprime$.
%

The fiducial volume considered in the analysis of the dilepton final state
(the left plot of Fig.~\ref{fig:xsec-relative}) is the same as that adopted in 
Ref.~\cite{ATLAS:2019erb} and is defined by the following set of kinematic 
cuts: $\eta_{\ell_{(e,\mu)}} < 2.5$, $E_T^\ell (p_T^\ell) > 30$ GeV and 
$M_{\ell^+ \ell^-} > \mzprime - 2\gammazprime$. Variations of fiducial 
dilepton rates with $\mzprime$ are then shown for two situations: (i) the pure 
BLSSM rate, i.e., when the contribution is solely due to the $Z'$-mediated 
diagram (represented by the symbol `\pentago'), and (ii) the overall BLSSM 
contribution, $\Delta_\text{BLSSM}$, i.e., the one that contains the pure BLSSM contribution from
$Z'$-mediation, as in (i), plus the contribution from the interference of the 
$Z'$-boson mediated  and the SM diagrams, mediated by the photon and the
$Z$-boson. Thus, $\Delta_\text{BLSSM}$ is equal to the difference 
between the total contribution (due to diagrams mediated by all of the
$Z'$-boson, photon and the $Z$-boson) and the pure SM contribution coming from 
the photon- and $Z$-boson mediated diagrams (represented by the symbol 
`\pentagofill')~\cite{Abdallah:2018kix}). The corresponding variation of the  
dilepton rate involving the SSM $Z'$-boson is  given by the solid blue curve. 
The solid purple and red curves indicate the 95\% CL upper limit on
$\sigma(pp \to \ell^+ \ell^-)$ as reported by the ATLAS 
experiment~\cite{ATLAS:2019erb} when $\gammazprime \over \mzprime$ is 3\% and 
10\%, respectively. This generically reveals that $\mzprime$ is ruled out up to a mass indicated by the point of intersection of any curve representing 
the variation of final state cross section for a particular scenario and that 
indicating the experimentally observed upper bound on the cross section in the 
same final state. Thus, for the SSM, one finds that $\mzprime \lesssim 5$~TeV 
is ruled out. This agrees with the finding of the ATLAS 
collaboration~\cite{ATLAS:2019erb}.

Further, it is observed that both the exclusive and overall contributions of 
the $Z'$-boson to the dilepton final state, in the BLSSM (for the values of 
$\gyb$ and $\gby$ as indicated in the left plot of
Fig.~\ref{fig:xsec-relative}), are smaller than the exclusive contribution of
$Z'_{\mathrm{SSM}}$ to the same final state, much so as $\mzprime$ grows. Also, 
one can see that the overall contribution to the dilepton final state in the 
BLSSM scenario is smaller than the pure $Z'$-boson  over the mass range
$2 \, \mathrm{TeV} \, < \mzprime < \, 3.5 \, \mathrm{TeV}$, which signifies 
destructive interference among the BLSSM and SM processes. The two curves in 
context intersect the limit curves roughly between 2.25~TeV and 2.35~TeV which 
thus represents the ballpark lower bound on $\mzprime$ in the current setup. 
Incidentally, this is a reasonable estimate given that 
${\gammazprime \over \mzprime} \lesssim 10\%$ about the intersection points, 
as can be read from the palette, and that the limit curves are also drawn for 
similar values of $\gammazprime \over \mzprime$. Clearly, in this case, the 
relaxation in the lower bound on $\mzprime$ (compared to the SSM case) is 
essentially due to weaker couplings of the $Z'$-boson to the electrons and 
muons. In addition, a little scrutiny of the plot reveals that  a destructive 
interference among $Z', Z$ and photon mediated processes (noted by comparing 
the bounds obtained from the curves drawn with the symbols `\pentago' and 
`\pentagofill')  also contributes to such a relaxation in a small way 
($\lesssim 50~\text{GeV}$). In Sec.~\ref{subsec:susy}, we will present a 
holistic account of a possible destructive interference and the extent of 
relaxation in the lower bound on $\mzprime$ that it could bring about under a 
realistic setup.

In the right plot of Fig.~\ref{fig:xsec-relative}, we present the variation 
of cross section for the process $pp \to W^+W^-$ mediated only by the
$Z'$-boson (following the relevant experimental analysis of 
Ref.~\cite{ATLAS:2020fry}) of the BLSSM scenario under consideration (the curve 
with points joined by a bold line) as a function of $\mzprime$ and compare the 
same with that of the experimentally observed $2\sigma$ upper limit on the 
same~\cite{ATLAS:2020fry}. Note that while Ref.~\cite{ATLAS:2020fry} reports a 
lower bound of $\approx 3.4$~TeV on the mass of the $Z'$-boson in the so-called 
Heavy Vector Triplet (HVT) scenario with $g_V=1$ (in which the heavy $Z'$- and 
$W'{^\pm}$-bosons have SM-like couplings), we find that the same data rule out 
$\mzprime$ values below $\approx 4.8$~TeV in the BLSSM for $\gyb=-0.3$ and
$\gby=0.005$. Further in comparison is the case for the $Z'$-boson of the SSM 
(with $\thetaprime=0.0026$~radian) which is illustrated by the solid blue line. 
As can be seen, there we find that the lower bound on $\mzprime$ exceeds 5~TeV.
%
\begin{figure}[t!]
\begin{center}
\hskip -25pt
\includegraphics[width=0.49\textwidth,height=4.5cm]{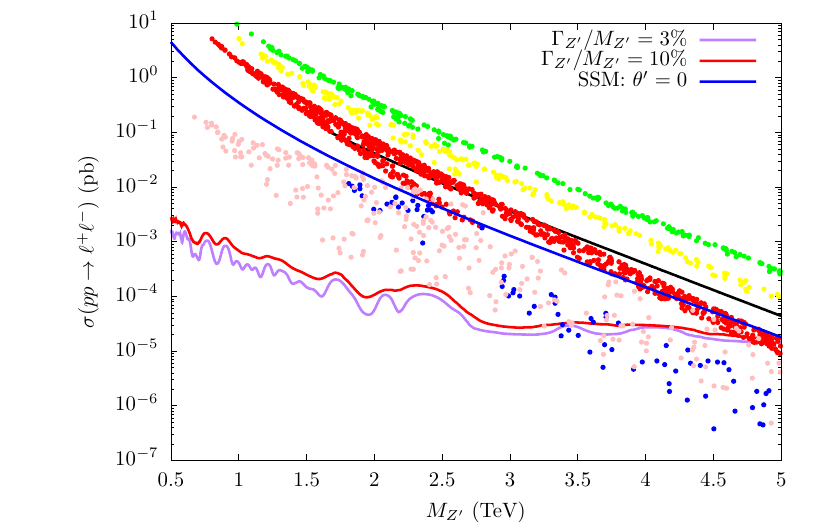}
\includegraphics[width=0.52\textwidth,height=4.55cm]{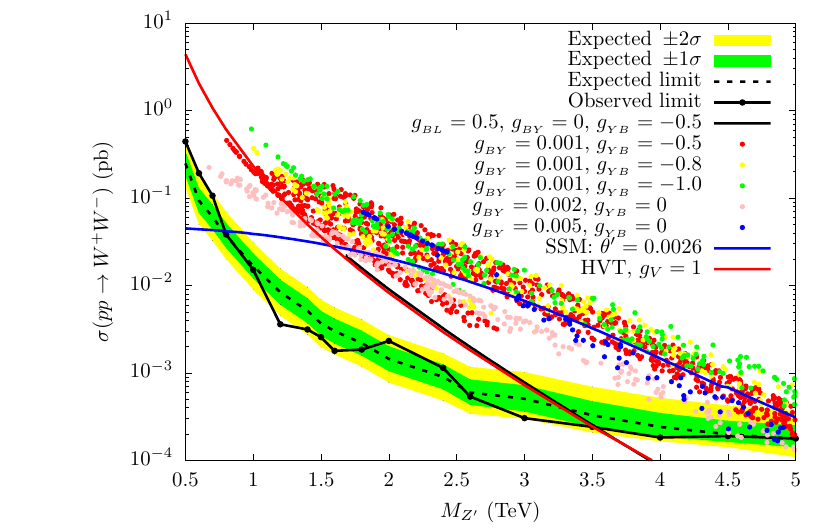} \\
\vskip 10pt
\hskip -25pt
\includegraphics[width=0.50\textwidth,height=4.5cm]{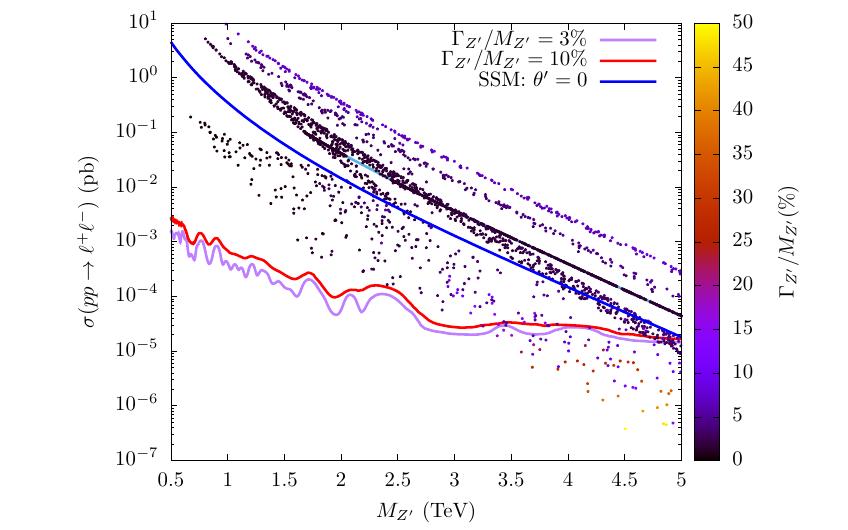}
\includegraphics[width=0.52\textwidth,height=4.5cm]{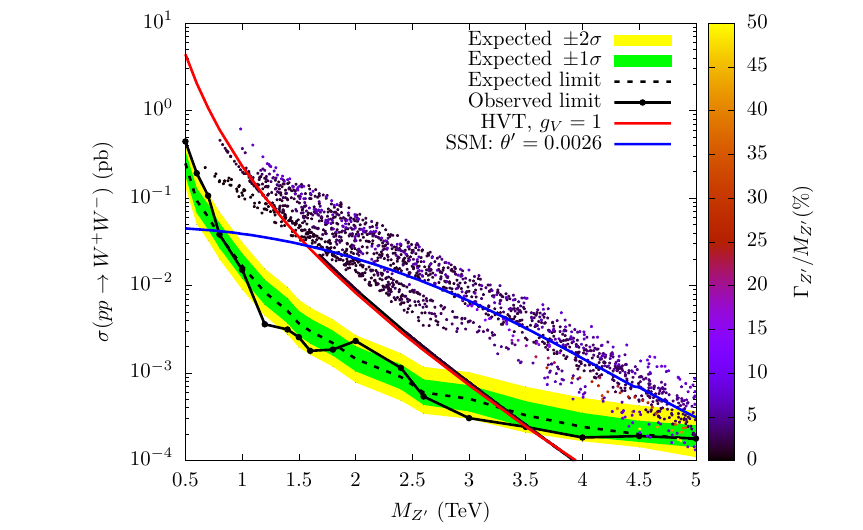}
\vskip 10pt
\hskip -25pt
\includegraphics[width=0.5\linewidth, height=4.5cm]{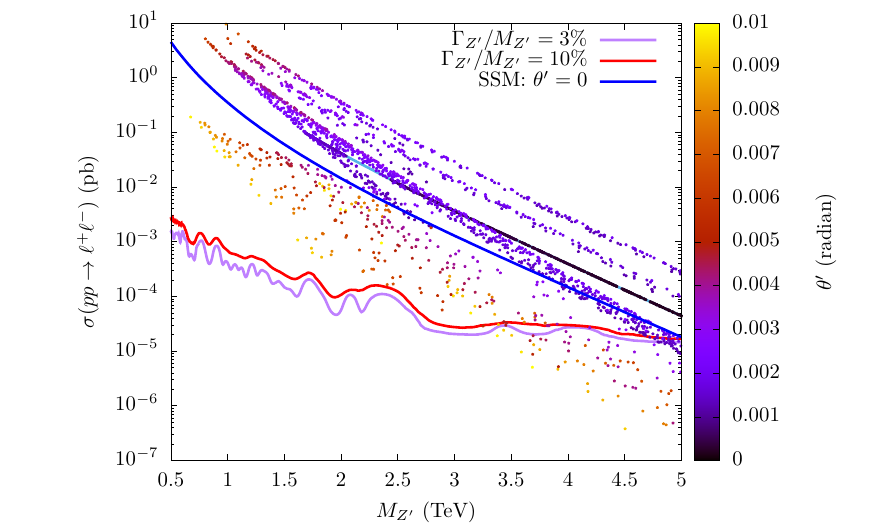}
\includegraphics[width=0.52\textwidth,height=4.5cm]{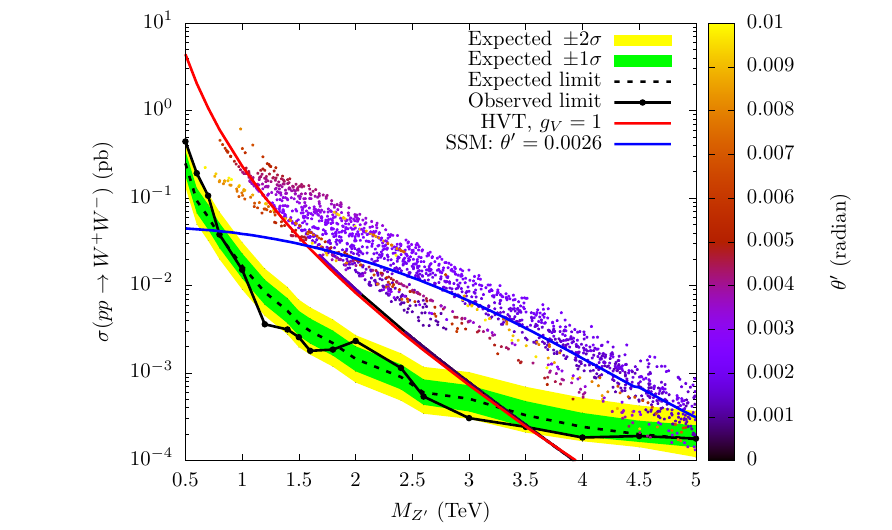}
\caption{
Scatter plots in the left (right) column showing variations of the fiducial 
(total) cross sections for the dilepton ($W^+W^-$) final states at the 
13~TeV LHC, as functions of $\mzprime$, and are generated by scanning over 
$\gbl \in [0.2, 0.8]$, and $v' \in [2 \, \text{TeV}, 25 \, \text{TeV}]$, for 
various representative combinations of values for $\gby$ and $\gyb$ (presented 
in different colors), as indicated in the top right plot, while requiring 
$\thetaprime \leq 0.01$~radian. In the plots in the middle (bottom) panel, 
colors for the same set of points indicate the magnitude of
$\gammazprime \over \mzprime$ ($\thetaprime$), as indicated by the respective 
palettes. See text for details.
}
\label{Fig:MZp-sigma}
\end{center}
\end{figure}
%

In Fig.~\ref{Fig:MZp-sigma} we extend the representative results of 
Fig.~\ref{fig:xsec-relative} to those with various discrete combinations of the 
couplings $\gby$ and $\gyb$ while we scan over
$\gbl \in [0.2,0.8]$ and $v' \in [2 \, \mathrm{TeV}, 25 \, \mathrm{TeV}]$.
For the dilepton final state (left column), we consider the fiducial cross 
section which now includes contributions coming from processes mediated by 
photon, $Z$- and $Z'$-bosons~\cite{ATLAS:2019erb}, while, for the $W^+W^-$ 
final state (right column), we continue to work with the total cross section by 
considering only the $Z'$-mediated process~\cite{ATLAS:2020fry}, following the 
respective experimental analyses, as indicated above. The plots in the top 
panel identify the scattered points that have specific combinations of $\gby$ 
and $\gyb$.

Referring to the dilepton final state (top left plot), it is seen that 
points with somewhat  larger values of $\gby$ (pink (blue) with
$\gby=0.002 \, (0.005)$) survive the latest LHC bounds and the lower bounds on 
$\mzprime$ could get relaxed to 3.7~TeV (3.4~TeV) (down from $\sim 5$~TeV, as 
obtained for the SSM scenario~\cite{ATLAS:2019erb}). As for the $W^+W^-$ final 
state (top right plot), the lower bound on $\mzprime$ is pushed down to
$\sim 3.9 $~TeV (again, down from $\sim 5$~TeV, as obtained for the SSM 
scenario~\cite{ATLAS:2020fry}) at $2\sigma$ level, for $\gby=0.002$. As 
discussed earlier, characteristically for this final state, increasing $\gby$ 
to 0.005 results in an enhanced cross section that overcompensates for any loss 
in cut efficiency and/or possible propagator suppression due to an enhanced $\gammazprime$.

The reasons behind the significant relaxations in the lower bounds of 
$\mzprime$ as seen in the dilepton final state can be traced to an enhanced 
$\gammazprime$, for such larger values of $\gby$. This can be gleaned from the 
middle left plot of Fig.~\ref{Fig:MZp-sigma} for which the same set of scatter 
points (as in the top left plot) are presented, but their colors now reveal the magnitude of $\gammazprime \over \mzprime$ as shown via the palette. Note that, for the points falling below the curves representing the experimental bound, the values of $\gammazprime$ are on the higher side of 10\%, reaching 50\% as $\mzprime$ grows, as a result of which the fiducial dilepton cross section suffers a significant drop due to eroding acceptance of the $M_{\ell^+\ell^-}$ cut originally tailored for the search for a narrow resonance and an enhanced propagator suppression, as pointed out in Sec.~\ref{subsec:mixing-mass-width}.
In this context, we must take note of the fact that the maximum value of 
$\gammazprime \over \mzprime$ considered by the relevant experiment, 
e.g., the ATLAS experiment~\cite{ATLAS:2019erb}, for the purpose, is 10\% and
that the upper bounds on the cross section in this final state would become 
further relaxed to higher values as larger values of
$\gammazprime \over \mzprime$ are considered. Hence, points with
${\gammazprime \over \mzprime} > 10\%$ that are currently found to fall above 
the red curve and therefore appear to have been ruled out may not necessarily 
be so when they are compared with the actual experimental limits obtained for 
similar (i.e., larger) values of $\gammazprime \over \mzprime$. 

For the $W^+W^-$ final state as well, an increase in
$\gammazprime \over \mzprime$ results in a suppression of its cross 
section. For a given value of $\mzprime$ (say around 4~TeV), this can now be 
seen from the middle right plot, which again uses the same scatter points 
appearing in the top right plot, but now reveals the values of
$\gammazprime \over \mzprime$. However, as discussed in the context of 
Fig.~\ref{fig:xsec-relative}, this suppression is counterpoised by a 
simultaneous increase in the direct contribution to the cross section thanks to 
a stronger $Z'W^+W^-$ interaction strength. This is the reason why, for a 
larger value of $\gammazprime \over \mzprime$, the relaxation in the lower 
bound on $\mzprime$ down to about 3.9~TeV, down from 5~TeV, in the
$W^+W^-$ final state, cannot get as drastic (i.e., down to about 2.5~TeV) as 
seen with the dilepton final state, at a comparable CL of around 95\%
($\sim 2\sigma$).

For both final states, the plots in the bottom panel present the same set of 
points as in the ones above, but now color-casted to indicate the values of  $\thetaprime$ via the 
palettes. These plots reveal a direct correlation between the magnitudes of 
$\gammazprime \over \mzprime$ (from the middle panel) and $\thetaprime$, as
expected theoretically. 

Thus, comparing the plots for the dilepton final state with those for the
$W^+W^-$ one, it can be inferred that ultimately it is the latter final state
that dictates the conservative lower bound on $\mzprime$ in the scenario under
consideration. We now bring into the picture possible SUSY decays of the
$Z'$-boson and study their implications for the upper bounds derived on 
$\mzprime$ from experiments. However, we first put together our important 
findings thus far.
%
\begin{itemize}
\item
It is not difficult to achieve a moderately large $\gammazprimeww$ that drives 
$\gammazprime \over \mzprime$ large enough, leading to a broad $Z'$-boson. The 
controlling factor being the magnitude of $\thetaprime$, it is necessary to be 
careful that it complies with the LEP and SLC precision data. We will soon find 
(see Sec.~\ref{subsec:susy}) how $\thetaprime$ draws the constraint from  those 
data and, in turn, how $\gammazprime \over \mzprime$ cannot grow beyond a 
point. These result in more conservative estimates of possible relaxations in 
the bounds mentioned above.
\item
Albeit a moderately large $\gammazprime \over \mzprime$ might help evade 
experimental bounds in the $Z'$ sector, as obtained from LHC analyses in the  
dilepton mode, it sharply invites bounds from LHC searches of the $Z'$-boson in 
the $W^+W^-$ final state.
\item
Conversely, if the $Z'$-boson is a narrow resonance, it could easily evade 
bounds obtained from the searches in the $W^+W^-$ final states, even for a not so heavy $Z'$-boson. However, in that case, the bounds derived from the dilepton data would quickly become rather restrictive.
\item
Hence, a heavy $Z'$-boson of the BLSSM scenario might have escaped current LHC 
searches in those regions of the parameter space where $\gammazprime$ is 
sufficiently large, thus making the narrow-width analysis of the dilepton 
data invalid while the event rate in the $W^+W^-$ final state (which is 
controlled by the same coupling strength as $\gammazprimeww$, the primary
contributor to $\gammazprime$) still remaining sufficiently low to evade the 
current (stronger) bound on $\mzprime$.
\item
Bounds obtained from searches of the $Z'$-boson in the $ZH_i$ final state are 
much weaker than those obtained from the dilepton or $W^+W^-$ final states. 
Hence, we ignore those.
\end{itemize}
%
\subsection{Implications of the $Z'$-boson decaying to SUSY and other BLSSM-specific states}
\label{subsec:susy}
In this section, we discuss the implications of the $Z'$-boson decaying to SUSY 
partners of the SM states and other BLSSM specific states like the extra Higgs 
bosons, the right-handed neutrinos, and their SUSY partners. A healthy 
BR of the $Z'$-boson in such states could straightaway alter 
(relax) the lower bounds on $\mzprime$ as obtained by the LHC experiments by 
analysing the dilepton and $W^+W^-$ final states. Further relaxation in 
these lower bounds is expected if, in addition, the $Z'$-boson is a reasonably 
broad resonance. As discussed earlier, the large width of such a resonance 
finds its dominant contribution in $\gammazprimeww$. Here, it is worth looking 
into whether, alongside $\gammazprimeww$, the combined partial width of the 
$Z'$-boson in such states could play a significant role in turning it into a 
broad resonance. For, a possibly significant share of the latter in 
$\gammazprime$ could help evade stronger bounds from the $W^+W^-$ data that 
arises when $\gammazprimeww$ is the only important contributor to 
$\gammazprime$. 

Under the circumstances, the maximal role that BLSSM-specific decays of the
$Z'$-boson could play in relaxing the reported bounds on $\mzprime$ is when the
right balance is struck between $\gammazprimeww$ and BR[$Z' \to W^+W^-$], while 
BR[$Z' \to$ BLSSM-specific states] is maximized. More specifically, the 
strategy that can be adopted to this end is to look for an optimally large 
$\gammazprimeww$ leading to, say, ${\gammazprime \over \mzprime} \gtrsim 10\%$, 
such that the same helps evade the mass bound derived from the dilepton mode 
to a large extent, while still not large enough to attract a serious 
constraint off the $W^+W^-$ mode. Then, one should strive to maximize the 
collective contribution of these BLSSM-specific states to $\gammazprime$ and 
the corresponding BR. This would further help evade bounds 
coming from both dilepton and $W^+W^-$ final states.

In the presence of BLSSM-specific decay modes of the $Z'$-boson which include various SUSY states, in particular, the EWinos, which can be naturally light, a host of Higgs-like scalars, and the right-handed neutrinos, which are all possibly not so heavy and are dominantly of $B$--$L$ origin, the bulk of $\gammazprime$, is shared alongside by its decays to two other 
decay modes: to $W^+W^-$ and $f\bar{f}$. However, decays of the $Z'$-boson to 
sfermions, including $\tilde{\nu}_R$'s, are kinematically prohibited given that
those states are taken to be much heavier, as mentioned earlier. Thus, the 
normalization of the $Z'$-boson's decay BRs is ensured by 
considering its three generic decays, viz., decays to a pair of SM fermions
($f \bar{f}$) and $W^+W^-$, which lead to canonical final states in which the 
recent searches of the $Z'$-boson at the LHC are mostly carried out, and the 
rest comprising its accessible BLSSM-specific decays, as mentioned above. 
These then allow us to present these three generic BRs of the 
$Z'$-boson in ternary scatter plots.

Towards this end, we now undertake a detailed scan of the BLSSM-specific 
coupling parameters over the ranges $\gbl \in [0.2.0.8]$,
$\gby \in [-0.005,0.005]$ and $\gyb \in [-0.5,0.5]$. Further, in this section, 
we now ensure that the value of $\thetaprime$ (which governs the size of 
$\gammazprime \over \mzprime$, the variable that is critically important for 
the present analysis) that we use strictly complies with the precision $Z$-pole 
data from the LEP and SLC experiments.
%
\begin{figure}[!t]
\begin{center}
\includegraphics[width=10cm,height=8cm]{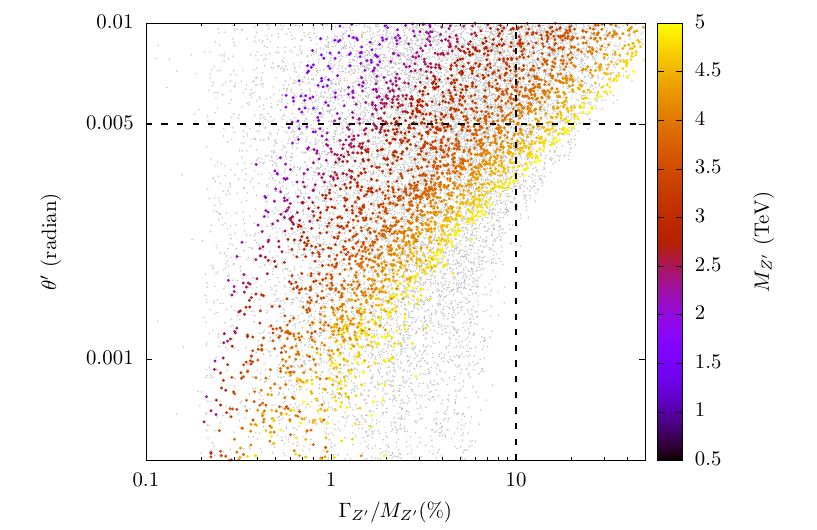}
\caption{Scatter plot in the ${\gammazprime \over \mzprime}$--$\thetaprime$
plane with $\mzprime$ in the color palette. BLSSM-specific coupling parameters 
over the following ranges: $\gbl \in [0.2.0.8]$, $\gby \in [-0.005,0.005]$ and 
$\gyb \in [-0.5,0.5]$. Also, $v'$ is varied over the range
$v' \in [2 \, \text{TeV}, \, 25 \, \text{TeV}]$. The points satisfy the 
somewhat weaker bound of Eq.~(\ref{eqn:oblique}) obtained from the global fit 
of the oblique parameters in the precision data from LEP and SLC experiments. 
Grey points are ruled out at 2$\sigma$ level by the dilepton data used in 
searches for a heavy $Z'$-boson at the 13~TeV LHC.
}
\label{Fig:MZp-thetap-GZpOMZp-1}
\end{center}
\end{figure}
%

The resulting distribution of points in the plane
$\frac{\gammazprime}{\mzprime}$--$\thetaprime$ is shown in
Fig.~\ref{Fig:MZp-thetap-GZpOMZp-1}, with associated values of $\mzprime$ 
indicated by the colors of the palette attached to it. These points satisfy the 
somewhat weaker bound of Eq.~(\ref{eqn:oblique}) obtained from the global fit 
of the oblique parameters in the precision data from LEP and SLC experiments. 
The upper right quadrant of this plot reveals that a value of
$\gammazprime \over \mzprime$ up to 20\% could be achieved for
$\mzprime \sim 5$~TeV, when we require a more stringent constraint of 
$\thetaprime \leq 0.005$~radian, to comply with the precision data from the
LEP and SLC experiments, as discussed in Sec.~\ref{subsec:bounds}. This 
constraint restricts us from achieving relatively large values of
$\gammazprime \over \mzprime$ ($\gtrsim 10\%$) for smaller values of 
$\mzprime$. Furthermore, points in grey are ruled out at $2\sigma$ level 
by the dilepton data used in searches of the $Z'$-boson at the 13~TeV LHC, 
similar to what was mentioned in the caption of
Fig.~\ref{Fig:MZp-thetap-GZpOMZp}. Note that
Fig.~\ref{Fig:MZp-thetap-GZpOMZp-1} displays scatter points in the 
same plane as the rightmost plot of Fig.~\ref{Fig:MZp-thetap-GZpOMZp} does. 
However, they differ in their appearances since the latter presents points only 
for discrete sets of values of various gauge couplings of the BLSSM scenario 
while, for the former, we scan over those coupling parameters. In what follows, 
we have imposed a somewhat stringent set of precision constraints in the form 
of $\thetaprime \leq 0.005$~radian and
$g_{\text{avg}}^\ell . \thetaprime \leq 0.0005$ (see Sec.~\ref{subsec:bounds}, 
and Fig.~\ref{Fig:gBL-gYB-contours} therein).
%
\begin{figure}[t]
\begin{center}
~\hspace*{-1.cm}\includegraphics[width=9.5cm,height=7.25cm]{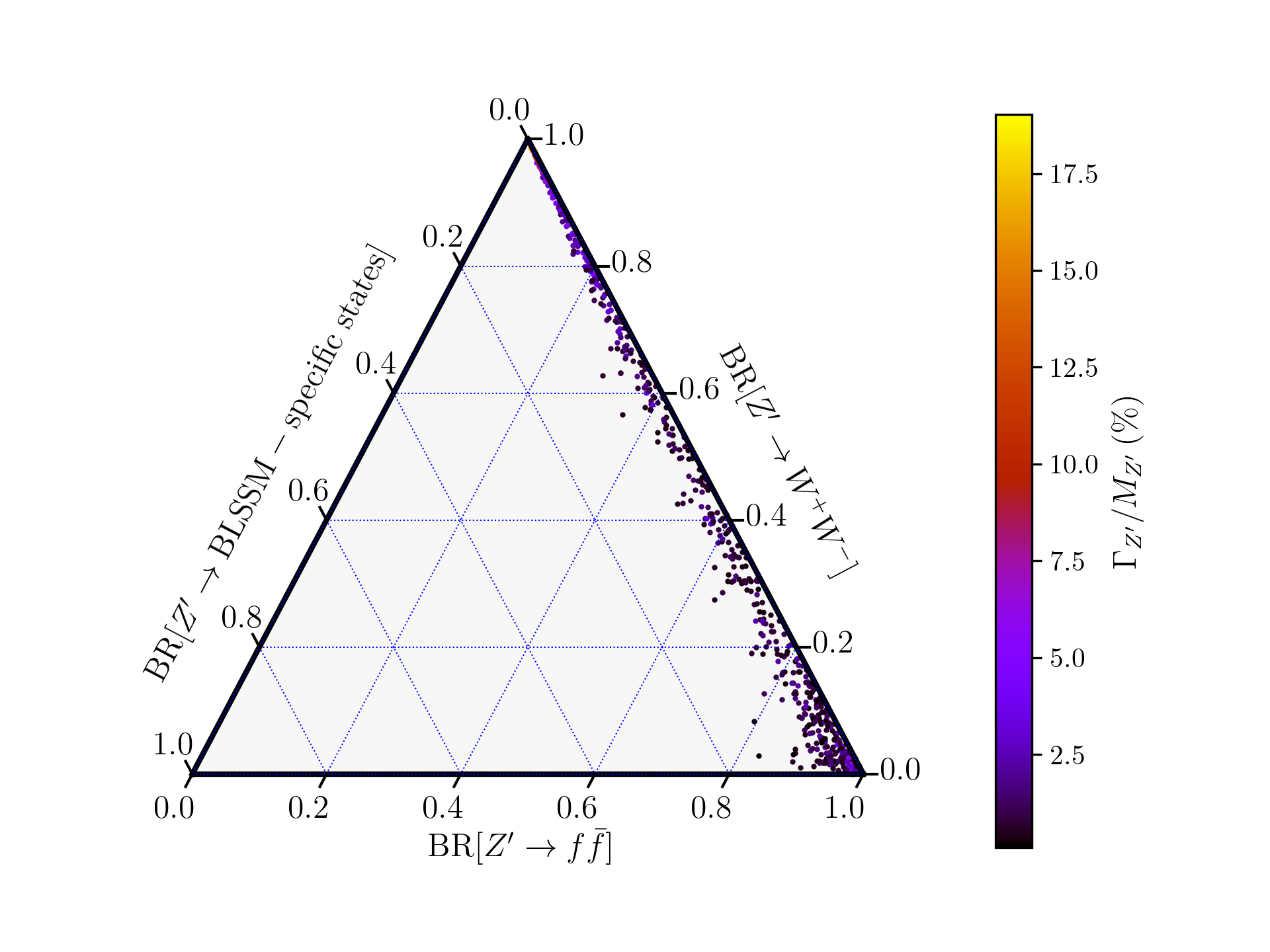}~\hspace*{-1cm}\includegraphics[width=9.5cm,height=7.25cm]{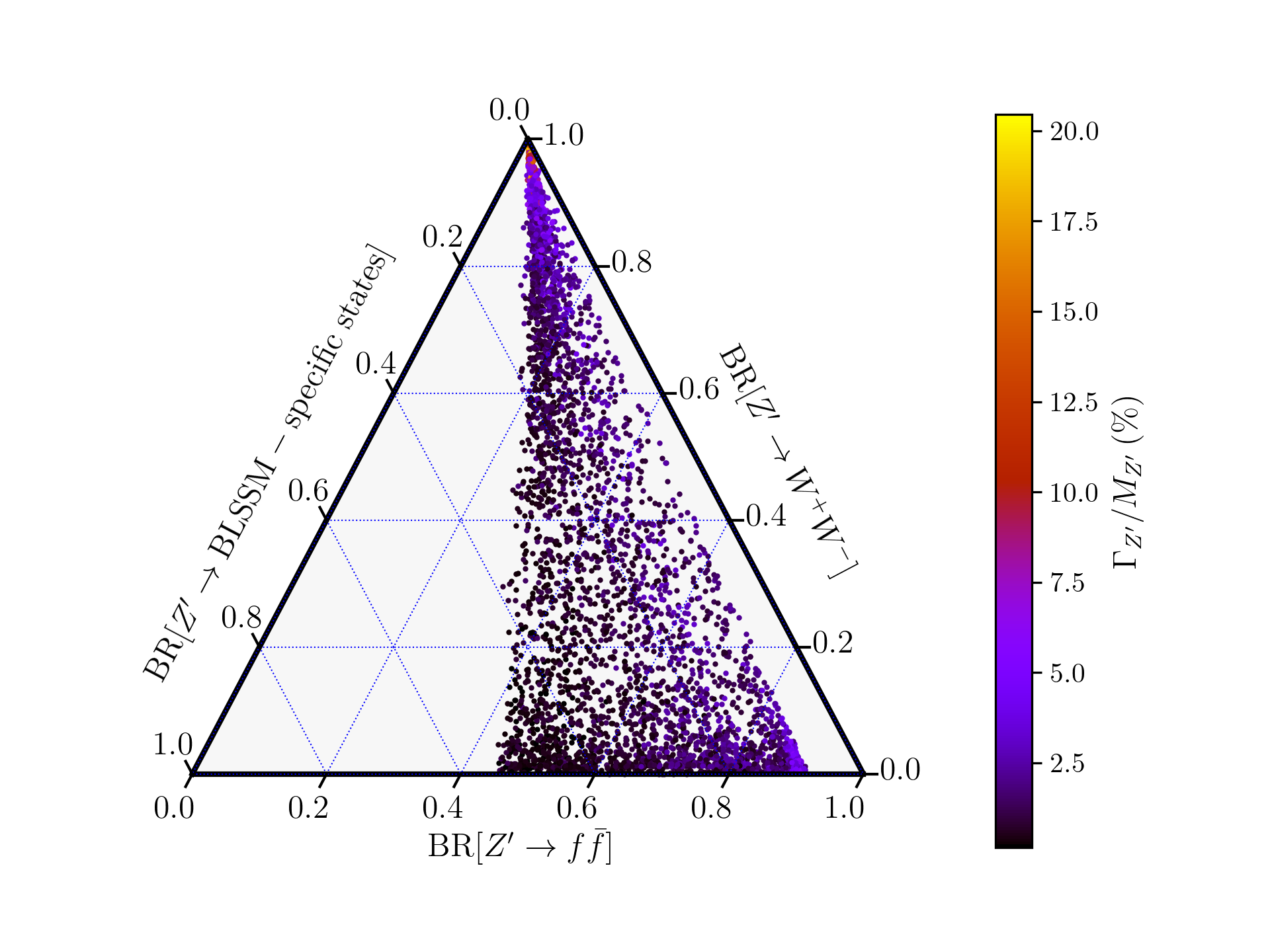}
\caption{Ternary scatter plots showing various BRs of the
$Z'$-boson for $M_1,M_2 < \mu,\mu'$ (i.e., the LSP and NLSP are not the 
bileptinos or higgsinos; the left plot) and $\mu,\mu' < M_1,M_2$ (i.e., the LSP 
and NLSP are bileptinos and/or higgsinos; the right plot). Points are generated 
by scanning the BLSSM-specific coupling parameters over the same ranges as in 
Fig.~\ref{Fig:MZp-thetap-GZpOMZp-1}. All sfermions are decoupled, while 
$M_{B'}$ is taken to be large ($|M_{B'}| \gg \mu,\mu', M_1,M_2$) and negative. 
Colors are indicative of values of $\gammazprime \over \mzprime$, as presented 
in the attached palettes. Various BLSSM-specific parameters are varied over the 
same ranges as in Fig.~\ref{Fig:MZp-thetap-GZpOMZp-1}. See text for details.
}
\label{Fig:MZp-BR-SUSY}
\end{center}
\end{figure}
%

In Fig.~\ref{Fig:MZp-BR-SUSY}, we present the ternary scatter plots with the
three dominant BRs along the arms of the triangles while the 
colors in the palettes indicate the magnitude of $\gammazprime \over \mzprime$.
It can be gleaned from Eqs.~(\ref{eqn:zpninjl})--(\ref{eqn:zpcicjr}) that the
$Z'$-boson mainly couples ($c_{\thetaprime}$-enhanced) to the bileptino and 
higgsino components of the neutralinos and the higgsino components of the 
charginos. Motivated by this, to keep the study tidy, we consider the following 
two broad scenarios for our immediate purpose, i.e., to understand how the SUSY 
final states in the decay of the $Z'$-boson share its BR with 
the other prominent decay modes of the $Z'$-boson, i.e., the $f\bar{f}$ and the 
$W^+W^-$ final states:
%
\begin{itemize}
\item 
$M_1,M_2 \ll \mu,\mu'$ (see the left plot of Fig.~\ref{Fig:MZp-BR-SUSY}), when the $Z'$-boson decays only to lighter EWinos that are mostly
gaugino-like with some higgsino admixtures, and 
\item
$\mu,\mu' \ll M_1,M_2$ (see the right plot of Fig.~\ref{Fig:MZp-BR-SUSY}, for which the $Z'$-boson could only decay to EWinos that are dominantly higgsinos and/or bileptinos.%
\footnote{We consider the off-diagonal element $M_{B'}$ appearing in the 
neutralino mass matrix of Eq.~(\ref{neutralino-mass-matrix}) to be negative and 
$|M_{B'}| \gg (\mu,\mu' \ll M_1,M_2)$. These help~\cite{Abdallah:2017gde} 
reduce the mass of the neutralino which is a mixture of bileptinos and BLino 
and allow the $Z'$-boson to decay into the same, thus enhancing its SUSY 
BR.}
\end{itemize}
In both cases, for the lighter EWino species, we consider sub-TeV masses.

The obvious difference between these two plots is that, in the right plot, the 
collective BR of the $Z'$-boson to various SUSY states 
(essentially bileptinos and higgsinos) and other scalars could reach 50\% (near 
the midpoint of the base of the equilateral triangle), when the rest 
of the BR is shared by BR[$Z' \to f\bar{f}$]. This is expected since, here, the lighter EWino states are 
dominated by bileptinos and the higgsinos and, at the same time, these are the 
components to which the $Z'$-boson does have couplings at the tree level. 
However, note that the SUSY contribution to $\gammazprime$, even when it is 
contributing maximally to the same, is not able to turn the $Z'$-boson fat 
enough to invalidate the use of NWA.

In contrast, in the left plot, the SUSY decays of the $Z'$-boson 
generally struggle to constitute even 20\% of $\gammazprime$ (though 
occasionally reaching up to 30\%). This is because the lighter EWinos 
are bino- and wino-dominated, the components to which the $Z'$-boson does not 
have a tree-level coupling, and the $Z'$-boson only interacts with such states 
through their subdominant bileptino and higgsino admixtures.

The next notable feature is the vertical edge of the populated region in the 
right plot. Along this edge
$\mathrm{BR}[Z' \to \mathrm{SUSY \; states}] \simeq \mathrm{BR} [Z' \to f\bar{f}]$.
For the populated region on the right of this edge, the former remains smaller 
than the latter. These features can be traced back to the fact that the 
specific $Z'$-boson couplings that induce these decays are very similar 
functions of the parameters $\gyb$ and $\gbl$
(see Eqs.~(\ref{eqn:zprime-ululbar})--(\ref{eqn:zprime-lrlrbar}) 
and~(\ref{eqn:zpninjl})--(\ref{eqn:zpcicjr})), only to be suppressed by the 
mixing factors in the neutralino/chargino sector, i.e., $N_{ij}$, $U_{ij}$ and 
$V_{ij}$ (with values $\leq 1$), for the SUSY decays. The blank wedge along the 
right arm of the triangle that shrinks as one moves towards the vertex appears 
due to the competing nature of fermionic and SUSY decays of the $Z'$-boson 
(which enables the SUSY decays of the $Z'$-boson to share a reasonable 
BR with BR[$Z' \to f \bar{f}$], in the limit of a vanishing 
$\gammazprimeww$, i.e., at the right end of the base arm), whose roles gradually 
diminish as $\gammazprimeww$ steadily increases in moving towards the vertex.

The most important message from the right plot of Fig.~\ref{Fig:MZp-BR-SUSY} 
is that a large $\gammazprime \over \mzprime$ (the reddish points near the 
vertex) necessarily associates with a large $\gammazprimeww$, thus pointing to 
the general fact that larger width to mass ratios for the $Z'$-boson are 
predominantly associated with larger $\gammazprimeww$. It should also be noted 
that unlike for the SUSY or fermionic decay widths of the $Z'$-boson, its 
partial width to $W^+W^-$ is dictated by $\thetaprime$ in a major way for small 
values of $\thetaprime$ (see Eq.~(\ref{eqn:gammaz})), which, in turn, is 
proportional to $\gby$ (see Eq.~(\ref{eqn:thetaprime2})).  The purplish band at 
the bottom right edge still signifies a borderline narrow $Z'$ resonance with 
its width mostly fed by $\gammazprimeff$.  The color starts to fade towards the 
red end as one moves towards the vertex, signifying an already moderate 
$\gammazprime \over \mzprime$ ($\sim10\%$) derived mostly from a fast growing  
$\gammazprimeww$. As for the left plot, it just conveys that when the lighter 
EWinos, which are the only SUSY states kinematically accessible in the 
decay of the $Z'$-boson, are gaugino-like, the total decay BR  
of the $Z'$-boson to those states remains small, barely exceeding 10\%.

Thus, we find that for larger values of $\gammazprime \over \mzprime$ 
($>10\%$), the significant contribution to it comes from  $\gammazprimeww$ and 
SUSY decays of the $Z'$-boson cannot become much instrumental in spawning 
further relaxation of the reported lower bounds on $\mzprime$ over and above 
what is already achieved due to a larger value of
$\gammazprime \over \mzprime$. In contrast, for a smaller
$\gammazprime \over \mzprime$, the SUSY decays of the $Z'$-boson could 
collectively play a significant role in depleting the BRs to 
both $\ell^+\ell^-$ and $W^+W^-$ final states. This helps evade the respective 
upper bounds on $\sigma \times \mathrm{BR}$ (and hence, the corresponding lower 
bounds on $\mzprime$) from the LHC when the $Z'$ resonance is not that broad. 
%

\begin{figure}[t!]
\begin{center}
\includegraphics[width=8.1cm,height=6cm]{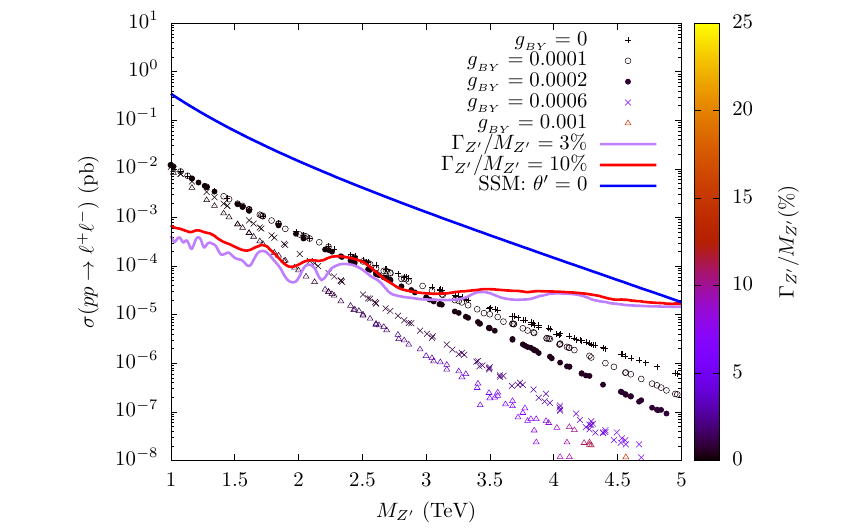}
\includegraphics[width=8.1cm,height=6cm]{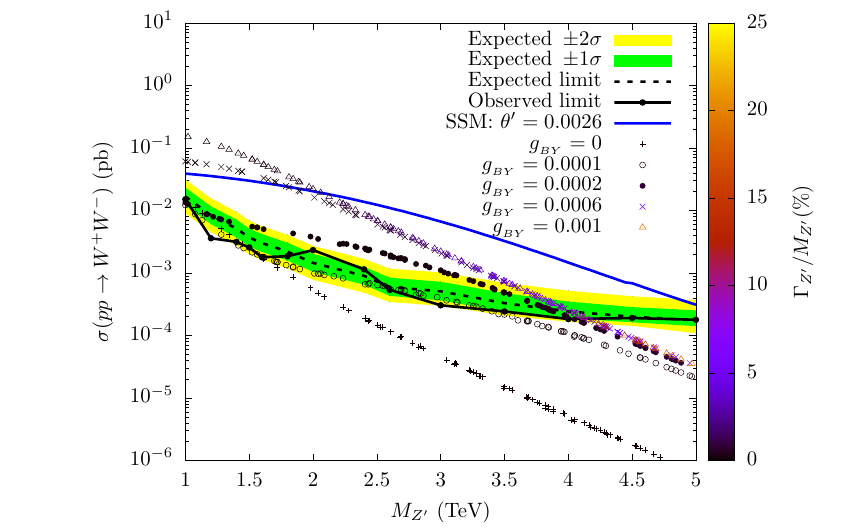}\\ 
\includegraphics[width=8.1cm,height=6cm]{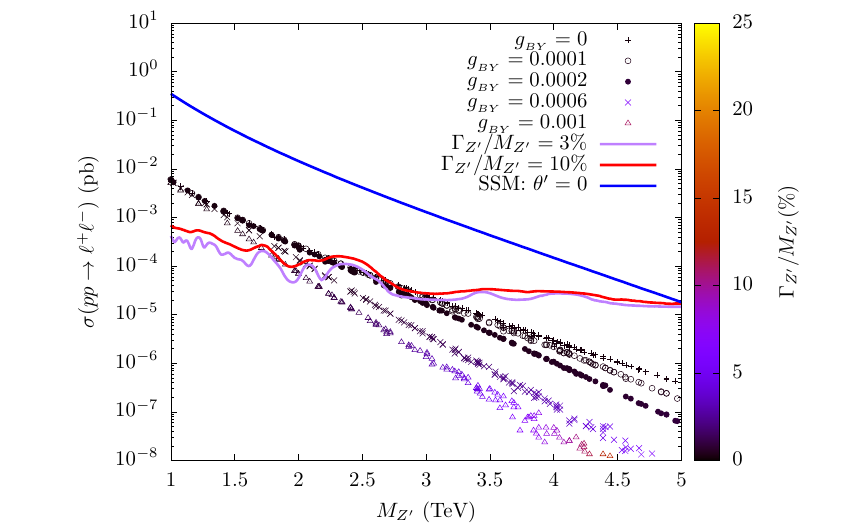}
\includegraphics[width=8.1cm,height=6cm]{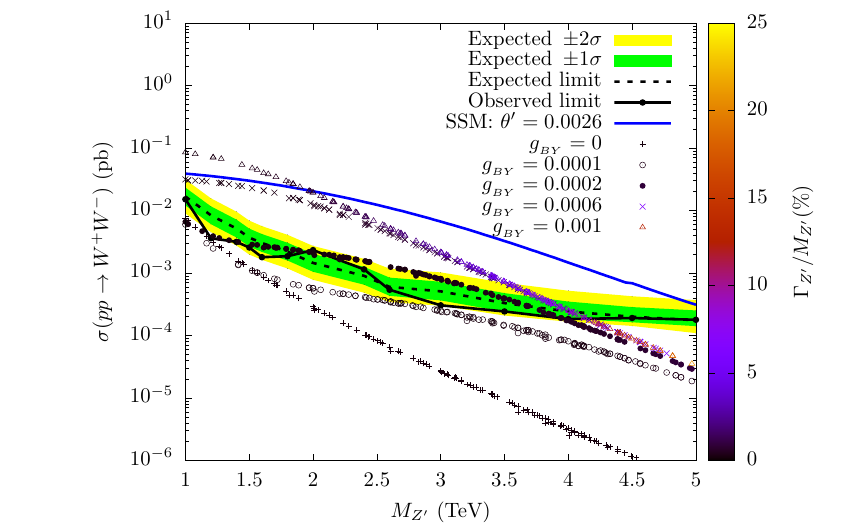} 
\caption{Variations of dilepton (the left column) and $W^+W^-$ (the right 
column) cross sections as functions of $\mzprime$, for five different values of 
$\gby=\{0, (1,2,6)\times 10^{-4},0.001\}$, and fixed values of $\gbl \,(=0.2)$ 
and $\gyb \, (=-0.12)$, with a varying $v'$ controlling the value of 
$\mzprime$. The top (bottom) panel corresponds to the case where
$\mu,\mu'\gg (\ll) \, M_1,M_2$. See text for details.
}
\label{Fig:MZp-sigma-SUSY-ll-WW-Bilepton}
\end{center}
\end{figure}
%

The above discussions can be put in context in the following way. As pointed 
out earlier, the loss in sensitivity that a sufficiently broad $Z'$-boson might 
otherwise have inflicted on the current searches of the $Z'$-boson in the 
$W^+W^-$ final state could be more than compensated for by an accompanying 
enhanced event rate in this final state. Thus, evading the lower bound on 
$\mzprime$ as obtained from searches in the $W^+W^-$ final state becomes 
difficult even for a moderately large value of $\gammazprime \over \mzprime$, 
or for that matter, $\thetaprime$, as permitted by $Z$-pole precision data.
Fig.~\ref{Fig:MZp-BR-SUSY} reveals that SUSY decays of the $Z'$-boson cannot 
contribute significantly to $\gammazprime$ even when
BR[$Z' \to \mathrm{SUSY \, states}$] reaches its maximum in the ballpark of 
50\%. So any weakening of the lower bound of $\mzprime$ that an enhanced rate 
of SUSY decays of the $Z'$-boson could lead to has to be via the resulting 
suppression of $Z'$-boson's BRs to $W^+W^-$ and $\ell^+\ell^-$ states, 
the $Z'$-boson still remaining a narrow resonance. In passing, it may be 
observed that BR[$Z' \to f \bar{f}$] receives contributions from
BR[$Z'\to q \bar{q}$] and BR[$Z'\to \ell \bar{\ell}$], the former having the 
larger share. However, it is well known that the dijet final state is prone to 
a large SM background. Hence, the dilepton final state offers a higher reach 
in $\mzprime$ even though it has a lower yield.

In Fig.~\ref{Fig:MZp-sigma-SUSY-ll-WW-Bilepton} we plot the variations of dilepton (left column) and $W^+W^-$ (right column) cross sections as functions 
of $\mzprime$, for five different values of
$\gby=\{0, (1,2,6)\times 10^{-4},0.001\}$, and fixed values of $\gbl \,(=0.2)$ 
and $\gyb \, (=-0.12)$, with a varying $v'$ controlling the value of 
$\mzprime$, for the cases $\mu, \mu' \gg (\ll) M_1,M_2$ presented in the upper 
(lower) panel, while the attached color palettes indicate the magnitude of 
$\gammazprime \over \mzprime$. These demonstrate the extent to which the lower 
bounds of $\mzprime$ could be relaxed in the presence of appreciable BRs of the $Z'$-boson to SUSY plus other BLSSM-specific states, both in 
the dilepton and $W^+W^-$ final states, for representative sets of
BLSSM-specific gauge coupling parameters.

The upper panel, for which the $U(1)$ and $SU(2)$ gauginos are the LSP and the 
NLSP, indicates that, for $\gby=\{1,2\} \times 10^{-4}$, there exist points 
allowed by both dilepton and $W^+W^-$ final states down to around
$\mzprime \sim 3.3$~TeV, while, in the lower panel, where higgsinos/bileptions 
are the lightest of the EWinos, one can find such allowed points down 
to $\mzprime \sim 2.5$~TeV (to see this, in both cases, one needs to track the 
points represented by circles with a dot inside and the filled circles). This 
clearly points to the fact that when the higgsinos and/or bileptinos are 
kinematically accessible in decays of the $Z'$-boson, thus resulting in a 
moderate to healthy BR of the same into such SUSY states, there 
would be a significant relaxation in the lower bound on $\mzprime$, as compared 
to when that is not the case. However, note that in both of these situations, 
the palettes indicate that the magnitudes of $\gammazprime \over \mzprime$ are 
on the smaller side, thus rendering the $Z'$-boson a narrow resonance. 

Also, note that the sequences that the values of $\gby$ follow, for which the 
points are ruled out or not, are just the opposite for the dilepton and 
diboson cases, i.e., for larger (smaller) values of $\gby$, expected yields in 
the dilepton final state tend to get allowed (disallowed) by the LHC 
analyses, and the situations are just the opposite for the $W^+W^-$ final 
state. In general, this is because $\gby$ dictates the magnitude of
$\thetaprime$, which, in turn, controls $\gammazprime \over \mzprime$, and, as 
we have already discussed, the magnitude of the latter affects the yields of 
these final states in opposite ways. These make of the presented $\gby$ values 
only $\gby =\{1,2\}\times 10^{-4}$ representative, as those only are allowed by 
searches in both dilepton and $W^+W^-$ final states. However, our (fixed) 
choices of other coupling parameters ($\gyb$ and $\gbl$) imply that when these 
parameters are also varied, there would be a continuous range of $\gby$ 
satisfying these LHC bounds (and other constraints) simultaneously. This 
prompts us to scan over all three relevant BLSSM-specific coupling parameters, 
$\gby$, $\gyb$ and $\gbl$. 
%

\begin{figure}[t!]
\begin{center}
\hspace{-1cm}\includegraphics[width=8.1cm,height=5.8cm]{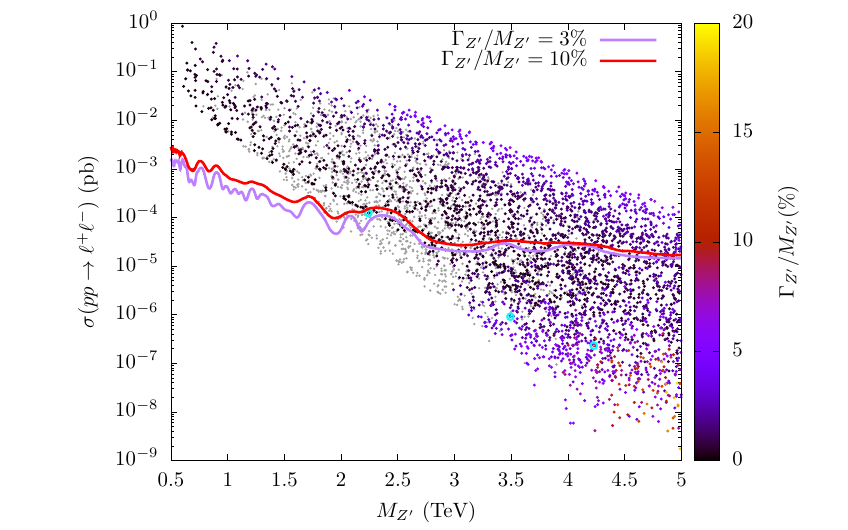}
\includegraphics[width=8.1cm,height=5.8cm]{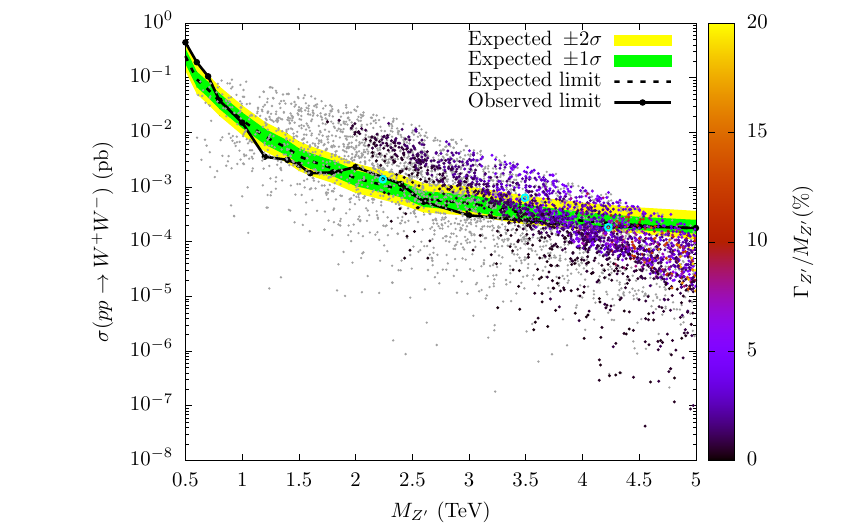}\\
\hspace{-0.48cm}\includegraphics[width=8.6cm,height=5.8cm]{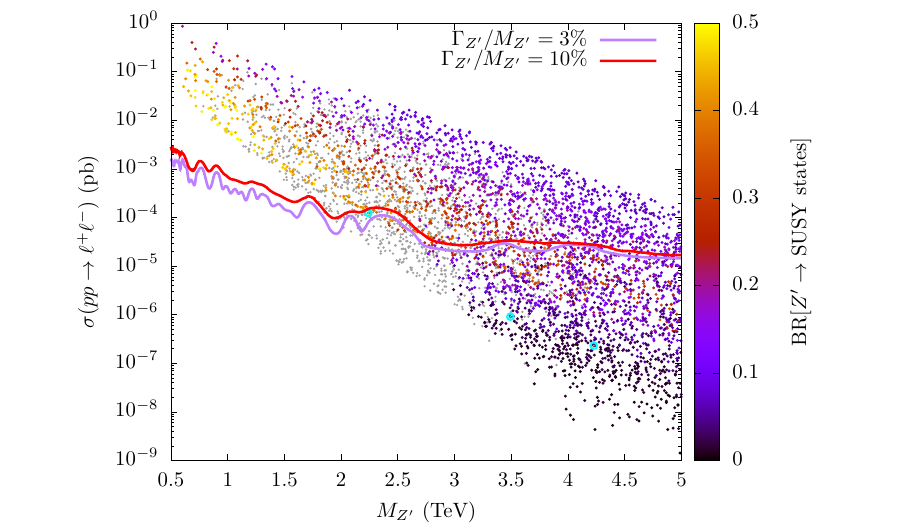}
\hspace{-0.4cm}\includegraphics[width=8.6cm,height=5.8cm]{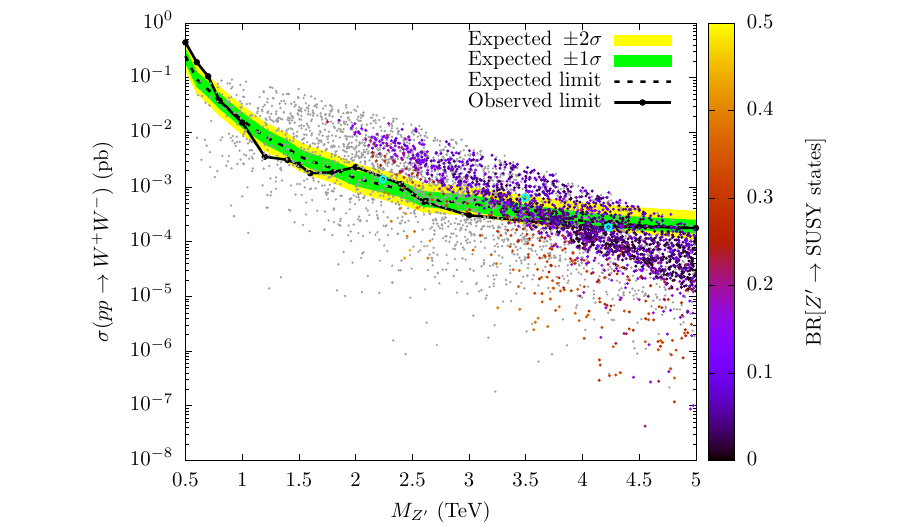}\\
\hspace{-0.85cm}\includegraphics[width=8.2cm,height=5.8cm]{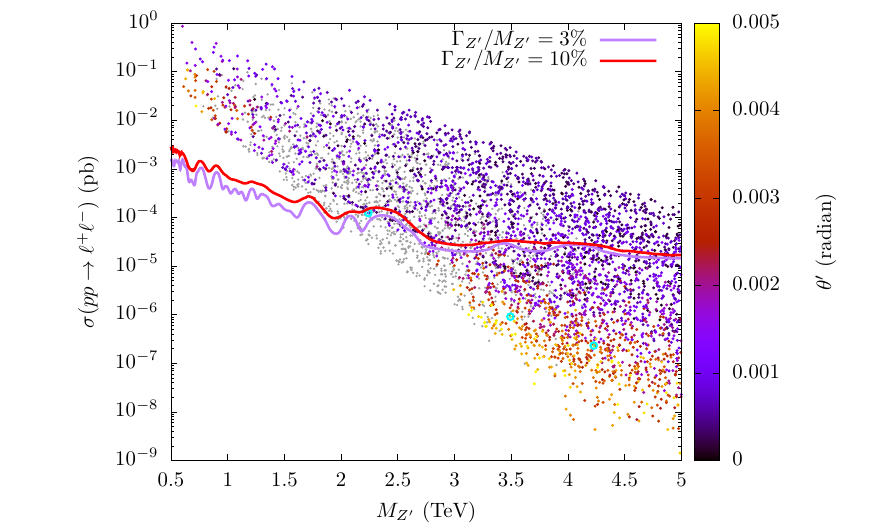}
\hspace{-0.07cm}\includegraphics[width=8.25cm,height=5.8cm]{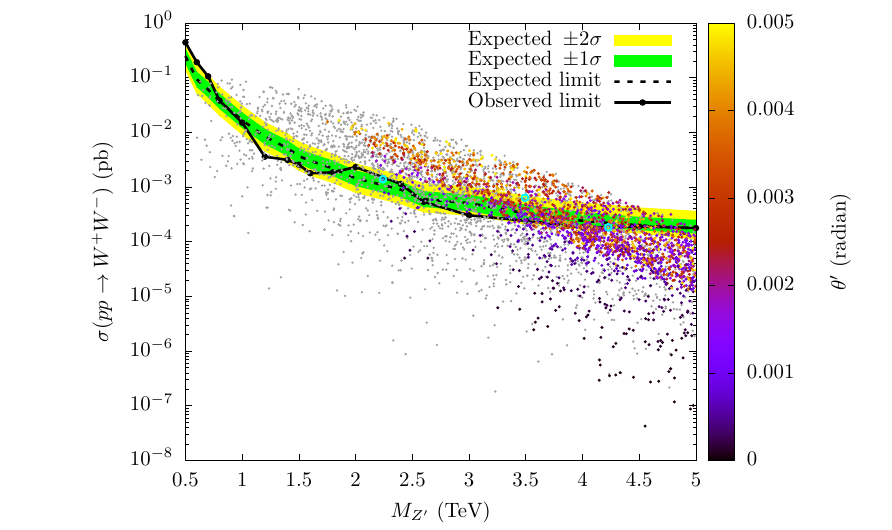} 
\caption{
Scatter plots showing dilepton (the left column) and $W^+W^-$ (the right 
column) cross sections against $\mzprime$, with the color palettes presenting 
$\gammazprime \over \mzprime$ (top panel), BR[$Z'\to$ SUSY states] (middle 
panel) and $\thetaprime$ (bottom panel), for $\mu,\mu'<M_1,M_2$, and 
contrasting them with the respective experimental upper bounds. Points in grey 
in the left (right) plots are excluded by the experimental upper bounds on 
$\sigma(pp \to Z') \times {\mathrm{BR}}(Z' \to W^+W^-)$ ($\sigma(pp \to Z') \times \mathrm{BR}(Z' \to \ell^+ \ell^-)$).
Various BLSSM-specific parameters are varied over the same ranges as in 
Fig.~\ref{Fig:MZp-thetap-GZpOMZp-1}. The cyan points are the ones that 
indicate the relaxed lower bounds on $\mzprime$ in various different 
circumstances (see Table~\ref{tab:relaxed-bounds}). See text for details.
}
\label{Fig:MZp-sigma-SUSY}
\end{center}
\end{figure}

In Fig.~\ref{Fig:MZp-sigma-SUSY} we present the same scanned points as in 
Fig.~\ref{Fig:MZp-thetap-GZpOMZp-1} and Fig.~\ref{Fig:MZp-BR-SUSY}, but only 
for the case $\mu,\mu' \ll M_1, M_2$. Hence, these plots should be seen as more 
detailed versions of the ones in the bottom panel of
Fig.~\ref{Fig:MZp-sigma-SUSY-ll-WW-Bilepton}. The plots in both upper and lower 
panels present the scanned points in the same planes as in
Fig.~\ref{Fig:MZp-sigma-SUSY-ll-WW-Bilepton}. However, while for the plots in 
the upper panel, the colors of the palettes are indicative of 
$\gammazprime \over \mzprime$, for the lower one, they correspond to
BR[$Z' \to$ SUSY states]. The points in light grey in the left (right) plots 
are disallowed by the LHC data in the $W^+W^-$ (dilepton) final state. 

The plots in the top panel of Fig.~\ref{Fig:MZp-sigma-SUSY}, when studied 
together, tells us that the scan does not reveal any notable relaxation of the 
lower bound on $\mzprime$ beyond what is indicated by the plots in the lower 
panel of Fig.~\ref{Fig:MZp-sigma-SUSY-ll-WW-Bilepton}, i.e.,
$\mzprime \gtrsim 2.5$~TeV, and re-emphasizes that such a relaxation could be 
possible only for smaller values of 
${\gammazprime \over \mzprime} \, (\lesssim 5\%)$, i.e., for a narrow
$Z'$-boson. For higher values of $\gammazprime \over \mzprime$, i.e., with an 
enhanced partial width $\gammazprimeww$, the lower bound on $\mzprime$ could at 
most be relaxed to 3.5~TeV, again in agreement with the finding from the plots 
in the lower panel of Fig.~\ref{Fig:MZp-sigma-SUSY-ll-WW-Bilepton}. Note that 
these two plots in the top panel of Fig.~\ref{Fig:MZp-sigma-SUSY} can be 
considered as the counterparts of the respective ones in the middle panel of 
Fig.~\ref{Fig:MZp-sigma}, after the imposition of stricter bounds from 
precision data and inclusion of BLSSM-specific decays of the $Z'$-boson, while 
scanning is done over the coupling parameters $\gby$ and $\gyb$, in addition to 
$\gbl$, the lone coupling parameter over which we scanned in 
Fig.~\ref{Fig:MZp-sigma}.

The plots in the middle panel of Fig.~\ref{Fig:MZp-sigma-SUSY} present the same 
set of points as the plots in the top panel, but now their colors represent the 
size of the BR of the $Z'$-boson decaying to SUSY states. These 
two plots point to  an additional, but important, finding that the size of this 
BR is anti-correlated to $\gammazprime \over \mzprime$ (see 
also the right plot of Fig.~\ref{Fig:MZp-BR-SUSY}). Furthermore, the plots on 
the left of the top and middle panels, when studied together, reveal that a 
relaxed lower bound on $\mzprime$ down to 2.5~TeV is obtained only for a
$Z'$-boson that is narrow and has a moderately large BR to SUSY 
states, which depletes the dilepton rate.

The plots in the bottom panel of Fig.~\ref{Fig:MZp-sigma-SUSY} contain the same 
set of points as in the top two panels, but colorcast to indicate the values of 
$\thetaprime$. As in Fig.~\ref{Fig:MZp-sigma}, these two plots again serve the 
purpose of consistency checks where values $\thetaprime$ are seen to be 
correlated (anti-correlated) to
$\gammazprime \over \mzprime$ (BR[$Z' \to \text{SUSY states}$]) from the plots 
in the top (middle) panel, something that is already expected on theoretical 
grounds.

In all the plots in Fig.~\ref{Fig:MZp-sigma-SUSY}, we have marked three points 
in cyan. Those indicate the relaxed lower bounds in $\mzprime$ in specific 
circumstances that we will soon elaborate on. Note that in all three cases the 
conservative lower bounds arise from studying the $W^+W^-$ final state, even as 
they all appear deep in the allowed region, as far as the dilepton final 
state is concerned.
%
\begin{figure}[t]
\begin{center}
\hspace{-1.cm}
\includegraphics[width=6.7cm,height=4.5cm]{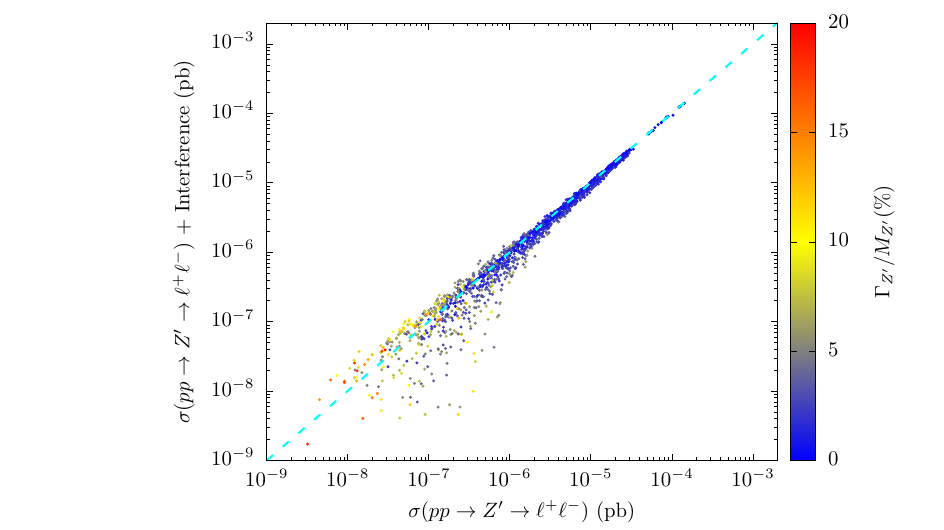}
\hspace{-0.5cm}
\includegraphics[width=5.5cm,height=4.5cm]{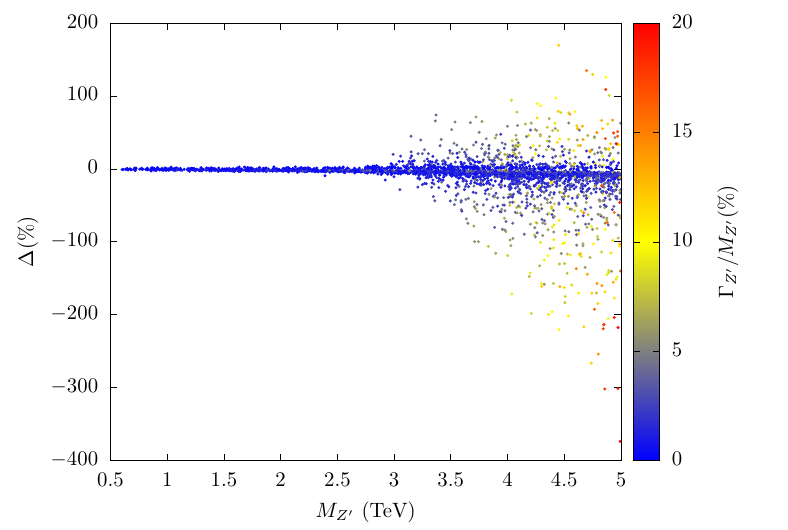}
\hspace{-0.8cm}
\includegraphics[width=5.9cm,height=4.5cm]{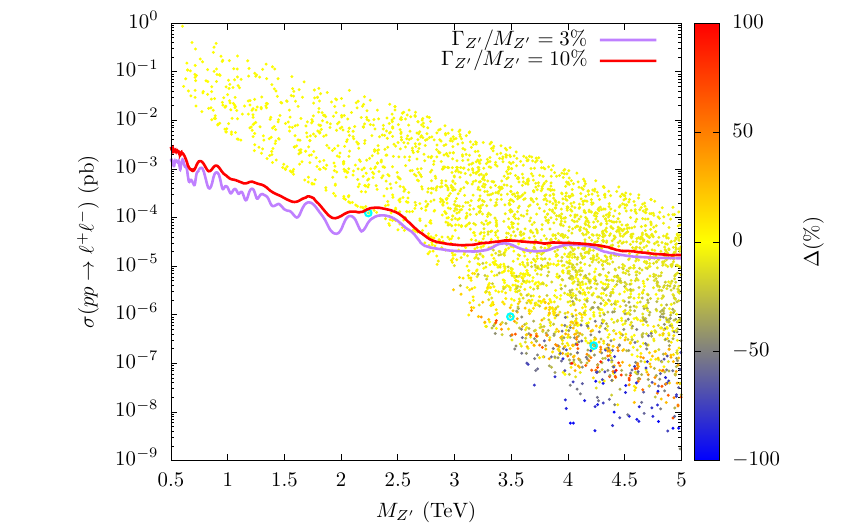} 
\caption{Left: Scatter plot in the plane of pure BLSSM (i.e., only
$Z'$-mediated) contribution to the cross sections for the dilepton 
production at the 13 TeV LHC (horizontal axis) and the same contribution plus 
the one arising from the interference of $Z'$-mediated and the SM (photon and 
$Z$-boson) diagrams (vertical axis), with $\gammazprime \over \mzprime$ (in \%) 
in the color palette. 
Middle: The same points as in the left plot presented in the 
$\mzprime$--$\Delta$ plane with $\gammazprime \over \mzprime$ in the color 
palette. 
Right: Scatter plot showing the same dilepton cross sections as functions of 
$\mzprime$ and contrasting them with the experimental upper bounds for two 
values of $\gammazprime \over \mzprime$, whereas the color palette indicates 
the magnitude of $\Delta$ (in \%, see text for details). Various BLSSM-specific 
parameters are varied over the same ranges as in 
Fig.~\ref{Fig:MZp-thetap-GZpOMZp-1}. The cyan points are the same ones that 
appear in Fig.~\ref{Fig:MZp-sigma-SUSY}.
}
\label{Fig:Interference}
\end{center}
\end{figure}

Before we summarise our findings, we briefly discuss the role of 
interference between the $Z'$-mediated diagram and the SM diagrams mediated by 
photon and the $Z$-boson in the process
$p p \stackrel[\gamma,Z]{Z'}{\longrightarrow} \ell^+ \ell^-$ (i.e., the 
dilepton final state), at the 13 TeV LHC, with the help of the plots in 
Fig.~\ref{Fig:Interference}. In the left plot, we present the scatter points of 
Fig.~\ref{Fig:MZp-sigma-SUSY} in the plane of pure $Z'$-boson contribution to 
the dilepton rate, i.e.,
$\sigma^{\ell\ell}_{Z'} = \sigma(pp \stackrel{Z'}{\longrightarrow} \ell^+\ell^-)$ (along the 
horizontal axis), and the same  plus the interference contribution from the 
$Z',\gamma, Z$ mediated processes, which is defined earlier in 
Sec.~\ref{subsec:prod-xsec}, in reference to Fig.~\ref{fig:xsec-relative}, as 
$\Delta_\text{BLSSM}$ (along the vertical axis). Note that 
$\Delta_\text{BLSSM}$ is the dominant effect at high $M_{\ell^+\ell^-}$ and 
determines the shape of the $Z'$-boson peak above the continuum SM background. 
By construction, the interference contributions vanish along the diagonal, 
while those are destructive (constructive) for points below (above) the 
diagonal. A possible destructive interference is interesting in the sense that 
it would tend to relax the lower bound of $\mzprime$.

The following set of information can be gleaned from the above-mentioned plot: 
(i)~by the time the interference effect becomes palpable, the BLSSM 
contribution to the dilepton cross section already falls below 
$\sim 10^{-6}$~pb, which is due to a heavier $Z'$-boson as the mediator,
(ii)~the magnitude of the destructive effect could be larger than its 
constructive counterpart, although such an effect becomes only pronounced when 
the cross sections themselves become smaller, and (iii) in general, the 
interference effects increase with the increasing magnitude of
$\gammazprime \over \mzprime$, i.e., for an even broader $Z'$ resonance, and 
for smaller values of the overall cross section, i.e., for larger $\mzprime$.

In the middle plot of Fig.~\ref{Fig:Interference}, we illustrate for the same 
set of points the actual magnitude and sign of the interference contributions
(`+' for constructive, and `$-$' for destructive) relative to the pure BLSSM 
($Z'$-boson) contribution $\sigma^{\ell\ell}_{Z'}$, i.e.,
$\Delta={{\Delta_\text{BLSSM} - \sigma^{\ell\ell}_{Z'}} \over \sigma^{\ell\ell}_{Z'}}$ (in \%).
We see that the $|\Delta|$ can become nominally very large for large values of 
$\mzprime$ and $\gammazprime \over \mzprime$. The plot on the right of
Fig.~\ref{Fig:Interference} is the actual limit plot, where the colors of the 
points indicate the magnitude and sign of the interference contribution, 
$\Delta$. It is clearly seen that $|\Delta|$ is larger at smaller cross 
sections that are well below the current limits shown via the red and purple 
lines. A closer inspection reveals that there are points in light grey carrying 
moderate contributions from destructive interference  ($\Delta \lesssim 50\%$) 
closely below the limit-contours for $\mzprime > 3$ TeV, which seem to be 
allowed solely due to such effects. 
%

\begin{table}
\begin{center}
\begin{tabular}{|w{c}{4.0cm}|w{c}{2.7cm}|w{c}{3.5cm}|w{c}{3.1cm}|w{c}{2.2cm}|}
\hline
 & \multicolumn{4}{c|}{Nature of the resonance} \\
 \cline{2-5}
\hspace*{-0.1cm}
Resonant & & & & \\
$Z'$ & ${\gammazprime \over \mzprime} \leq 5\%$ 
    & ${\gammazprime \over \mzprime} \in [5\%, 10\%]$
    & ${\gammazprime \over \mzprime}  \in [10\%, 20\%]$
    & ${\gammazprime \over \mzprime} \lesssim 50\%$ \\
decays & {\footnotesize (Narrow)} & {\footnotesize (Broad)} & {\footnotesize (Fat)} & {\footnotesize (Obese)} \\
 & & & & {\scriptsize * Not feasible} \\
\hline
& & & & \\
SUSY/BLSSM  & 3.15~{\rm TeV} & 3.49~{\rm TeV} & 4.23~{\rm TeV} & 4.5~{\rm TeV} \\
decays playing no role & \scriptsize(leptophobic effect) & \scriptsize (leptophobia \!+ \!\!\! width \!effect \!\!) & \scriptsize (mostly width effect) & \scriptsize (width effect) \\
 & \scriptsize $(pp\to \ell^+\ell^-)$ & \scriptsize $(pp\to \ell^+\ell^-,W^+W^-)$  &\scriptsize $(pp\to W^+W^-)$ & \scriptsize $(pp\to W^+W^-)$ \\
  &  &   & &   \scriptsize * See Fig.~\ref{Fig:MZp-sigma}\\[-0.5cm]
  & & & & \\
  & & & & \\
SUSY/BLSSM & 2.24~{\rm TeV} & $\times$ & $\times$ & $\times$ \\
decays playing a role & \scriptsize (SUSY decays of $Z'$ & & & \\
& \scriptsize + \!\!\! leptophobic effect) &  & & \\
& & & & \\[-0.5cm]
 & \scriptsize $(pp\to \ell^+\ell^-)$ &   & &  \\
  & & & & \\
\hline
\end{tabular}
\caption{
\label{tab:relaxed-bounds}
Relaxed lower bounds of $\mzprime$ (down from $\simeq 5$~TeV) for different 
regimes of $\gammazprime \over \mzprime$, and when the SUSY/BLSSM decays of the 
$Z'$-boson does not and does play a role. See text for details.
} 
\end{center}
\end{table}

In Table~\ref{tab:relaxed-bounds}, we collect the relaxed lower bounds 
mentioned above on $\mzprime$ under various situations based on whether the 
SUSY/BLSSM decays of the $Z'$-boson have any role to play and the magnitudes of
$\gammazprime \over \mzprime$. All entries for the relaxed lower bounds on 
$\mzprime$, except for the one in the rightmost column, are based on 
Fig.~\ref{Fig:MZp-sigma-SUSY}, i.e., for realistic scenarios having 
$\thetaprime \leq 0.005$ and $g_\text{avg}^\ell .\thetaprime \, \leq 0.0005$. 
In the rightmost column, we present the corresponding bound for a much larger 
value of ${\gammazprime \over \mzprime} \lesssim 50\%$, as gleaned from 
Fig.~\ref{Fig:MZp-sigma}. Note that such a large value of
$\gammazprime \over \mzprime$ is not realistic, since, in this case, we do not 
impose the above-mentioned bounds of $\thetaprime$ and
$g_\text{avg}^\ell .\thetaprime$, and the column is introduced for the sole 
purpose of comparing the effect of an increasing magnitude of 
$\gammazprime \over \mzprime$. These numbers demonstrate how, with increasing 
$\gammazprime \over \mzprime$, strengthening of the bound from the $W^+W^-$ 
final state starts to overwhelm a greater relaxation of the lower bound on 
$\mzprime$ that comes from the dilepton final state. It is also clear from 
Table~\ref{tab:relaxed-bounds} how the presence of SUSY/BLSSM decays of a 
somewhat leptophobic $Z'$-boson, and not an enhanced
$\gammazprime \over \mzprime$, brings about the maximum possible relaxation in 
the lower bound on $\mzprime$, as reported by the LHC experiments. We also mention the final states (dilepton and/or $W^+W^-$) from which each of the relaxed lower bounds on $\mzprime$ is derived.
%

\begin{figure}[t!]
\begin{center}
\hspace*{-0.50cm}
\includegraphics[width=0.290\linewidth,height=4.5cm]{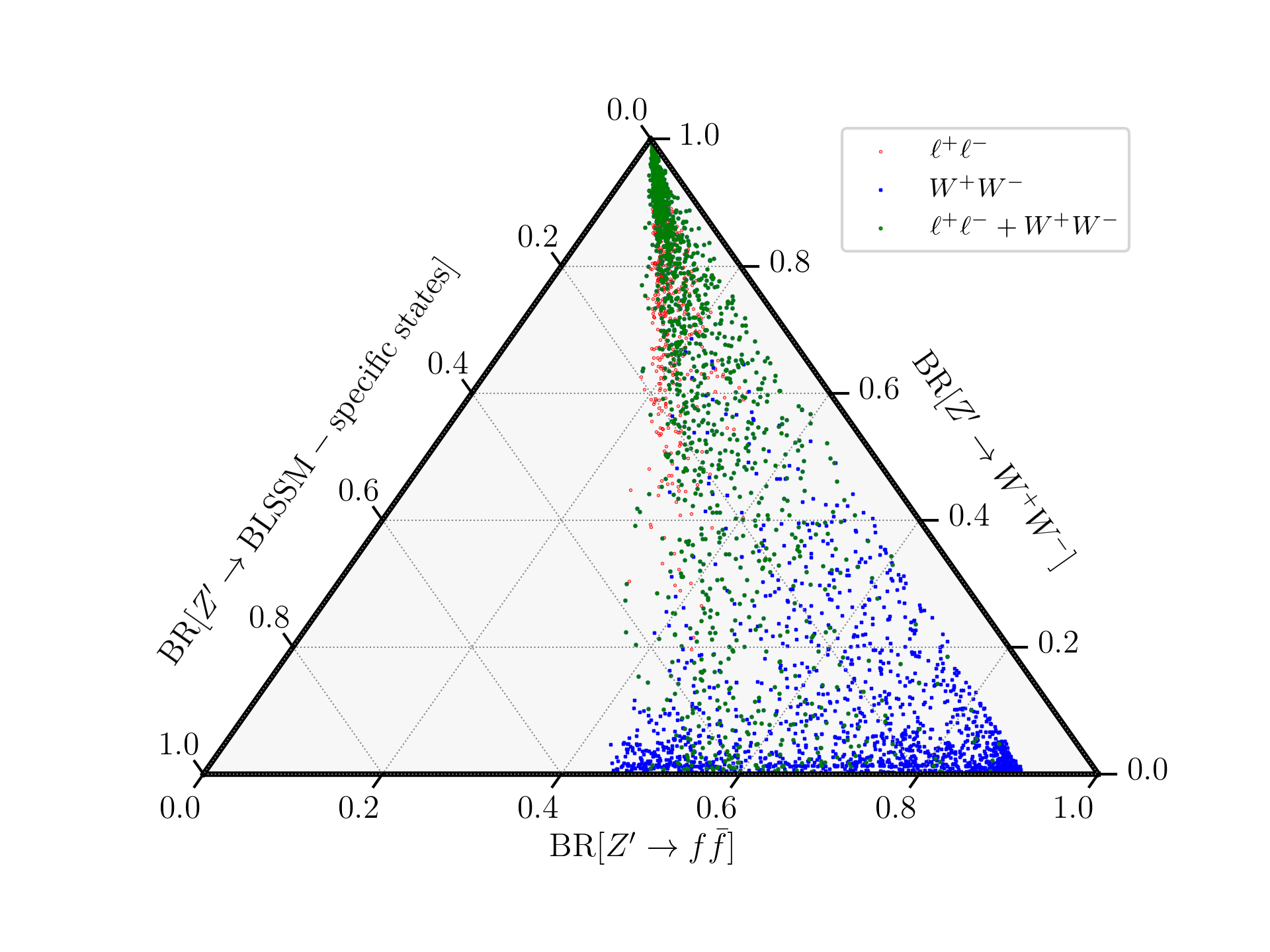}
\includegraphics[width=0.370\linewidth,height=4.5cm]{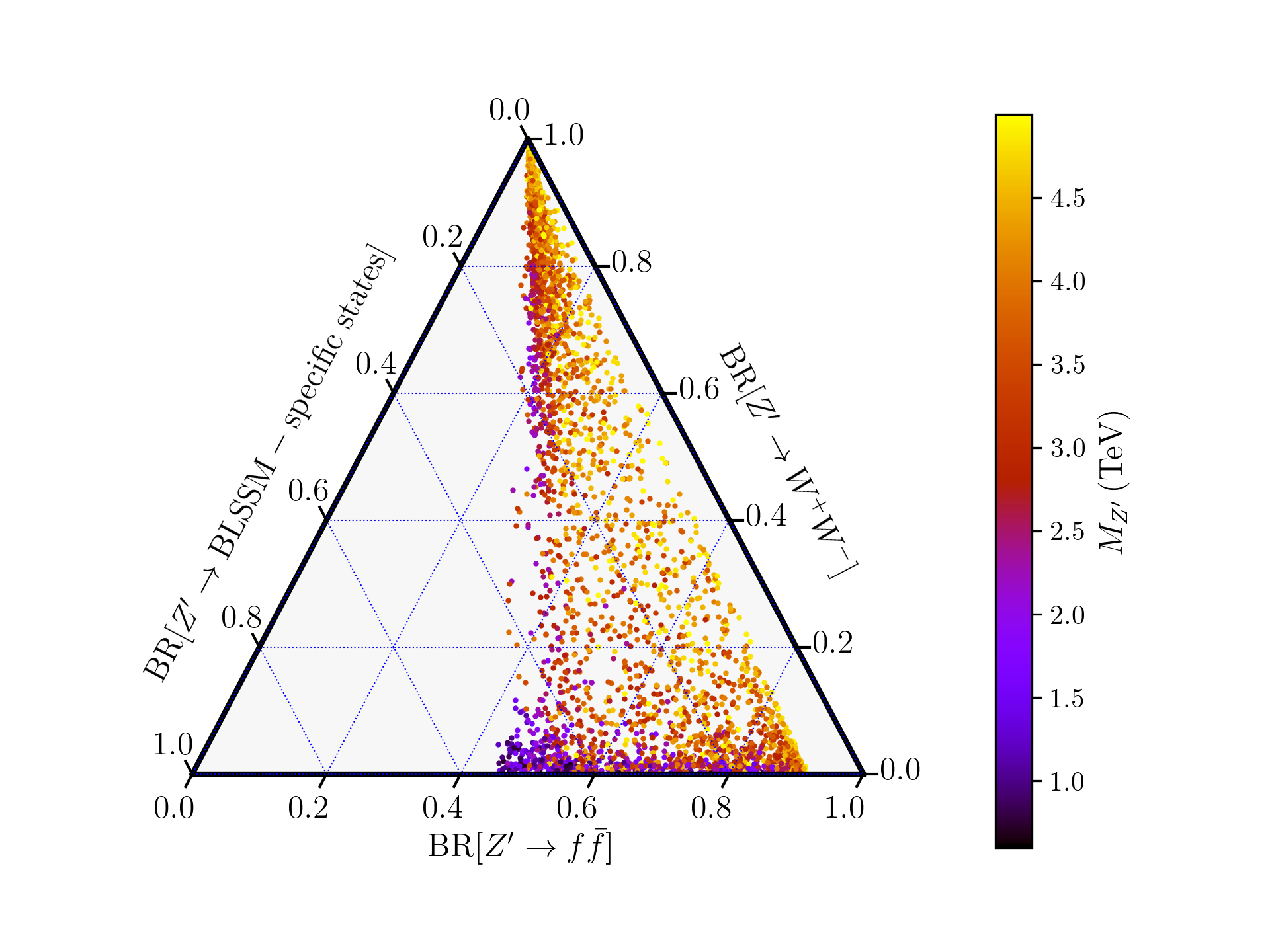}
\hspace*{-0.5cm}
\includegraphics[width=0.370\linewidth,height=4.5cm]{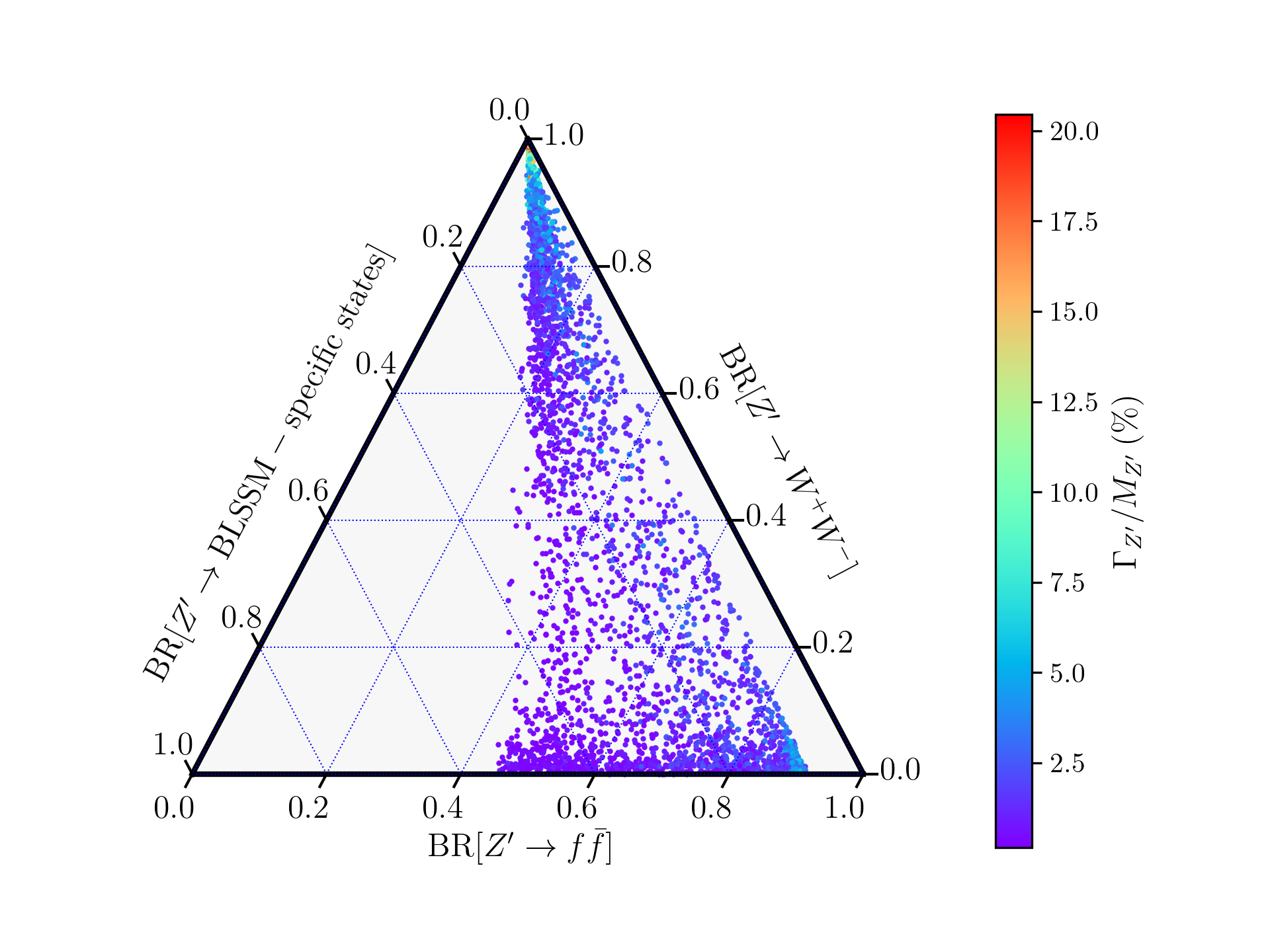}
\caption{
Ternary scatter plots showing the sharing of the BRs of the
$Z'$-boson in various modes (shown along the arms of the triangles) for points 
from Fig.~\ref{Fig:MZp-sigma-SUSY}  for which rates in the dilepton (in 
red), in $W^+W^-$ (in blue) and in both of these final states (in green) 
survive the corresponding bounds from the LHC experiments (the left plot); the 
same points with their colors indicating the values of $\mzprime$ (middle 
plot), and $\gammazprime \over \mzprime$ (the right plot) via the palettes. 
}
\label{Fig:BRs-xsec}
\end{center}
\end{figure}

Finally, we move on to study the sharing of the three primary branching 
fractions of the $Z'$-boson as described earlier in this section in reference 
to Fig.~\ref{Fig:MZp-BR-SUSY}. In the leftmost ternary plot of 
Fig.~\ref{Fig:BRs-xsec}, we present a scatter plot of these BRs
whose rates survive the concerned bounds on $\sigma \times \mathrm{BR}$ are
presented using scatter points. As may be expected, points survive bounds in 
the dilepton (in red) or the $W^+W^-$ (in blue) final states when the $Z'$-
boson BRs to these respective final states are on the smaller 
side. The green points are the sought after ones for which the reported upper 
bounds on $\sigma \times \mathrm{BR}$, for each of the final states, are 
averted once $Z'$'s decays to SUSY states are considered, thus implying a 
relaxation of the reported lower bounds on~$\mzprime$.

A priori, it might seem a little curious as to how there could be a
significant population of green points in the neighborhood of the tip of the
left ternary plot of Fig.~\ref{Fig:BRs-xsec} implying that experimental bounds
on $\sigma \times \mathrm{BR}|_{W^+W^-}$ could still be evaded even when
BR[$Z' \to W^+W^-$] is large. This could be understood by closely inspecting 
the middle and right ternary plots of Fig.~\ref{Fig:BRs-xsec}. These two plots 
indicate that for such points, $\mzprime \gtrsim 5$~TeV (middle plot) and 
${\gammazprime \over \mzprime} \gtrsim 20\%$ (right plot). As we have 
already discussed, a heavier $Z'$-boson and a larger
$\gammazprime \over \mzprime$, both affect the experimental sensitivities of 
the searches. When working in tandem, such a collective suppression could 
prevail over the gain in $\sigma \times \mathrm{BR}|_{W^+W^-}$ resulting solely 
from an enhanced $Z'W^+W^-$ coupling. This helps evade the experimental upper 
bounds on $\sigma \times \mathrm{BR}|_{W^+W^-}$ thus leaving a population of 
green points for large values of BR[$Z' \to W^+W^-$].
%

\vspace{-0.4cm}
\section{Conclusions}
\label{Sec:4}
We have studied the phenomenology of the $Z'$-boson of the BLSSM scenario at 
the 13~TeV LHC in two different, hitherto unexplored regimes, with an emphasis 
on how the recent lower bounds on $\mzprime$ ($\gtrsim 5$~TeV) could be evaded: 
first, when it is a broad resonance, and then, in a situation where it has an 
appreciable decay BR to possibly light SUSY states, viz., the 
EWinos.

Regarding a broad $Z'$-boson, we first study how `fat' it could get in the 
BLSSM. We find that the broadness of the $Z'$-boson, parametrized by the 
quantity  $\gammazprime \over \mzprime$, is mainly determined by the partial 
width $\gammazprimeww$ which is controlled by the $Z$--$Z'$ mixing angle,
$\thetaprime$. We have analysed in detail how the much relevant observables 
like $\mzprime$, $\gammazprime \over \mzprime$ and $\thetaprime$ vary as 
functions of the BLSSM-specific gauge coupling parameters like $\gbl$, 
$\gby$ and $\gyb$. Further, we have studied the mutual relationships of these 
observables over the space of the said coupling parameters. However, in the 
BLSSM, the $Z'$-boson cannot be indefinitely wide given a non-negotiable upper 
bound on $\thetaprime$ arising from the measured value of the oblique
$T$-parameter in the precision $Z$-pole data obtained at the LEP and SLC 
experiments.

However, to gain a thorough understanding of the mutual relationship of the 
said observables as the BLSSM gauge couplings vary, we have first considered a 
relaxed upper bound on $\thetaprime \, \leq 0.01$, before adopting the stricter 
bound (as mentioned above) of $\thetaprime \, \leq 0.005$, for presenting our 
final results. At the same time, to comply with the upper bound on the strength 
of the effective leptonic coupling of $Z'$-boson, as obtained from various
$Z$-pole observables at the LEP and SLC experiments, we have required 
${g_\text{avg}^\ell}$.$\thetaprime \leq 0.0005$. Further, we have exploited a 
somewhat leptophobic $Z'$-boson (i.e., a smaller ${g_\text{avg}^\ell}$) that is 
attainable in the BLSSM, which allows us to use a $\thetaprime$ value up to 
0.005, thus aiding a wider $Z'$-boson. Note that while such a leptophobic
$Z'$-boson could help it evade the current lower bound on its mass as obtained 
from the dilepton final state at the LHC, the enhanced value of 
$\thetaprime$ that we can otherwise afford now, and hence a broader $Z'$-boson, 
would start drawing constraints from the $W^+W^-$ final state because of its 
enhanced rate.

Thus, the dilepton and $W^+W^-$ final states play complementary roles in the 
search for the $Z'$ resonance in the BLSSM scenario, i.e., while for smaller
values of $\gammazprime \over \mzprime$, more stringent lower bounds on the 
$\mzprime$ come from searches of the $Z'$-boson in the dilepton final state, 
the same would arise from searches in the $W^+W^-$ final state when 
$\gammazprime \over \mzprime$ is on the larger side. Conversely, a larger 
(smaller) relaxation of the current lower bound on $\mzprime$ occurs in the 
dilepton ($W^+W^-$) final state for a larger (smaller) value of
$\gammazprime \over \mzprime$. These compel us to ensure simultaneous 
compliance of the reported bounds on the rates in both the final states, and 
consider the larger of the minimum values of $\mzprime$ that are still allowed 
in these final states to be the relaxed lower bound on $\mzprime$, which is 
smaller than the lower bound on it ($\sim 5$~TeV, at 95\% CL ($\sim 2\sigma$)), 
as reported by the LHC experiments from their analyses in each of the two final 
states.

Thus, in the absence of any appreciable branching of a resonant $Z'$-boson to 
SUSY and/or BLSSM-specific states, we find the relaxed lower bounds on 
$\mzprime$ to be on the ballparks of 3.15~TeV, 3.49~TeV and 4.23~TeV, when the 
values  of $\gammazprime \over \mzprime$ range over $\leq 5\%$, 5\%--10\% and 
10\%--20\%, respectively. In the first case with
${\gammazprime \over \mzprime} \leq 5\%$, for which the $Z'$ resonance is 
customarily treated as a narrow one, the strongest lower bound on $\mzprime$ 
comes from the dilepton final state, where the relaxation results from the 
weakened (leptophobic) coupling of the $Z'$-boson to the leptons. In the other 
two cases with larger values of $\gammazprime \over \mzprime$, the relaxation 
in the lower bound on $\mzprime$ becomes progressively weaker. This is because 
these bounds arise from the $W^+W^-$ final state, for which the loss in 
sensitivity due to broadness of the resonance is partially mitigated by 
increased $W^+W^-$ production rates since, for a given $\mzprime$, both the 
broadness ($\gammazprime$) and the said rate depend directly on the $Z'W^+W^-$ 
coupling. 

In contrast, the maximum relaxation in the lower bound on $\mzprime$, 
down to $\sim 2.24$~TeV, is realized in the presence of appreciable 
branchings of a resonant $Z'$-boson to SUSY and/or BLSSM-specific states, for 
a moderately leptophobic $Z'$-boson. A finite/modest BR of the 
$Z'$-boson to these states, which could go up to $\sim 50\%$, would already 
suppress its leptonic BR and hence the dilepton yield. On 
top of this, presence of moderate leptophobia in the $Z'$-boson would further 
diminish the same, thus weakening the lower bound on $\mzprime$ even further.
It should also be noted that a meaningful sharing of the BRs of 
the $Z'$-boson between the leptonic final states and the BLSSM-specific ones 
only happens for a narrow $Z'$ resonance, i.e., when $\thetaprime$ and hence 
the $Z'W^+W^-$ coupling is small. As a result, such a drastic relaxation in the 
lower bound of $\mzprime$ is possible exclusively for a narrow $Z'$-boson. A destructive interference between the SM and BLSSM processes could at best play a mild to moderate role in such relaxations, only for not so heavy a $Z'$-boson.

We thus conclude that relaxations in the lower bound on $\mzprime$ to different 
extents are possible in the BLSSM scenario, which could accommodate a broad
$Z'$-boson, or a leptophobic one and/or realistic BLSSM spectra which allow for 
new massive states to which a heavier $Z'$-boson could decay to and are not yet 
constrained by experiments including the LHC. These not only open up the 
possibility of existence of a $Z'$-boson with a mass significantly below the 
current LHC limit which it might have missed, but also simultaneously seeing 
the same decaying to new states from beyond the SM.
%
\section*{Acknowledgements}
AD thanks M. C. Kumar and Anurag Tripathi for useful discussions. WA and AD 
acknowledge the use of the High Performance Computing facility at HRI and thank 
Chandan Kanaujiya and Ravindra Yadav for technical help. The work of WA is 
partially supported by the Science, Technology and Innovation Funding Authority 
(STDF) under grant number 50806. SM is supported in part through the NExT
Institute and the Science and Technology Facilities Council (STFC) Consolidated Grant ST/X000583/1. 
%
\appendix*
\section{Couplings of the $Z'$-boson to Higgs and SUSY states of the BLSSM}
We list below the tree-level couplings of the $Z'$-boson to various Higgs and 
SUSY states of interest. We extract these forms from the Lagrangian of the 
BLSSM scenario as implemented in the package {\tt SARAH}~\cite{Staub:2015kfa}.
\begin{enumerate}
\item Couplings of the $Z'$-boson to various neutral ($CP$-even ($H$) and
$CP$-odd ($A$)) physical Higgs states of the BLSSM.
\begin{itemize}
\item
Couplings of the $Z'$-boson to $Z$-boson and $H_i$ \big($Z'$-$Z$-$H_i$\big):
\begin{align}
Z'Z H_i & : -\frac{1}{4}  \bigg[ \Big\{ \big(g_1 s_{\theta_W}+g_2 c_{\theta_W} \big)^2-\gyb^2 \Big\} s_{2\thetaprime} \nonumber \\
& \hskip 30pt +2 \gyb  \big(g_1 s_{\theta_W}+g_2 c_{\theta_W} \big)c_{2\thetaprime} \bigg] \Big(v_1 Z^H_{i1}+v_2 Z^H_{i2} \Big) \nonumber \\
&~~  +  \Big[ \big(\gbl^2 -\gby^2 s^2_{\theta_W} \big) s_{2\thetaprime} -2 \gbl \gby s_{\theta_W} c_{2\thetaprime} \Big] \Big(v'_1 Z^H_{i3}+v'_2 Z^H_{i4}\Big) \; ,
\end{align}
where, $Z_{ij}^H \; (i, j = 1,2,3, 4)$ stands for the elements of the mixing matrix $Z^H$ which diagonalizes the $CP$-even mass-squared matrix of
Eq.~(\ref{eqn:cpeven-matrix}).
\item Couplings of the $Z'$-boson to $A_i$ and $H_j$ \big($Z'$-$A_i$-$H_j$\big):
\begin{align}
Z'A_i H_j & : \frac{i}{2} 
\bigg[\{  (g_1 s_{\theta_W}  + 
        g_2 c_{\theta_W}) s_{\thetaprime} +\gyb c_{\thetaprime}\} \left(Z^A_{i1} Z^H_{j1}-Z^A_{i2} Z^H_{j2}\right) \nonumber \\
&~~   + 2 ( \gby s_{\theta_W} s_{\thetaprime} +\gbl c_{\thetaprime} ) \left(Z^A_{i3} Z^H_{j3} - Z^A_{i4} Z^H_{j4}\right)\bigg],
\end{align}
where, $Z_{ij}^A \; (i, j = 1,2,3, 4)$ stands for the elements of the mixing matrix $Z^A$ which diagonalizes the $CP$-odd mass-squared matrix of Eq.~(\ref{eqn:cpodd-matrix}).
%
\end{itemize}
\item Couplings of the $Z'$-boson to the EWinos ($\chi$) follow from the Lagrangian term
\begin{equation}
{\cal{L}}_{Z' \chi_i \chi_j} \equiv \bar{\chi}_i \gamma^\mu \big(c_{L_{{f}_{ij}}} P_L + c_{R_{f_{ij}}} P_R \big) \chi_j Z'_\mu \, ,
\end{equation}
where the projection operators $P_{L,R}$ are given by
$P_{L,R} \equiv (1 \mp \gamma_5)/2$. In the following, we present the
coefficients $c_{L,R}$ for $Z'$-boson's couplings to different genres of 
fermions.
%
\begin{itemize}
\item Couplings of the $Z'$-boson to the neutralinos
\big($Z'$-$\tilde \chi^0_i$-$\tilde \chi^0_j$ \big):
\begin{align} 
c_{L\tilde \chi^0_{ij}} =& \frac{1}{2} \Bigg[N^*_{j 3} \Big\{ \big(g_1   s_{\theta_W}   + g_2 c_{\theta_W}  \big)s_{\thetaprime}   + \gyb  c_{\thetaprime}  \Big\} N_{{i 3}} \nonumber \\ 
 &-N^*_{j 4} \big(g_1 s_{\theta_W}  s_{\thetaprime}   + g_2 c_{\theta_W}  s_{\thetaprime}   + \gyb c_{\thetaprime}  \big)N_{{i 4}} \nonumber \\ 
 &+2 \big(\gby s_{\theta_W}s_{\thetaprime}+
 \gbl c_{\thetaprime} \big)   \big(N^*_{j 6} N_{{i 6}}  - N^*_{j 7} N_{{i 7}} \big) \Bigg]\, ,
 \label{eqn:zpninjl}\\ 
 c_{R\tilde \chi^0_{ij}} =&  \,-\frac{1}{2} \Bigg[N^*_{i 3} \Big\{ \big(g_1 s_{\theta_W}   + g_2 c_{\theta_W}  \big)s_{\thetaprime}   + \gyb c_{\thetaprime}  \Big\} N_{{j 3}} \nonumber \\ 
 &-N^*_{i 4} \big(g_1 s_{\theta_W}  s_{\thetaprime}   + g_2 c_{\theta_W}  s_{\thetaprime}   + \gyb c_{\thetaprime}  \big)N_{{j 4}} \nonumber \\ 
 &+2 \big( \gby s_{\theta_W} s_{\thetaprime}+\gbl c_{\thetaprime} \big)  \big(N^*_{i 6} N_{{j 6}}  - N^*_{i 7} N_{{j 7}} \big)\Bigg] \, ,
 \label{eqn:zpninjr}
 \end{align} 
where, $N_{kl}$ are the elements of the symmetric, unitary $7 \times 7$ matrix
`$N$' that diagonalizes the neutralino mass matrix of Eq.~(\ref{neutralino-mass-matrix}).
%
\item Couplings of the $Z'$-boson to the charginos \big($Z'$-$\tilde \chi^+_i$-$\tilde \chi^-_j$\big):
\begin{align} 
c_{L\tilde \chi^\pm_{ij}} =\, - \frac{1}{2} \Bigg[2 g_2 U^*_{j 1} c_{\theta_W}  s_{\thetaprime}  U_{{i 1}}
 -U^*_{j 2} \Big\{ \big( g_1 s_{\theta_W}   - g_2 c_{\theta_W}  \big) s_{\thetaprime}   + \gyb c_{\thetaprime}  \Big\} U_{{i 2}} \Bigg] \, , 
\label{eqn:zpcicjl} \\ 
c_{R\tilde \chi^\pm_{ij}}  =  \,-\frac{1}{2} \Bigg[2 g_2 V^*_{i 1} c_{\theta_W}  s_{\thetaprime}  V_{{j 1}} 
 -V^*_{i 2} \Big\{ \big( g_1 s_{\theta_W}   - g_2 c_{\theta_W} 
 \big) s_{\thetaprime}   + \gyb c_{\thetaprime} \Big\} V_{{j 2}} \Bigg] \, ,
\label{eqn:zpcicjr}
\end{align}
where, $U_{kl}$ and $V_{mn}$ are unitary $2 \times 2$ matrices that diagonalize 
the asymmetric chargino mass matrix of Eq.~(\ref{chargino-mass-matrix}).
\end{itemize}
%
\item Couplings of the $Z'$-boson to the sfermions follow from the Lagrangian term
\begin{equation}
{\cal L}_{Z' \tilde{f} \tilde{f}^*} \equiv c_{\tilde f,ij} \tilde f_i  \tilde f_j^* 
\big(p_{\tilde{f}_i}^{\mu}- p_{\tilde{f}^*_j}^{\mu} \big) Z'_\mu\, ,
\end{equation}
where, $p_{\tilde{f}_i}$ and $p_{\tilde{f}^*_j}$ are the four-momenta of the 
sfermions. 

In the following, we define the coefficients $c_{\tilde f,ij}$ for various such 
couplings, wherein $Z^{\tilde{f}}$ denotes the ($6 \times 6$) matrix needed to 
diagonalize the mass-squared matrix of the sfermion $\tilde{f}$ written in the 
generic basis $\big\{ \tilde{f_i}_L, \tilde{f_i}_R \big\}$, where
$i \,(= 1,2,3)$ is the generation index and $L$ and $R$ stand for the left and 
right chiralities, respectively.
%
\begin{itemize}
\item Couplings of the $Z'$-boson to the charged sleptons \big($Z'$-$\tilde{\ell}_i$-$\tilde{\ell}^*_j$\big):
\begin{align}
c_{\tilde{\ell},ij} =&\frac{1}{2} \Bigg[\Big\{\big[\big(\gby+g_1\big) s_{\theta_W}- g_2 c_{\theta_W}  \big] s_{\thetaprime}   + \big(\gyb + \gbl \big)c_{\thetaprime} \Big\} \sum_{a=1}^{3}Z^{\tilde{\ell}^*}_{i a} Z_{j a}^{\tilde{\ell}}  \nonumber \\ 
 &+\Big\{ \big(\gby+2 g_1\big) s_{\theta_W}  s_{\thetaprime}   + \big (2 \gyb  + \gbl \big)c_{\thetaprime}  \Big\} \sum_{a=1}^{3}Z^{\tilde{\ell}^*}_{i, 3 + a} Z_{j, 3 + a}^{\tilde{\ell}}  \Bigg]\, . 
\end{align}
\item Couplings of the $Z'$-boson to the pseudoscalar and scalar sneutrinos \big($Z'$-$\tilde \nu^P_i$- $\tilde \nu_j^{S*}$\big):
\begin{align} 
c_{\tilde\nu, ij} =&-\frac{i}{2} \Bigg[ \Big\{ \big[\big(\gby+g_1\big) s_{\theta_W}   + g_2 c_{\theta_W}  \big] s_{\thetaprime}   + \big( \gyb + \gbl \big) c_{\thetaprime}  \Big\} \sum_{a=1}^{3}Z^{\tilde{\nu}^{P*}}_{i a} Z^{\tilde{\nu}^{S*}}_{j a}  \nonumber \\ 
 &+ \big(\gby s_{\theta_W}  s_{\thetaprime} +\gbl c_{\thetaprime}\big) \sum_{a=1}^{3}Z^{\tilde{\nu}^{P*}}_{i, 3 + a} Z^{\tilde{\nu}^{S*}}_{j, 3 + a}  \Bigg]\,.
\end{align}  
\item Couplings of the $Z'$-boson to the up-type squarks \big($Z'$-$\tilde u_{i \alpha}$-$\tilde u_{j \beta}^*$\big):
\begin{align} 
c_{\tilde u,ij} =&-\frac{1}{6} \delta_{\alpha \beta} \Bigg[ \Big\{ \big[\big(\gby+g_1\big) s_{\theta_W}-3 g_2 c_{\theta_W}   \big] s_{\thetaprime}  + \big( \gyb + \gbl \big) c_{\thetaprime}  \Big\} \sum_{a=1}^{3}Z^{U*}_{i a} Z_{j a}^{U}  \nonumber \\ 
 &+\Big\{\big(\gby+ 4 g_1\big) s_{\theta_W}  s_{\thetaprime}   + \big( 4 \gyb  + \gbl \big) c_{\thetaprime}  \Big\} \sum_{a=1}^{3}Z^{U*}_{i, 3 + a} Z_{j, 3 + a}^{U}  \Bigg]\, ,
\end{align} 
where, $\alpha$ and $\beta$ are the color indices.
\item Couplings of the $Z'$-boson to the down-type squarks \big($Z'$-$\tilde d_{i \alpha}$-$\tilde d_{j \beta}^*$\big):
\begin{align} 
c_{\tilde d,ij} =&-\frac{1}{6} \delta_{\alpha \beta} \Bigg[ \Big\{ \big[\big(\gby+g_1\big) s_{\theta_W}+3 g_2 c_{\theta_W}   \big] s_{\thetaprime}   + \big( \gyb + \gbl \big)c_{\thetaprime}  \Big\} \sum_{a=1}^{3}Z^{D*}_{i a} Z_{{j a}}^{D}  \nonumber \\ 
 &+\Big\{ \big(\gby- 2 g_1\big) s_{\theta_W}  s_{\thetaprime}   + \big( -2 \gyb  + \gbl \big) c_{\thetaprime}   \Big\} \sum_{a=1}^{3}Z^{D*}_{i, 3 + a} Z_{j, 3 + a}^{D}  \Bigg]\,.\end{align} 
\end{itemize}
%
 \end{enumerate}
 %

\end{document}